\documentclass[twocolumn,showkeys,showpacs,preprintnumbers,prd,superscriptaddress,nofootinbib]{revtex4-1}

\usepackage{graphicx}
\usepackage{epsf}
\usepackage{bm}
\usepackage{amsmath}
\usepackage{amsfonts}
\usepackage{amssymb}
\usepackage{epstopdf}
\usepackage{natbib}
\usepackage{color}
\usepackage{verbatim}
\usepackage{multirow}
\usepackage{bm}
\usepackage{hyperref}
\usepackage{xcolor}
\usepackage{mathrsfs}
\usepackage{alphalph}
\hypersetup{colorlinks   = true,
            urlcolor     = blue,
            citecolor    = blue,
            linkcolor    = blue,
            menucolor    = blue,
            anchorcolor  = blue,
            filecolor    = blue}
            
\date{\today}

\makeatletter
\newcommand{\fmarki}{\ensuremath{\alpha}}
\newcommand{\fmarkii}{\ensuremath{\beta}}
\newcommand{\fmarkiii}{\ensuremath{\gamma}}
\newcommand{\fmarkiv}{\ensuremath{\delta}}
\newcommand{\fmarkv}{\ensuremath{\epsilon}}
\newcommand{\fmarkvi}{\ensuremath{\zeta}}
\newcommand{\fmarkvii}{\ensuremath{\eta}}
\newcommand{\fmarkviii}{\ensuremath{\theta}}
\newcommand{\fmarkix}{\ensuremath{\iota}}
\newcommand{\fmarkx}{\ensuremath{\kappa}}
\newcommand{\fmarkxi}{\ensuremath{\lambda}}
\newcommand{\fmarkxii}{\ensuremath{\mu}}
\newcommand{\fmarkxiii}{\ensuremath{\nu}}
\newcommand{\fmarkxiv}{\ensuremath{\xi}}
\newcommand{\fmarkxv}{\ensuremath{o}}
                
\def\@fnsymbol#1{{\ifcase#1\or \fmarki\or \fmarkii\or \fmarkiii\or \fmarkiv\or \fmarkv\or \fmarkvi\or \fmarkvii\or \fmarkviii\or \fmarkix\or \fmarkx\or \fmarkxi\or \fmarkxii\or \fmarkxiii\or \fmarkxiv\or \fmarkxv\or \else\@ctrerr\fi}}
\makeatother

\makeatletter
\def\thickhline{%
  \noalign{\ifnum0=`}\fi\hrule \@height \thickarrayrulewidth \futurelet
   \reserved@a\@xthickhline}
\def\@xthickhline{\ifx\reserved@a\thickhline
               \vskip\doublerulesep
               \vskip-\thickarrayrulewidth
             \fi
      \ifnum0=`{\fi}}
\makeatother

\newlength{\thickarrayrulewidth}
\begin{document}

\preprint{\hfill UCI-HEP-TR-2022-07}

\title{Horizon-scale tests of gravity theories and fundamental physics\\from the Event Horizon Telescope image of Sagittarius A$^*$}

\author{Sunny Vagnozzi}
\email{sunny.vagnozzi@unitn.it}
\thanks{Corresponding author}
\affiliation{Department of Physics, University of Trento, 38123 Povo (TN), Italy}
\affiliation{Trento Institute for Fundamental Physics and Applications-INFN, 38123 Povo (TN), Italy \looseness=-1}
\affiliation{Kavli Institute for Cosmology, University of Cambridge, Cambridge CB3 0HA, United Kingdom \looseness=-1}

\author{Rittick Roy}
\email{20210190048@fudan.edu.cn}
\affiliation{Center for Field Theory and Particle Physics, Fudan University, 200438 Shanghai, P.\ R.\ China}
\affiliation{Department of Physics, Fudan University, 200438 Shanghai, P.\ R.\ China}

\author{Yu-Dai Tsai}
\email{yudait1@uci.edu}
\affiliation{Department of Physics and Astronomy, University of California, Irvine, CA 92697-4575, USA \looseness=-1}

\author{Luca Visinelli}
\email{luca.visinelli@sjtu.edu.cn}
\affiliation{Tsung-Dao Lee Institute, Shanghai Jiao Tong University, 201210 Shanghai, P.\ R.\ China}
\affiliation{School of Physics and Astronomy, Shanghai Jiao Tong University, 200240 Shanghai, P.\ R.\ China}

\author{Misba Afrin}
\email{misba@ctp-jamia.res.in}
\affiliation{Centre for Theoretical Physics, Jamia Millia Islamia, New Delhi 110025, India}

\author{Alireza Allahyari}
\email{alireza.al@khu.ac.ir}
\affiliation{Department of Astronomy and High Energy Physics, Kharazmi University, 15719-14911, Tehran, Iran \looseness=-1}
\affiliation{School of Astronomy, Institute for Research in Fundamental Sciences, P. O. Box 19395-5531, Tehran, Iran}

\author{Parth Bambhaniya}
\email{parthbambhaniya.as@charusat.ac.in}
\affiliation{International Center for Cosmology, Charusat University, Anand, Gujarat 388421, India}

\author{Dipanjan Dey}
\email{dy930229@dal.ca}
\affiliation{Department of Mathematics and Statistics, Dalhousie University, Halifax, Nova Scotia B3H 3J5, Canada}
\affiliation{International Center for Cosmology, Charusat University, Anand, Gujarat 388421, India}

\author{Sushant G. Ghosh}
\email{sghosh2@jmi.ac.in}
\affiliation{Centre for Theoretical Physics, Jamia Millia Islamia, New Delhi 110025, India}
\affiliation{School of Mathematics, Statistics, and Computer Science, University of KwaZulu-Natal, Durban 4000, South Africa \looseness=-3}

\author{Pankaj S. Joshi}
\email{pankaj.joshi@ahduni.edu.in}
\affiliation{International Centre for Space and Cosmology, Ahmedabad University, Ahmedabad, Gujarat 380009, India \looseness=-1}
\affiliation{International Center for Cosmology, Charusat University, Anand, Gujarat 388421, India}

\author{Kimet Jusufi}
\email{kimet.jusufi@unite.edu.mk}
\affiliation{Physics Department, State University of Tetovo, 1200 Tetovo, North Macedonia \looseness=-1}

\author{Mohsen Khodadi}
\email{m.khodadi@hafez.shirazu.ac.ir}
\affiliation{Department of Physics, College of Sciences, Shiraz University, Shiraz 71454, Iran}
\affiliation{Biruni Observatory, College of Sciences, Shiraz University, Shiraz 71454, Iran}
\affiliation{School of Astronomy, Institute for Research in Fundamental Sciences, P. O. Box 19395-5531, Tehran, Iran}
\affiliation{School of Physics, Institute for Research in Fundamental Sciences, P. O. Box 19395-5531, Tehran, Iran}

\author{Rahul Kumar Walia}
\email{rahulkumar@arizona.edu}
\affiliation{Department of Physics, University of Arizona, Tucson, AZ 85721, USA \looseness=-1}
\affiliation{School of Mathematics, Statistics, and Computer Science, University of KwaZulu-Natal, Durban 4000, South Africa \looseness=-3}

\author{Ali \"{O}vg\"{u}n}
\email{ali.ovgun@emu.edu.tr}
\affiliation{Physics Department, Eastern Mediterranean University, Famagusta, 99628 North Cyprus via Mersin 10, Turkey \looseness=-1}

\author{Cosimo Bambi}
\email{bambi@fudan.edu.cn}
\affiliation{Center for Field Theory and Particle Physics, Fudan University, 200438 Shanghai, P.\ R.\ China}
\affiliation{Department of Physics, Fudan University, 200438 Shanghai, P.\ R.\ China}

\begin{abstract}
\noindent Horizon-scale images of black holes (BHs) and their shadows have opened an unprecedented window onto tests of gravity and fundamental physics in the strong-field regime. We consider a wide range of well-motivated deviations from classical General Relativity (GR) BH solutions, and constrain them using the Event Horizon Telescope (EHT) observations of Sagittarius A$^*$ (Sgr A$^*$), connecting the size of the bright ring of emission to that of the underlying BH shadow and exploiting high-precision measurements of Sgr A$^*$'s mass-to-distance ratio. The scenarios we consider, and whose fundamental parameters we constrain, include various regular BHs, string-inspired space-times, violations of the no-hair theorem driven by additional fields, alternative theories of gravity, novel fundamental physics frameworks, and BH mimickers including well-motivated wormhole and naked singularity space-times. We demonstrate that the EHT image of Sgr A$^*$ places particularly stringent constraints on models predicting a shadow size larger than that of a Schwarzschild BH of a given mass, with the resulting limits in some cases surpassing cosmological ones. Our results are among the first tests of fundamental physics from the shadow of Sgr A$^*$ and, while the latter appears to be in excellent agreement with the predictions of GR, we have shown that a number of well-motivated alternative scenarios, including BH mimickers, are far from being ruled out at present.
\end{abstract}

\maketitle

\section{Introduction}
\label{sec:introduction}

Black holes (BHs) are among the most extreme regions of space-time~\cite{Schwarzschild:1916uq,Penrose:1964wq,Bambi:2017iyh}, and are widely believed to hold the key towards unraveling various key aspects of fundamental physics, including the behavior of gravity in the strong-field regime, the possible existence of new fundamental degrees of freedom, the unification of quantum mechanics and gravity, and the nature of space-time itself~\cite{Hawking:1976ra,Barack:2018yly}. We have now been ushered into an era where BHs and their observational effects are witnessed on a regular basis and on a wide range of scales. Perhaps the most impressive example in this sense are horizon-scale images of supermassive BHs (SMBHs), delivered through Very Long Baseline Interferometry (VLBI), and containing information about the space-time around SMBHs. The first groundbreaking horizon-scale SMBH images were delivered by the Event Horizon Telescope (EHT), a millimeter VLBI array with Earth-scale baseline coverage~\cite{Fish:2016jil}, which in 2019 resolved the near-horizon region of the SMBH M87$^*$~\cite{EventHorizonTelescope:2019dse,EventHorizonTelescope:2019uob,EventHorizonTelescope:2019jan,EventHorizonTelescope:2019ths,EventHorizonTelescope:2019pgp, EventHorizonTelescope:2019ggy}, before later revealing its magnetic field structure~\cite{EventHorizonTelescope:2021bee,EventHorizonTelescope:2021srq}. This was recently followed by the EHT's first images of Sagittarius A$^*$ (Sgr A$^*$), the SMBH located at the Milky Way center~\cite{EventHorizonTelescope:2022xnr,EventHorizonTelescope:2022vjs,EventHorizonTelescope:2022wok,EventHorizonTelescope:2022exc,EventHorizonTelescope:2022urf,EventHorizonTelescope:2022xqj,EventHorizonTelescope:2022gsd,EventHorizonTelescope:2022ago,EventHorizonTelescope:2022okn,EventHorizonTelescope:2022tzy}.

The main features observed in VLBI horizon-scale images of BHs are a bright emission ring surrounding a central brightness depression, with the latter related to the BH shadow (see e.g.\ Refs.~\cite{Synge:1966okc,Luminet:1979nyg,Virbhadra:1999nm,Falcke:1999pj,Claudel:2000yi,Bozza:2001xd,Virbhadra:2002ju,Bozza:2002zj,Takahashi:2004xh,Virbhadra:2008ws} for some of the earliest calculations in this sense, Refs.~\cite{Abdujabbarov:2015xqa,Tsupko:2017rdo,Medeiros:2019cde,Farah:2020jkv,Wang:2022kvg} for more recent studies on the shapes of BH shadows and related observables, and higher-order photon rings in Ref.~\cite{Tsupko:2022kwi}). On the plane of a distant observer, the boundary of the BH shadow marks the apparent image of the photon region (the boundary of the region of space-time which supports closed spherical photon orbits),~\footnote{In the case of spherically symmetric space-times, such as with Schwarzschild BHs, the photon region is simply referred to as photon sphere.} and separates capture orbits from scattering orbits: for detailed reviews on BH shadows, see e.g.\ Refs.~\cite{Cunha:2018acu,Dokuchaev:2019jqq,Perlick:2021aok,Chen:2022scf,AbhishekChowdhuri:2023ekr}. Under certain conditions and after appropriate calibration, the radius of the bright ring can serve as a proxy for the BH shadow radius, with very little dependence on the details of the surrounding accretion flow~\cite{Falcke:1999pj,EventHorizonTelescope:2019pgp,EventHorizonTelescope:2019ggy,Narayan:2019imo,Volkel:2020xlc,Bronzwaer:2021lzo,Lara:2021zth,Ozel:2021ayr,Younsi:2021dxe,Kocherlakota:2022jnz}.~\footnote{Some concerns on whether VLBI horizon-scale images are really seeing the image of the BH shadow surrounded by the photon ring were raised in Refs.~\cite{Gralla:2019xty,Gralla:2019drh,Gralla:2020yvo,Gralla:2020srx,Gralla:2020pra}, but have been addressed in most of the above works. See also theoretical concerns on the possibility of testing gravity using BH images raised in Ref.~\cite{Glampedakis:2021oie}. Finally, see Refs.~\cite{Atamurotov:2015nra,Babar:2020txt,Atamurotov:2021cgh} for further relevant work.} This is possible if \textit{i)} a bright source of photons is present and strongly lensed near the horizon, and especially \textit{ii)} the surrounding emission region is geometrically thick and optically thin at the wavelength at which the VLBI network operates. Most SMBHs we know of, including Sgr A$^*$ and M87$^*$~\cite{Narayan:2019imo}, operate at sub-Eddington accretion rates and are powered by radiatively inefficient advection-dominated accretion flows, as a result satisfying both the previous conditions.

The possibility of connecting the ring and BH shadow angular radii opens up the fascinating prospect of using BH shadows to test fundamental physics, once the BH mass-to-distance ratio is known~\cite{Johannsen:2015hib,Psaltis:2018xkc}. For a Schwarzschild BH of mass $M$ located at distance $D$, the shadow is predicted to be a perfect circle of radius $r_{\rm sh} = 3\sqrt{3}G_NM/c^2$, therefore subtending (within the small angle approximation) an angular diameter $\theta_{\rm sh} = 6\sqrt{3}G_NM/(c^2D) = 6\sqrt{3}\theta_g$, where $\theta_g=G_NM/(c^2D)$ is the angular gravitational radius of the BH.~\footnote{In what follows, we shall work in units where $4\pi\epsilon_0=\hbar=c=G_N=1$. At times we will find it convenient to work in units of BH mass, and therefore will set $M=1$, while rescaling all other parameters accordingly.} For Kerr BHs, the shadow is slightly asymmetric along the spin axis, with $r_{\rm sh}$ being marginally smaller compared to the Schwarzschild case, but still depending predominantly on the BH mass $M$, and only marginally on the (dimensionless) spin $a_{\star}$ and observer's inclination angle $i$. Crucially, the values of $\theta_{\rm sh}$ and $r_{\rm sh}$ can change considerably for other metrics, including those describing BHs in alternative theories of gravity, the effects of corrections from new physics possibly leading to violations of the no-hair theorem~\cite{Israel:1967wq,Israel:1967za,Carter:1968rr,Carter:1971zc,Robinson:1975bv}, or ``BH mimickers'', i.e.\ (possibly horizonless) compact objects other than BHs~\cite{Bambi:2008jg,Ohgami:2015nra,Shaikh:2018kfv,Cardoso:2019rvt,Bambi:2019tjh,Joshi:2020tlq,Guo:2020tgv,Visinelli:2021uve,Herdeiro:2021lwl,Saurabh:2022jjv,Cardoso:2022fbq,Tahelyani:2022uxw}. This paves the way towards tests of fundamental physics from the angular sizes of BH shadows of known mass-to-distance ratio.

\begin{figure}
\centering
\includegraphics[width=1.0\linewidth]{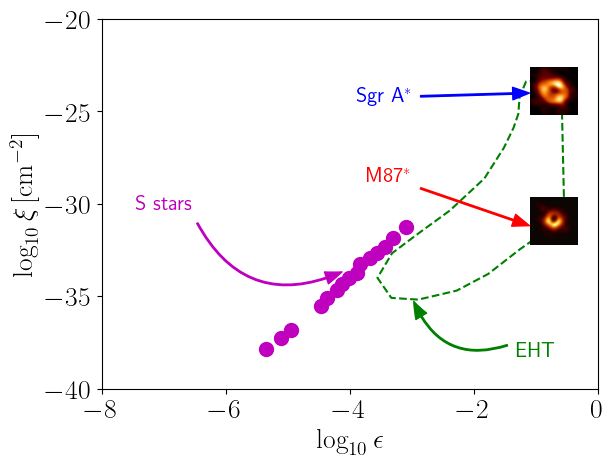}
\caption{Regimes probed by various gravitational systems of interest, represented on the Baker-Psaltis-Skordis ``curvature-potential'' diagram. The shadows of Sgr A$^*$ and M87$^*$ are indicated by the respective icons, the region bounded by the green dashed curve encloses other potential more distant EHT sources, and the magenta points correspond to the S-stars orbiting close to Sgr A$^*$. Credits for Sgr A$^*$ and M87$^*$'s images are due to the \textcopyright EHT Collaboration and \textcopyright ESO (sources: ``\href{https://eventhorizontelescope.org/blog/astronomers-reveal-first-image-black-hole-heart-our-galaxy}{Astronomers Reveal First Image of the Black Hole at the Heart of Our Galaxy}'' and ``\href{https://www.eso.org/public/images/eso1907a/}{First Image of a Black Hole}'' respectively -- both images are licensed under a \href{https://creativecommons.org/licenses/by/4.0/}{Creative Commons Attribution 4.0 International License}).}
\label{fig:potential_curvature_plot}
\end{figure}

With the EHT image of M87$^*$, the prospects of testing fundamental physics with BH shadows have become reality, as demonstrated by a large and growing body of literature devoted to such tests, mostly focusing on the size of M87$^*$'s shadow, but in some cases also considering additional observables such as the shadow circularity and axis ratio, as well as future prospects (see e.g.\ Refs.~\cite{Atamurotov:2013sca,Giddings:2019jwy,Held:2019xde,Wei:2019pjf,Tamburini:2019vrf,Konoplya:2019nzp,Shaikh:2019fpu,Contreras:2019nih,Bar:2019pnz,Jusufi:2019nrn,Vagnozzi:2019apd,Roy:2019esk,Long:2019nox,Zhu:2019ura,Contreras:2019cmf,Qi:2019zdk,Neves:2019lio,Javed:2019rrg,Tian:2019yhn,Cunha:2019ikd,Banerjee:2019nnj,Shaikh:2019hbm,Kumar:2019pjp,Allahyari:2019jqz,Li:2019lsm,Jusufi:2019ltj,Rummel:2019ads,Kumar:2020hgm,Li:2020drn,Narang:2020bgo,Liu:2020ola,Konoplya:2020bxa,Guo:2020zmf,Pantig:2020uhp,Wei:2020ght,Kumar:2020owy,Islam:2020xmy,Chen:2020aix,Sau:2020xau,Jusufi:2020dhz,Kumar:2020oqp,Chen:2020qyp,Zeng:2020dco,Neves:2020doc,Ovgun:2020gjz,Badia:2020pnh,Jusufi:2020cpn,Khodadi:2020jij,Belhaj:2020nqy,Jusufi:2020agr,Kumar:2020yem,Jusufi:2020odz,Belhaj:2020mlv,Kruglov:2020tes,EventHorizonTelescope:2020qrl,Ghasemi-Nodehi:2020oiz,Ghosh:2020spb,Khodadi:2020gns,Lee:2021sws,Contreras:2021yxe,Shaikh:2021yux,Afrin:2021imp,Addazi:2021pty,EventHorizonTelescope:2021dqv,Shaikh:2021cvl,Badia:2021kpk,Wang:2021irh,Khodadi:2021gbc,Cai:2021uov,Frion:2021jse,Wei:2021lku,Rahaman:2021web,Walia:2021emv,Afrin:2021wlj,Jusufi:2021fek,Cimdiker:2021cpz,Pal:2021nul,Li:2021ypw,Roy:2021uye,Afrin:2021ggx,Pantig:2022toh,He:2022yse,Jha:2022ewi,Meng:2022kjs,Daas:2022iid,Rosa:2022tfv,Bogush:2022hop,Chen:2022ynz,Chen:2022nbb,Narzilloev:2022bbs,He:2022opa,Zhu:2022shb,Belhaj:2022ipb,Zhang:2022osx,Hendi:2022qgi,Rayimbaev:2022hca,Tang:2022hsu,Karmakar:2022idu,Hou:2022eev,Yu:2022yyv,Li:2022eue,Lobos:2022jsz}). The new horizon-scale image of Sgr A$^*$ delivered by the EHT offers yet another opportunity for performing tests of gravity and fundamental physics in the strong-field regime, which we shall here exploit. Even though such tests have already been performed with M87$^*$, there are very good motivations, or even advantages, for independently carrying them out on the shadow of Sgr A$^*$ (see also Refs.~\cite{Psaltis:2010ca,Johannsen:2011mt,Psaltis:2015uza,Johannsen:2015mdd,Goddi:2016qax}):
\begin{enumerate}
\item Sgr A$^*$'s proximity to us makes it significantly easier to calibrate its mass and distance, and therefore its mass-to-distance ratio, whose value is crucial to connect the observed angular size of its ring to theoretical predictions for the size of its shadow within different fundamental physics scenarios. This is a distinct advantage with respect to M87$^*$, whose mass is the source of significant uncertainties, with measurements based on stellar dynamics~\cite{Gebhardt:2011yw} or gas dynamics~\cite{Walsh:2013uua} differing by up to a factor of $2$ (see also Sec.~IIIA of Ref.~\cite{Vagnozzi:2020quf}).
\item Sgr A$^*$'s mass in the ${\cal O}(10^6)M_{\odot}$ range is several orders of magnitude below M87$^*$'s mass in the ${\cal O}(10^9)M_{\odot}$ range, allowing us to probe fundamental physics in a completely different and complementary strong curvature regime (see more below and in Fig.~\ref{fig:potential_curvature_plot}). For the same reason, Sgr A$^*$'s shadow can potentially set much tighter constraints than M87$^*$'s shadow on dimensionful fundamental parameters which scale as a positive power of mass in the units we adopt (as is the case for several BH charges or ``hair parameters'').
\item Finally, independent constraints from independent sources are always extremely valuable in tests of fundamental physics, and there is, therefore, significant value in performing such tests on Sgr A$^*$ independently of the results obtained from M87$^*$. As we shall explicitly show, unlike M87$^*$, the EHT image of Sgr A$^*$ sets particularly stringent constraints on theories and frameworks which predict a shadow radius \textit{larger} than $3\sqrt{3}M$. We will consider many such examples in this work.
\end{enumerate}

To elaborate more on point 2.\ above, we consider the Baker-Psaltis-Skordis ``curvature-potential'' diagram introduced in Ref.~\cite{Baker:2014zba} to place, link, and compare on a common set of axes the regimes explored by a wide range of tests of gravity. Specifically, on the $x$ axis we plot $\log_{10}\epsilon$, with $\epsilon$ the magnitude of the Newtonian gravitational potential, quantifying deviations from the Minkowski metric. For the relevant case of a test particle orbiting a central object of mass $M$ (such as a BH) at a distance $r$, $\epsilon$ is given by (re-introducing SI units):
\begin{eqnarray}
\epsilon \equiv \frac{G_NM}{c^2r}\,.
\label{eq:epsilon}
\end{eqnarray}
The larger $\epsilon$, the stronger the gravitational field probed by an observer or a test particle, with the limit $\epsilon \to 0.5$ corresponding to a test particle approaching the event horizon -- loosely speaking, $\epsilon$ measures the ``strength'' of the metric, and in the following we will refer to it as ``potential'' (note that $\epsilon$ is simply the inverse of the radial coordinate in units of gravitational radius, and is sometimes referred to as ``compactness parameter''). On the $y$ axis we instead plot $\log_{10}\xi$, with $\xi$ the (square root of the) Kretschmann scalar which, for a Schwarzschild BH of mass $M$, reads:
\begin{eqnarray}
\xi = \left ( R_{\mu\nu\rho\sigma}R^{\mu\nu\rho\sigma} \right ) ^{\frac{1}{2}} = \frac{4\sqrt{3}G_NM}{c^2r^3}\,.
\label{eq:xi}
\end{eqnarray}
The larger $\xi$, the larger the total curvature due to both local and distant masses -- loosely speaking, $\xi$ measures the ``strength'' of the Riemann tensor, and in the following we will refer to it as ``curvature''. For illustrative purposes, given that the EHT probes the space-time around the BH photon ring, we compute Eqs.~(\ref{eq:epsilon},\ref{eq:xi}) at $r=3GM/c^2$, which is the location of the photon ring for a Schwarzschild BH, finding:
\begin{eqnarray}
\epsilon = \frac{1}{3}\,,\quad \xi \approx 1.2 \times 10^{-11} \left ( \frac{M}{M_{\odot}} \right ) ^{-2}\,{\rm cm^{-2}}\,.
\label{eq:epsilonxi}
\end{eqnarray}
While the shadows of Sgr A$^*$ and M87$^*$ probe (by construction) the same potential regime ($\log_{10}\epsilon \approx -0.48$), they probe two completely different curvature regimes, the former in the $\xi \sim {\cal O}(10^{-25})\,{\rm cm^{-2}}$ region and the latter in the much weaker $\xi \sim {\cal O}(10^{-31})\,{\rm cm}^{-2}$ region. We illustrate this visually in Fig.~\ref{fig:potential_curvature_plot}, where we show the region of potential-curvature parameter space probed by both Sgr A$^*$ and M87$^*$, as indicated by the respective icons. Following Refs.~\cite{Baker:2014zba,Johannsen:2012vz}, the region bounded by the green dashed curve encloses other potential more distant EHT sources, for which the expected resolution in units of gravitational radii is of course much lower compared to Sgr A$^*$ and M87$^*$ (hence the lower values of $\epsilon$ probed). Finally, the magenta points correspond to the S-stars orbiting close to Sgr A$^*$, following Refs.~\cite{Baker:2014zba,Gillessen:2008qv}. Fig.~\ref{fig:potential_curvature_plot} clearly allows us to appreciate how the shadows of Sgr A$^*$ and M87$^*$ probe gravity and, more generally, fundamental physics, in two completely different strong gravitational field regimes, reaffirming once more the value in performing the tests we shall conduct in this paper. We note that the tests we shall perform are parametric tests of theories of gravity and fundamental physics scenarios. In other words, such tests are performed either on specific known metric solutions of such theories, or anyhow on explicit metrics which are introduced phenomenologically (and whose origin in some cases is well rooted into specific underlying theories or frameworks). Another possibility, which we shall not pursue here, is to perform similar tests on theory-agnostic metrics which, while not arising from any specific theory, can be mapped to different theories and thus may effectively constrain a number of theories simultaneously (see e.g.\ Refs.~\cite{Nampalliwar:2021oqr,Nampalliwar:2021ytm,Kocherlakota:2022jnz} in a related context).

The rest of this paper is then organized as follows. In Sec.~\ref{sec:methodology} we discuss the methodology and assumptions entering into the computation of the sizes of BH shadows, and give a very brief overview of the fundamental physics scenarios we consider. Sec.~\ref{sec:results} is divided into a large number of subsections (from Sec.~\ref{subsec:rn} to Sec.~\ref{subsec:others}), one for each of the models and fundamental physics scenarios considered (which in some cases contain various sub-classes), for which we report constraints obtained adopting the previously discussed methodology. In Sec.~\ref{sec:briefdiscussion} we provide a brief overall discussion of our results, and a brief overview of complementary probes and future prospects. Finally, in Sec.~\ref{sec:conclusions} we draw concluding remarks and outline future directions.



\section{Methodology}
\label{sec:methodology}

The methodology we shall follow relies on comparing the observed angular radius of the ring-like feature in the EHT horizon-scale image of Sgr A$^*$ with the theoretically computed angular radius of the shadows of BHs (or other alternative compact objects) within each fundamental physics scenario we consider, with prior knowledge of Sgr A$^*$'s mass-to-distance ratio. Requiring consistency between the two quantities, within the uncertainty allowed by the EHT observations, allows us to set constraints on the parameters describing the space-time in question. This simple methodology has been discussed several times in the past (see for example Refs.~\cite{Johannsen:2015hib,Psaltis:2018xkc}) and has been successfully applied to the EHT image of M87$^*$ by the EHT collaboration themselves in Ref.~\cite{EventHorizonTelescope:2021dqv}, though for reasons discussed earlier in Sec.~\ref{sec:introduction}, the application to Sgr A$^*$ is in principle more robust. Note that, much as all related works in the literature, we will be using the observationally provided values for the mass-to-distance ratio and shadow radius implicitly assuming that their inferred values are independent of both the underlying space-time metric and/or theory of gravity. This allows us to use the same values for these quantities when moving across the different scenarios we consider.

To apply this methodology, we require two ingredients. The first is Sgr A$^*$'s mass-to-distance ratio. The mass and distance to Sgr A$^*$, $M$, and $D$, have been studied in detail over the past decades exploiting stellar cluster dynamics, and in particular the motion of S-stars, individual stars resolved within $1''$ of the Galactic Center. A significant role has been played by S0-2 which, with a $K$-band magnitude of $\sim 14$, period of $\sim 16\,{\rm yrs}$, and semimajor axis of $\sim 125\,{\rm mas}$, is the brightest star with a relatively close orbit and short period close to the Galactic Center. Its orbit has been exquisitely tracked by two sets of instruments/teams,~\footnote{The motion of the S0-2 star has been used to set constraints on various aspects of fundamental physics, see e.g.\ Refs.~\cite{deMartino:2021daj,Jusufi:2021lei,Benisty:2021cmq,Borka:2021omc,Yuan:2022nmu}.} which we shall refer to as ``Keck'' and ``VLTI'' (standing for ``Very Large Telescope Interferometer''), respectively. Following Ref.~\cite{EventHorizonTelescope:2022xqj}, we adopt the (correlated) mass and distance estimates given in Table~\ref{tab:massdistance}~\cite{Do:2019txf,GRAVITY:2020gka}, where uncertainties are quoted at $68\%$ confidence level (C.L.), reporting (where available) systematic uncertainties. We refer the reader to Sec.~2.1 of Ref.~\cite{EventHorizonTelescope:2022xqj} for detailed discussions on these measurements.
\begin{table}[tb!]
\def\arraystretch{1.5}
\begin{tabular}{cccc}
\hline\hline
Survey & $M\,(\times 10^6M_\odot)$ & $D\,$(kpc) & Reference \\
\hline
Keck & $3.951 \pm 0.047$ & $7.953 \pm 0.050 \pm 0.032$ & \cite{Do:2019txf} \\ 
VLTI & $4.297 \pm 0.012 \pm 0.040$ & $8.277 \pm 0.009 \pm 0.033$ & \cite{GRAVITY:2020gka} \\ 
\end{tabular}
\caption{Mass and distance to Sgr A$^*$ as inferred from the Keck and VLTI instruments.}
\label{tab:massdistance}
\end{table}

The second ingredient we require is a calibration factor connecting the size of the bright ring of emission with the size of the corresponding shadow, which quantifies how safe it is to use the size of the bright ring of emission as a proxy for the shadow size. This calibration factor depends on the near-horizon physics of image formation and, as already anticipated earlier, is expected to be very close to unity for optically thin, geometrically thick radiatively inefficient advection-dominated accretion flows, such as the one surrounding Sgr A$^*$. In practice, the calibration factor accounts multiplicatively for various sources of uncertainty, ranging from formal measurement uncertainties, to fitting/model uncertainties, to theoretical uncertainties pertaining to the emissivity of the plasma (see Sec.~3 of Ref.~\cite{EventHorizonTelescope:2022xqj}).

Detailed studies of various sources of uncertainty have been conducted by the EHT in Ref.~\cite{EventHorizonTelescope:2022xqj} and used to determine the above calibration factor. Folding in the calibration factor with uncertainties in Sgr A$^*$'s mass-to-distance ratio, and the angular diameter of Sgr A$^*$'s bright ring of emission, the EHT inferred $\delta$, the fractional deviation between the inferred shadow radius $r_{\rm sh}$ and the shadow radius of a Schwarzschild BH of angular size $\theta$ ($\theta_{\rm sh}=3\sqrt{3}\theta_g$), $r_{\rm sh,Schwarzschild}=3\sqrt{3}M$. In practice, $\delta$ is given by:
\begin{eqnarray}
\delta \equiv \frac{r_{\rm sh}}{r_{\rm sh,Schwarzschild}}-1=\frac{r_{\rm sh}}{3\sqrt{3}M}-1\,.
\label{eq:definitiondelta}
\end{eqnarray}
The inferred value of $\delta$ depends on the mass-to-distance ratio assumed, with the Keck and VLTI measurements resulting in the following estimates~\cite{EventHorizonTelescope:2022xqj}:
\begin{itemize}
\item \textrm{Keck}: $\delta=-0.04^{+0.09}_{-0.10}$;
\item \textrm{VLTI}: $\delta=-0.08^{+0.09}_{-0.09}$,
\end{itemize}
with a good agreement between the two bounds.

In order to be as conservative as possible, and to simplify our later discussion, we take the average of the \textrm{Keck}- and \textrm{VLTI}-based estimates of $\delta$, treating them as uncorrelated, as they are obtained from two different instruments/surveys. This leads to the following estimate of $\delta$ which we shall adopt throughout this work:
\begin{eqnarray}
\delta \simeq -0.060 \pm 0.065\,,
\label{eq:delta}
\end{eqnarray}
which, under the assumption of Gaussianity (itself supported by the shapes of the posteriors shown in Fig.~12 of Ref.~\cite{EventHorizonTelescope:2022xqj}), trivially leads to the following $1\sigma$ and $2\sigma$ intervals for $\delta$:
\begin{eqnarray}
-0.125 \lesssim \delta \lesssim 0.005 \quad &(1\sigma)&\,, \nonumber \\
-0.19 \lesssim \delta \lesssim 0.07 \quad &(2\sigma)&\,.
\label{eq:intervalsdelta}
\end{eqnarray}
As is clear from the previous \textrm{Keck}- and \textrm{VLTI}-based estimates which we have averaged, there is overall a very slight preference for Sgr A$^*$'s shadow being \textit{marginally smaller} than the prediction $3\sqrt{3}M$ for a Schwarzschild BH of given mass, with a $\lesssim 68\%$~C.L. preference for $\delta<0$. At the level of shadow radius $r_{\rm sh}$ and once more assuming Gaussian uncertainties, after expressing $r_{\rm sh}/M=3\sqrt{3}(\delta+1)$ by inverting Eq.~(\ref{eq:definitiondelta}), it is easy to see that the bounds reported in Eqs.~(\ref{eq:intervalsdelta}) translate to the following $1\sigma$ constraints on $r_{\rm sh}/M$:
\begin{eqnarray}
4.55 \lesssim r_{\rm sh}/M \lesssim 5.22\,,
\label{eq:1sigma}
\end{eqnarray}
as well as the following $2\sigma$ constraints:
\begin{eqnarray}
4.21 \lesssim r_{\rm sh}/M \lesssim 5.56\,.
\label{eq:2sigma}
\end{eqnarray}
These constraints are in good agreement with those reported in Ref.~\cite{EventHorizonTelescope:2022xqj}, though with slightly smaller uncertainties (by a factor of $\approx \sqrt{2}$) as a result of having taken the average between the \textrm{Keck}- and \textrm{VLTI}-based estimates. In what follows, we shall use the bounds in Eqs.~(\ref{eq:1sigma},\ref{eq:2sigma}) to constrain the parameters governing space-times beyond the Schwarzschild BH.

\begin{figure}
\centering
\includegraphics[width=1.0\linewidth]{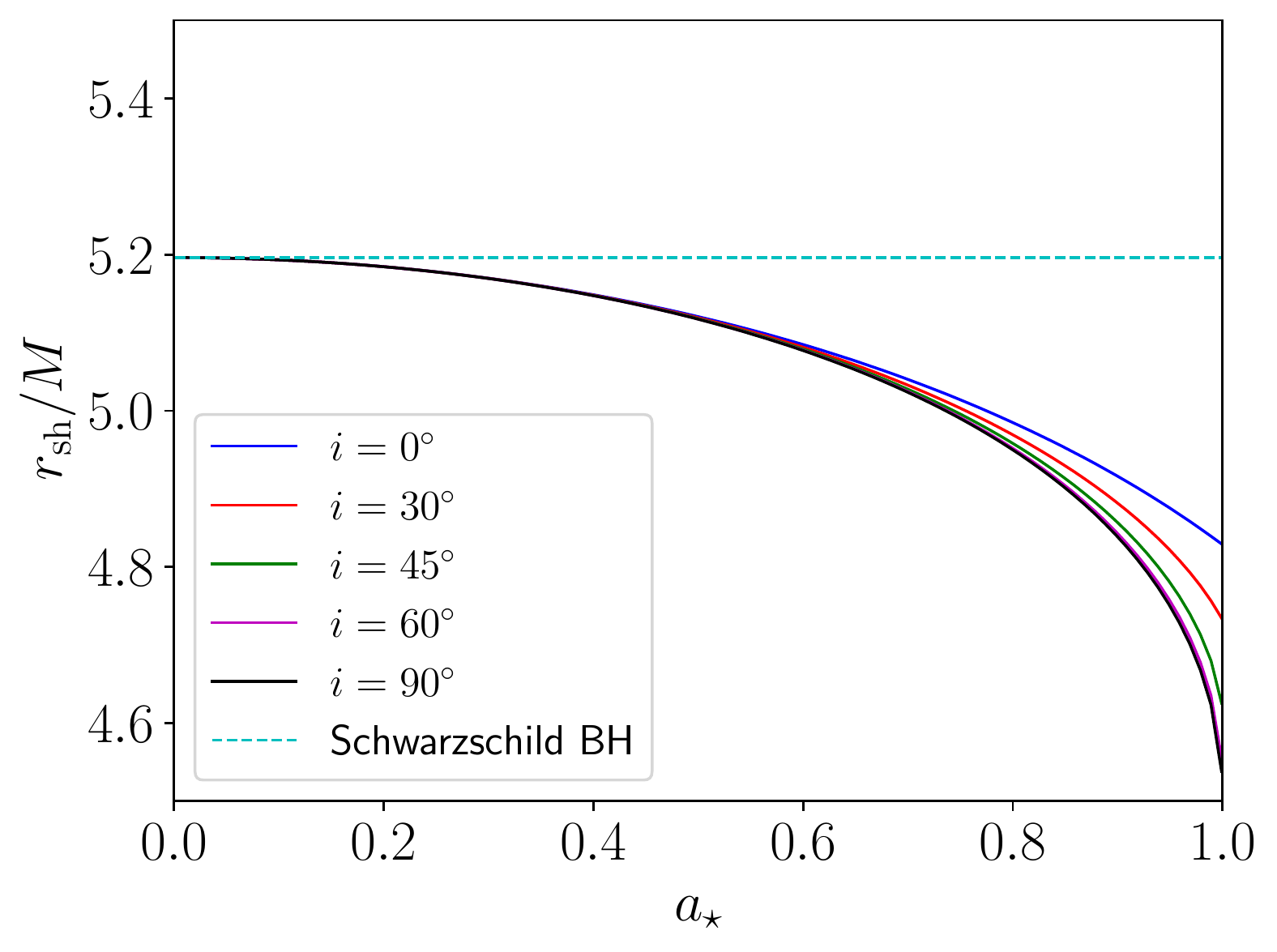}
\caption{Shadow radius in units of BH mass, $r_{\rm sh}/M$, for a Kerr BH (solid curves), as a function of the dimensionless spin $a_{\star}$, for various values of the observer's inclination angle $i$ as given by the color coding (with $i=0^{\circ}$ and $i=90^{\circ}$ corresponding to face-on and edge-on viewing respectively). The dashed cyan line indicates the size of a Schwarzschild BH, i.e.\ the limit $a_{\star} \to 0$, for which $r_{\rm sh}/M=3\sqrt{3}$. It is clear that the effect of spin on the BH shadow radius is small, particularly at low values of the spin.}
\label{fig:radius_spin_shadow}
\end{figure}

For simplicity, throughout this work we shall restrict ourselves to static spherically symmetric metrics, i.e.\ neglecting the effect of spin. The reason behind this choice is two-fold. First, the effect of spin on the shadow radius is small: we illustrate this in Fig.~\ref{fig:radius_spin_shadow} for the case of a Kerr BH, where we plot the predicted shadow radius as a function of the dimensionless spin $a_{\star}$, for different values of the observer's inclination angle $i$. Clearly, the effect is more pronounced for high inclination angles (i.e.\ for almost edge-on viewing), but remains small ($\lesssim 12\%$).~\footnote{It is worth remarking that this is strictly speaking true for the Kerr space-time. We will adopt the assumption (implicitly also made in Ref.~\cite{EventHorizonTelescope:2021dqv}) that the effect of spin remains small even in alternative metrics. Whether this is actually true should be checked on a case-by-case basis, although for many of the cases we will study, the rotating version of the metric is not available.}

Second, and perhaps most importantly, there is at present no clear consensus on the value of Sgr A$^*$'s spin and inclination angle. The EHT images are in principle consistent with large spin and low inclination angle, but are far from being inconsistent with low spin and large inclination angle~\cite{EventHorizonTelescope:2022xnr,EventHorizonTelescope:2022urf}. To be precise, however, this is strictly speaking only true for the Kerr metric, which has been assumed in deriving these results. Independent works based on radio, infrared, and X-ray emission, as well as millimeter VLBI, exclude extremal spin ($1-a_{\star} \ll 1$), but have been unable to place strong constraints otherwise~\cite{Broderick:2008sp,2010MNRAS.403L..74K,Broderick:2011mk}. Estimates based on semi-analytical models, magnetohydrodynamics simulations, or flare emissions, have reported constraints across a wide range of spin values~\cite{Huang:2009sq,Shcherbakov:2010ki}. One of the most recent dynamical estimates of Sgr A$^*$'s spin was reported in Ref.~\cite{Fragione:2020khu} by Fragione and Loeb, based on the impact on the orbits of the S-stars of frame-dragging precession, which would tend to erase the orbital planes in which the S-stars formed and are found today: observations of the alignment of the orbital planes of the S-stars today require Sgr A$^*$'s spin to be very low, $a_{\star} \lesssim 0.1$ (see also Ref.~\cite{Fragione:2022oau}). Some of these estimates rely on an assumed metric, although it is worth pointing out that the result of Ref.~\cite{Fragione:2020khu} does not. Constraints on the inclination angle are even more uncertain. The inconsistency across different estimates of Sgr A$^*$'s spin prompts us to a conservative approach where we neglect the effect of spin, while taking the very recent estimate of Ref.~\cite{Fragione:2020khu} as indication that the spin may be low: for $a_{\star} \lesssim 0.1$, as Fig.~\ref{fig:radius_spin_shadow} clearly shows, the effect of spin on the shadow radius is negligible at all inclination angles.

Finally, simple shadow-based observables which are most sensitive to the spin are those related to the shadow's circularity, such as the deviation from circularity $\Delta C$ discussed in Ref.~\cite{Bambi:2019tjh} (and adopted by several works which used the shadow of M87$^*$ to constrain fundamental physics), which for a Kerr BH should be $\lesssim 4\%$, increasing with spin and inclination angle. However, the sparse interferometric coverage of the EHT's 2017 observations of Sgr A$^*$, the associated significant uncertainties in circularity measurements, and the short variability timescale of Sgr A$^*$, prevented the collaboration from quantifying the circularity of the shadow based on observational features~\cite{EventHorizonTelescope:2022exc,EventHorizonTelescope:2022xqj,Vagnozzi:2022tba}. In closing, we note that most of the space-times considered in Ref.~\cite{EventHorizonTelescope:2022xqj} are also spherically symmetric, for reasons similar to the ones we outlined, providing an independent validation of our approach (although with all the caveats discussed above).

\subsection{Black hole shadow radius in spherically symmetric space-times}
\label{subsec:bhshadowradius}

Here we briefly review the calculation of the shadow radius in spherically symmetric space-times. If such a space-time possesses a photon sphere, the gravitationally lensed image thereof as viewed by an observed located at infinity will constitute the BH shadow.~\footnote{While a space-time with a photon sphere can cast a shadow, the existence of a photon sphere is strictly speaking not a mandatory requirement for a space-time to cast a shadow, see e.g.\ Ref.~\cite{Joshi:2020tlq}.} Let us consider a generic static, spherically symmetric, asymptotically flat space-time, i.e.\ one admitting a global, non-vanishing, time-like, hypersurface-orthogonal Killing vector field, with no off-diagonal components in the matrix representation of the metric tensor. Without loss of generality, the space-time line element in Boyer-Lindquist coordinates can be expressed as:
\begin{eqnarray}
{\rm d}s^2 = -A(r){\rm d}t^2 + B(r){\rm d}r^2+C(r){\rm d}\Omega^2\,,
\label{eq:staticsphericallysymmetricasymptoticallyflat}
\end{eqnarray}
where $d\Omega$ is the differential unit of solid angle.~\footnote{We are implicitly assuming that $r$ matches the quantity $r$ provided by astrophysical determinations (which is mathematically consistent with our later treatment of the shadow radius).} In the following, we shall henceforth refer to the function $A(r)=-g_{tt}(r)$ as the ``metric function'', and occasionally to $C(r)=g_{\theta\theta}=g_{\phi\phi}/\sin^2\theta$ as the ``angular metric function'' (though we note that the latter terminology is not a widespread one). If the space-time is asymptotically flat, $A(r) \to 1$ as $r \to \infty$ in Eq.~(\ref{eq:staticsphericallysymmetricasymptoticallyflat}). Note also that in most cases of interest $C(r)=r^2$, although here we choose to keep the discussion as general as possible.

Let us then define the function $h(r)$ (see e.g.\ Ref.~\cite{Perlick:2021aok}):
\begin{eqnarray}
h(r) \equiv \sqrt{\frac{C(r)}{A(r)}}\,.
\label{eq:hr}
\end{eqnarray}
If the metric in Eq.~(\ref{eq:staticsphericallysymmetricasymptoticallyflat}) possesses a photon sphere, the radial coordinate thereof, $r_{\rm ph}$, is given by the solution to the following implicit equation:
\begin{eqnarray}
\frac{\rm d}{{\rm d}r} \left [ h^2(r_{\rm ph}) \right ] =0\,,
\label{eq:photonspherehr}
\end{eqnarray}
which can be arranged to the following form:
\begin{eqnarray}
C'(r_{\rm ph})A(r_{\rm ph})-C(r_{\rm ph})A'(r_{\rm ph})=0\,.
\label{eq:photonspherecomplete}
\end{eqnarray}
Here, the prime denotes a derivative with respect to $r$. If, as in the majority of cases we will consider, the angular metric function is $C(r)=r^2$, then Eq.~(\ref{eq:photonspherecomplete}) reduces to the following well-known expression:
\begin{eqnarray}
A(r_{\rm ph})-\frac{1}{2}r_{\rm ph}A'(r_{\rm ph})=0\,.
\label{eq:photonsphere}
\end{eqnarray}
The shadow radius $r_{\rm sh}$ corresponds to the gravitationally lensed image of the surface defined by $r_{\rm ph}$, and is therefore given by (see e.g.\ Refs.~\cite{Psaltis:2007rv,Cunha:2018acu,Dokuchaev:2019jqq,EventHorizonTelescope:2020qrl,Perlick:2021aok}):
\begin{eqnarray}
r_{\rm sh} = \sqrt{\frac{C(r)}{A(r)}}\Bigg\vert_{r_{\rm ph}}\,.
\label{eq:rshcomplete}
\end{eqnarray}
For the ``standard'' angular metric function $C(r)=r^2$ Eq.~(\ref{eq:rshcomplete}) reduces to the following well-known expression:
\begin{eqnarray}
r_{\rm sh} = \frac{r}{\sqrt{A(r)}}\Bigg\vert_{r_{\rm ph}}\,.
\label{eq:rsh}
\end{eqnarray}
In the case of the Schwarzschild BH, whose metric function is $A(r)=1-2M/r$ and whose angular metric function is $C(r)=r^2$~\cite{Schwarzschild:1916uq}, combining Eqs.~(\ref{eq:photonsphere},\ref{eq:rsh}) leads to the well-known result $r_{\rm sh}=3\sqrt{3}M$. For obvious symmetry reasons, the BH shadow for a spherically symmetric space-time is a circle of radius $r_{\rm sh}$ on the plane of a distant observer, irrespective of the inclination angle.

Note that the computation we have outlined is strictly valid for \textit{asymptotically flat} space-times only. For other metrics, including but not limited to those matching on to the cosmological accelerated expansion at large distances and hence asymptotically de Sitter, the size of the BH shadow can explicitly depend on the radial coordinate of the distant observer, and on whether the observer is static or comoving (see e.g.\ Refs.~\cite{Kumar:2017tdw,Perlick:2021aok} for extensive discussions on this point). An example of such a situation occurs with the Kottler metric~\cite{Kottler:1918ghw}. Given Sgr A$^*$'s proximity to us, for practical purposes we focus on a static observer (i.e.\ one which is not in the Hubble flow with respect to the BH) located at a distance $r_O$. For simplicity we also focus on the case where the angular metric function is $C(r)=r^2$. Then, the angular size of the BH shadow $\alpha_{\rm sh}$ is given by (see e.g.\ Ref.~\cite{Perlick:2021aok}):
\begin{eqnarray}
\sin^2\alpha_{\rm sh} = \frac{r^2_{\rm ph}}{A(r_{\rm ph})}\frac{A(r_O)}{r_O}\,.
\label{eq:alphashnotasymptoticallyflat}
\end{eqnarray}
In the physically relevant small-angle approximation, it is easy to see that the shadow size is given by (see Ref.~\cite{Perlick:2021aok} for more detailed discussions):
\begin{eqnarray}
r_{\rm sh} = r_{\rm ph}\sqrt{\frac{A(r_O)}{A(r_{\rm ph})}}\,.
\label{eq:rshnotasymptoticallyflat}
\end{eqnarray}
The explicit dependence of the shadow size on the observer's position is clear in Eq.~(\ref{eq:rshnotasymptoticallyflat}). It is trivial to see that for a distant observer viewing a BH described by an asymptotically flat metric, Eq.~(\ref{eq:rshnotasymptoticallyflat}) reduces to Eq.~(\ref{eq:rsh}), since $A(r_O) \approx 1$ as long as the observer is sufficiently distant from the BH. For instance, for the Schwarzschild metric this condition is satisfied as long as $r_O \gg M$, i.e.\ the distance between the BH and the observer is much larger than the BH gravitational radius. This condition is satisfied for the case of Sgr A$^*$, whose distance from us of $\approx 8\,{\rm kpc}$ should be compared to its gravitational radius $r_g \approx {\cal O}(10^{-7})\,{\rm pc}$.

As a further caveat, it is also worth noting that the computation outlined in Eqs.~(\ref{eq:hr}--\ref{eq:rsh}) does not (necessarily) hold in theories with electromagnetic Lagrangian other than the Maxwell one (for instance non-linear electrodynamics), since in this context photons do not necessarily move along null geodesics of the metric tensor, but along null geodesics of an effective geometry~\cite{Novello:1999pg,DeLorenci:2000yh}, a fact which has only been recently appreciated in the literature. We will consider certain classes of non-linear electrodynamics theories in this work, carefully accounting for the effective geometry.

As a final caveat, it is worth pointing out that the ``mathematical'' result for the shadow radius obtained through the procedure discussed previously is ``astrophysically'' relevant only if the resulting shadow size is located outside the event horizon for a BH, or the throat for a wormhole. In the following, we have explicitly checked that this is indeed the case for all space-times discussed, or added caveats as appropriate.

In what follows, we will compute the shadow radius of various space-times of interest, and then compare them to the image of Sgr A$^*$ by imposing the bounds of Eqs.~(\ref{eq:1sigma},\ref{eq:2sigma}).~\footnote{For each space-time, we typically explicitly provide only $A(r)=-g_{tt}$ [and $C(r)$ if $\neq r^2$], as this is the metric component directly entering in the shadow radius calculation. For most of the space-times we study, the condition $g_{tt}g_{rr}=-1$ is fulfilled: as discussed in Ref.~\cite{Jacobson:2007tj}, this occurs if and only if the radial null-null components of $R_{\mu\nu}$ vanish or, equivalently, if the restriction of $R_{\mu\nu}$ to the $t$-$r$ subspace is proportional to $g_{\mu\nu}$ (which on solutions of the Einstein equations implies that the radial pressure is equal to minus the energy density). The only space-times we study for which this condition is violated are: the Morris-Thorne traversable wormhole (Sec.~\ref{subsec:morristhorne}), the Damour-Solodukhin wormhole (Sec.~\ref{subsec:damoursolodukhin}), the Joshi-Malafarina-Narayan naked singularity (Sec.~\ref{subsec:jmn1}), the naked singularity surrounded by a thin shell of matter (Sec.~\ref{subsec:tsmns}), the BH in Clifton-Barrow $f(R)$ gravity (Sec.~\ref{subsec:cliftonbarrow}), the Sen BH (Sec.~\ref{subsec:sen}), the Einstein-Maxwell-dilaton-1 BH (Sec.~\ref{subsec:emd1}), the BH in Loop Quantum Gravity (Sec.~\ref{subsec:lqg}), the DST BH (Sec.~\ref{subsec:dst}), the BH in bumblebee gravity (Sec.~\ref{subsubsec:bumblebee}), and the Casimir wormhole (Sec.~\ref{subsubsec:casimirwh}).} These, as discussed earlier, already take into account various observational, theoretical, and modeling uncertainties, including the potential multiplicative offset between the radius of the bright ring of emission and the underlying shadow radius. We will show that the EHT image of Sgr A$^*$ sets particularly stringent constraints on theories and frameworks which predict a shadow radius \textit{larger} than the Schwarzschild radius $3\sqrt{3}M$. Although the effect of new physics in many of the cases we will consider is that of decreasing the shadow radius with respect to the Schwarzschild case, a few will actually turn out to increase it, and these will be the models whose parameters are most tightly constrained. The key difference between these two classes of scenarios can easily be understood in terms of the effective potential experienced by photons which, assuming spherical symmetry and considering motion in the equatorial plane ($\theta=\pi/2$, which simplifies the calculations and is consistent with the assumed symmetries), is proportional to $V_{\rm eff}(r) \propto A(r)/r^2$. The shadow radius is then directly related to the maximum of the effective potential, and this can easily be shown to move towards the left/right within the scenarios leading to a decrease/increase of the shadow radius with respect to the Schwarzschild case.

\subsection{Brief overview and motivation of scenarios and space-times considered}
\label{subsec:fundamental}

Before moving on to actually performing our analysis, we will give a very brief overview of the space-times we consider, along with motivation for moving beyond the Schwarzschild space-time. We will consider a wide range of gravity theories, fundamental physics scenarios, and space-times beyond those of the Schwarzschild BH, for which $r_{\rm sh}=3\sqrt{3}M$. The no-hair theorem (NHT)~\cite{Israel:1967wq,Israel:1967za,Carter:1968rr,Carter:1971zc,Robinson:1975bv} states that the only possible stationary, axisymmetric, and asymptotically flat BH solutions of the 4D electrovacuum Einstein-Maxwell equations are described by the Kerr-Newman family of metrics~\cite{Newman:1965tw,Newman:1965my}. Recall that the latter describes electrically charged, rotating BHs, and reduces to the Kerr metric in the absence of electric charge, to the Reissner-Nordstr\"{o}m metric in the absence of rotation, and finally to the Schwarzschild metric in the absence of both charge and rotation. As John Archibald Wheeler phrased it, ``\textit{black holes have no hair}'', with the term ``hair'' referring generically to parameters other than the BH mass $M$, spin $J$, and electric charge $Q$, which are required for a complete description of the BH solution.

At present, there is no sign of tension between the NHT and observations of astrophysical BHs. Why, then, is going beyond the Kerr-Newman family of metrics a well-motivated endeavor? The first reason is tied to the fact that a wide variety of theoretical and observational issues hint towards the possibility that our understanding of gravity as provided by General Relativity (GR) is likely to be incomplete. These hints range from the quest for unifying gravity and quantum mechanics (QM), to the conflict between unitary evolution in QM and Hawking BH radiation as encapsulated in the BH information paradox~\cite{Hawking:1975vcx}, and finally to cosmological observations requiring the existence of dark matter and dark energy, likely a phase of cosmic inflation, and possibly a phase of early dark energy around the time of recombination. In the presence of new physics which eventually will address these issues, it is perfectly reasonable to assume that the NHT may only be an approximation, valid to the current level of precision. In fact, several theoretical approaches towards addressing the above issues posit the existence of new fields (particularly scalar fields) or modifications to GR (often introducing effective scalar degrees of freedom), see e.g.\ Refs.~\cite{Guth:1980zm,Linde:1981mu,Wetterich:1987fm,Ratra:1987rm,Caldwell:1997ii,Benaoum:2002zs,Arkani-Hamed:2008hhe,Cai:2009zp,Cyr-Racine:2012tfp,Cline:2013gha,Petraki:2014uza,Rinaldi:2014yta,Foot:2014uba,Foot:2014osa,Rinaldi:2015iza,Cai:2015emx,Sola:2016jky,Nojiri:2017ncd,Capozziello:2017buj,Benisty:2018qed,Poulin:2018cxd,Heckman:2019dsj,Niedermann:2019olb,Sakstein:2019fmf,Vagnozzi:2019kvw,DiLuzio:2020wdo,Vagnozzi:2020gtf,Vagnozzi:2021gjh,Pati:2021ach,Benetti:2021uea,Benevento:2022cql,Narawade:2022jeg,Reeves:2022aoi,Gadbail:2023suo} for an inevitably incomplete selection of examples. All of these scenarios can very naturally lead to violations of the NHT.~\footnote{Note that, although one would in principle expect the effects of dark energy and modified gravity models invoked to explain cosmic acceleration to appear on scales $\gtrsim 150\,{\rm Mpc}$, whereas Sgr A$^*$ is $8\,{\rm kpc}$ distant from us, in practice these models almost always result in unwanted effects (e.g.\ fifth forces) on local scales, unless screening mechanisms are invoked~\cite{Brax:2021wcv}. These screening mechanisms typically kick in at scales intermediate between cosmological [${\cal O}({\rm Gpc})$] and local [${\cal O}({\rm pc})$] ones, so quite naturally in the ${\cal O}({\rm kpc})$ range, lending motivation to the search for novel phenomenological signatures in this regime. In addition, it is worth noting that several ``quintessential inflation'' models exist wherein inflation and cosmic acceleration are unified as the result of a single scalar degree of freedom~\cite{Peebles:1998qn,deHaro:2021swo}.} Moving beyond astrophysical systems, the study of controlled violations of the NHT is well-motivated by developments in our understanding of the gauge/gravity duality~\cite{Maldacena:1997re}, with important applications to various condensed matter systems~\cite{Gubser:2005ih}, including holographic superconductors~\cite{Horowitz:2009ij} and quantum liquids~\cite{Liu:2009dm}.

The second reason is possibly even more fundamental, and is related to the well-known Penrose-Hawking singularity theorems, i.e.\ the fact that in GR continuous gravitational collapse leads to the inevitable, but at the same time arguably undesirable, appearance of singularities~\cite{Penrose:1964wq,Hawking:1970zqf}. For instance, the Kerr-Newman family of metrics possesses a well-known physical (non-coordinate) singularity at $r=0$. The cosmic censorship conjecture~\cite{Penrose:1969pc,Wald:1997wa} notwithstanding, the mere existence of singularities has prompted a long-standing search for ``regular'' BH solutions, which regularize the central singularity (see e.g.\ Refs.~\cite{Roman:1983zza,Borde:1996df,AyonBeato:1998ub,AyonBeato:1999rg,Bronnikov:2005gm,Iso:2006ut,Berej:2006cc,Bronnikov:2012ch,Rinaldi:2012vy,Bambi:2013ufa,Sebastiani:2013fsa,Frolov:2014jva,Toshmatov:2014nya,Bardeen:2014uaa,Stuchlik:2014qja,Schee:2015nua,Johannsen:2015pca,Junior:2015fya,Myrzakulov:2015qaa,Schee:2016mjd,Abdujabbarov:2016hnw,Fan:2016hvf,Frolov:2017rjz,Toshmatov:2017zpr,Chinaglia:2017uqd,Chinaglia:2017wih,Frolov:2017dwy,Jusufi:2018jof,Carballo-Rubio:2018pmi,Cano:2018aod,Bardeen:2018frm,Stuchlik:2019uvf,Ovgun:2019wej,Han:2019lfs,Rodrigues:2019xrc,Panotopoulos:2019qjk,Schee:2019gki,Jusufi:2019caq,Ndongmo:2019ywh,Bertipagani:2020awe,Nashed:2021pah,Simpson:2021dyo,Zhao:2022gxl,Javed:2022rrs,Riaz:2022rlx,Sebastiani:2022wbz,Franzin:2022wai,Singh:2022dqs,Chataignier:2022yic,Zhou:2022yio,Olmo:2022cui,Bonanno:2022jjp,Carballo-Rubio:2022nuj,Carballo-Rubio:2023mvr,Olmo:2023lil} for examples). Therefore, testing the metrics of regular BHs, or BH mimickers which address at least in part the existence of singularities, is a very well-motivated direction, particularly at present time with the availability of horizon-scale BH images. Note that many of these regular BH solutions can be obtained as solutions to the Einstein field equations coupled to a suitable non-linear electrodynamics source. Finally, as the nature of dark matter, dark energy, and the origin of structure (whether through a phase of inflation or an alternative mechanism) is not fully understood at present, one should be open-minded to the possibility that one or more of these phenomena may be connected to potential modifications of GR.

With these considerations in mind, in this work we will consider a wide range of space-times beyond the Schwarzschild BH, testing them against the EHT horizon-scale image of Sgr A$^*$. The space-times we consider broadly speaking fall within these (by no means mutually exclusive) categories:
\begin{enumerate}
\item Regular BHs, whether arising from specific theories or constructed in a phenomenological setting;
\item BHs in modified theories of gravity, modified electrodynamics, and string-inspired settings;
\item BHs in theories with additional matter fields (usually scalar fields), typically bringing about violations of the no-hair theorem;
\item BH mimickers such as wormholes (or effective wormhole geometries) and naked singularities;
\item Modifications induced by novel fundamental physics frameworks (e.g.\ generalized or extended uncertainty principles, non-commutative geometries, or entropy laws beyond the Bekenstein-Hawking entropy).
\end{enumerate}
Of course, several space-times we will consider fall within more than one of the above categories at the same time (especially 1. and 2., or 1. and 3.). Most of these space-times are described by one or more extra parameters, which in most cases we shall generically refer to as ``hair'', ``hair parameters'', or ``charges'', in the latter case following the language adopted by Refs.~\cite{EventHorizonTelescope:2021dqv,EventHorizonTelescope:2022xqj}.

Before moving on to the results, one final clarification is in order. In all the cases we will consider, the hair parameters discussed above essentially fall into one of two categories. In the first case, they are ``universal'', i.e.\ the same for each BH in the Universe regardless of its other hairs (mass, spin, electric charge). In the second case, they are ``specific'', i.e.\ they can vary from BH to BH. With a slight abuse of language, we shall refer to these as being \textit{universal hairs} and \textit{specific hairs} respectively: note that the two should not be confused with the concepts of primary hair and secondary hair, although they are to some extent related.

An example of universal hair could be one which is exclusively determined by a parameter (e.g.\ a coupling) of the underlying Lagrangian of the theory (which is therefore uniquely determined), or by another fundamental parameter introduced at a phenomenological level. A very simple example of specific hair is the electric charge of a BH appearing in the well-known Reissner-Nordstr\"{o}m metric: this is a conserved charge which essentially emerges as an integration constant, and which can therefore take different values from BH to BH. As far as we can tell, the distinction between these two types of hairs has not been sufficiently emphasized in the literature on tests of fundamental physics from BH shadows.

We note that the comparison between constraints on specific hairs obtained from BH shadows and other astrophysical probes (for instance gravitational waves, stellar motions, and cosmology) is meaningful only if the constraints are referred to the same source. On the other hand, in the case of universal hairs such a comparison can be made even without reference to any specific source. In what follows, for each BH solution we will study, we will clearly indicate whether the additional hair parameter which controls deviations from the Schwarzschild metric is a specific hair or an universal hair.

\section{Results}
\label{sec:results}

\subsection{Reissner-Nordstr\"{o}m black hole and naked singularity}
\label{subsec:rn}

The first geometry we consider beyond the Schwarzschild BH, and arguably the simplest extension thereof, is the Reissner-Nordstr\"{o}m (RN) metric describing an electrically charged, non-rotating BH. Explicitly introducing the BH mass (which we will later set to $M=1$ for convenience), and recalling that we are setting $4\pi\epsilon_0=\hbar=c=G_N=1$, the metric function of the RN BH with electric charge $Q$ is given by~\cite{1916AnP...355..106R,1917AnP...359..117W,1918KNAB...20.1238N,1921RSPSA..99..123J}:~\footnote{For completeness, given the importance of the Reissner-Nordstr\"{o}m space-time, we note that the metric function in SI units reads $A(r) = 1-2G_NM/c^2r+GQ^2/4\pi\epsilon_0c^4r^2$.}
\begin{eqnarray}
A(r) = 1-\frac{2M}{r}+\frac{Q^2}{r^2}\,,
\label{eq:metricrn}
\end{eqnarray}
where $Q$ describes a specific hair. Henceforth, if not explicitly stated, the angular metric function will be implicitly assumed to be given by $C(r)=r^2$. Setting $M=1$,~\footnote{This operation effectively corresponds to rescaling the Boyer-Lindquist radial coordinate and electric charge as $\widetilde{r}=r/M$ and $\widetilde{Q}=Q/M$ respectively. Throughout this work we will drop the tildes, and whether we are referring to the original or rescaled quantities should be evident from the context.} it is easy to show that this space-time possesses an event horizon only for $Q \leq 1$. However, even for larger values of the electric charge the space-time, which is now a naked singularity, can still possess a photon sphere (and thereby cast a shadow-like feature), provided $1<Q \leq \sqrt{9/8} \approx 1.06$: this region of parameter space describes what is commonly referred to as the ``RN naked singularity''.

Using Eqs.~(\ref{eq:photonsphere},\ref{eq:rsh}), we straightforwardly obtain the following expression for the radius of the shadow cast by the RN space-time, which is valid within both the BH and naked singularity regimes:
\begin{eqnarray}
r_{\rm sh} = \frac{\sqrt{2} \left ( 3+\sqrt{9-8Q^2} \right ) }{\sqrt{4+\frac{\sqrt{9-8Q^2}-3}{Q^2}}}\,.
\label{eq:shadowsizern}
\end{eqnarray}
It is at first glance not obvious that Eq.~(\ref{eq:shadowsizern}) does indeed reduce to the Schwarzschild result of $r_{\rm sh}=3\sqrt{3}$ in the $Q \to 0$ limit, due to the apparent divergence in the denominator. However, the proper way of taking this limit is to Taylor-expand the denominator: when doing so, one indeed finds that the denominator of Eq.~(\ref{eq:shadowsizern}) tends to $\sqrt{8/3}$, whereas the numerator straightforwardly tends to $6\sqrt{2}$, leading overall to the correct limit of $3\sqrt{3}$.
\begin{figure}
\centering
\includegraphics[width=1.0\linewidth]{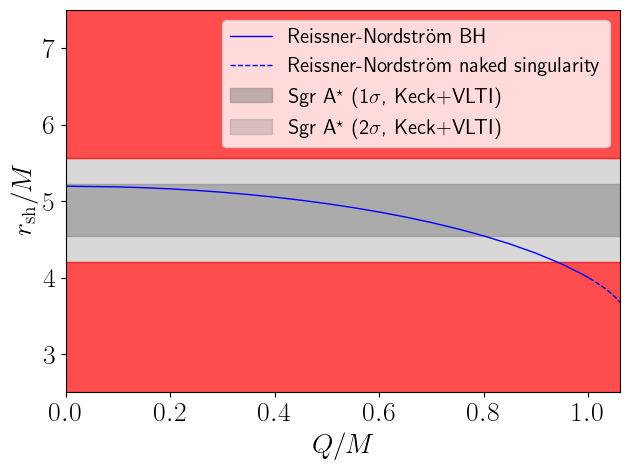}
\caption{Shadow radius $r_{\rm sh}$ of the Reissner-Nordstr\"{o}m black hole (solid curve) and naked singularity (dashed curve) with metric function given by Eq.~(\ref{eq:metricrn}) and in units of the BH mass $M$, as a function of the normalized electric charge $Q/M$, as discussed in Sec.~\ref{subsec:rn}. The dark gray and light gray regions are consistent with the EHT horizon-scale image of Sgr A$^*$ at $1\sigma$ and $2\sigma$ respectively, after averaging the \textrm{Keck} and \textrm{VLTI} mass-to-distance ratio priors for Sgr A$^*$ [see Eqs.~(\ref{eq:1sigma},\ref{eq:2sigma})]. The red regions are instead excluded by the same observations at more than $2\sigma$.}
\label{fig:shadow_reissner_nordstrom_bh_ns}
\end{figure}

The evolution of the shadow radius as a function of the electric charge is shown in Fig.~\ref{fig:shadow_reissner_nordstrom_bh_ns}, for both the BH (solid) and naked singularity (dashed) regimes of the RN space-time, alongside the observational constraints imposed by the EHT image of Sgr A$^*$ [Eqs.~(\ref{eq:1sigma},\ref{eq:2sigma})]. We see that as the electric charge increases, the shadow radius decreases, which can be understood by studying how the electric charge affects the effective potential felt by test particles. This is a feature which is common to a number of extensions of the Schwarzschild metric, as we shall see throughout the paper: most extensions will lead to a shadow radius which \textit{decreases} with increasing charge/hair parameter (although notable exceptions exist, which we shall also discuss in this paper).

From Fig.~\ref{fig:shadow_reissner_nordstrom_bh_ns}, we observe that the EHT observations set the $1\sigma$ upper limit $Q \lesssim 0.8M$, and the $2\sigma$ upper limit $Q \lesssim 0.95M$. Therefore, within more than 2 standard deviations, the EHT observations rule out the possibility of Sgr A$^*$ being an extremal RN BH ($Q=M$). Of course, the naked singularity regime ($M<Q \leq \sqrt{9/8}M$) is completely ruled out: therefore, the EHT observations rule out the possibility of Sgr A$^*$ being one of the simplest possible naked singularities (although other naked singularity space-times are allowed as we shall see later). Finally, re-inserting SI units, we see that the previous $2\sigma$ upper limit translates to $Q \lesssim 0.95\sqrt{4\pi\epsilon_0G}M \sim 6.5 \times 10^{26}\,{\rm C}$. In practice, this limit is significantly weaker than the limit $Q \lesssim 3 \times 10^8\,{\rm C}$, or equivalently $Q \lesssim {\cal O}(10^{-19})$ in units of mass of Sgr A$^*$, obtained from astrophysical considerations in Refs.~\cite{Zajacek:2018ycb,Zajacek:2018vsj} (see also Refs.~\cite{Wald:1974np,Ray:2003gt,Iorio:2012dbo,Zakharov:2014lqa,Zajacek:2019kla}).

\subsection{Bardeen magnetically charged regular black hole}
\label{subsec:bardeen}

We now consider regular BH solutions, which are free of the $r=0$ core singularity present within the Kerr-Newman family of metrics. Several approaches towards constructing regular BH metrics replace the Kerr-Newman core with a different type of core, such as a de Sitter or Minkowski core, or phenomenologically smearing the core over a larger surface. The Bardeen space-time is one of the first regular BH solutions to the Einstein equations ever proposed~\cite{Bardeen:1968ghw}, and describes the space-time of a singularity-free BH with a de Sitter core~\cite{Sakharov:1966aja}. The Bardeen space-time carries a magnetic charge $q_m$, and can be thought of as a magnetic monopole satisfying the weak energy condition~\cite{Ayon-Beato:2000mjt}. It also emerges in the context of GR coupled to a very specific non-linear electrodynamics Lagrangian~\cite{Ayon-Beato:2004ywd}, and can further be interpreted as a quantum-corrected Schwarzschild BH~\cite{Maluf:2018ksj}.

\begin{figure}
\centering
\includegraphics[width=1.0\linewidth]{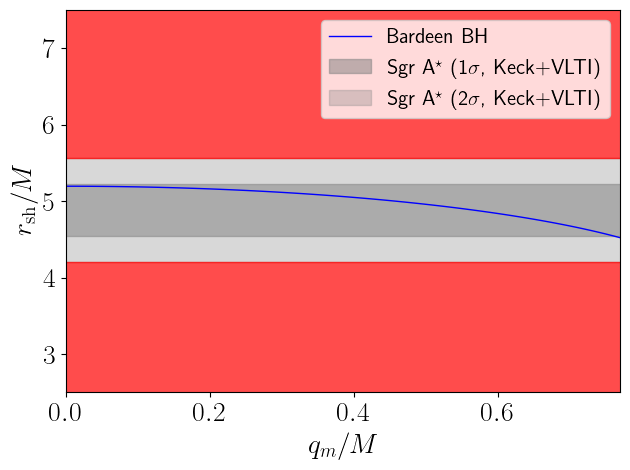}
\caption{Same as in Fig.~\ref{fig:shadow_reissner_nordstrom_bh_ns} for the Bardeen magnetically charged regular BH with metric function given by Eq.~\eqref{eq:metricbardeen}, as discussed in Sec.~\ref{subsec:bardeen}.}
\label{fig:shadow_bardeen_bh}
\end{figure}

The metric function for the Bardeen magnetically charged BH is given by~\cite{Bardeen:1968ghw}:
\begin{eqnarray}
A(r) = 1-\frac{2Mr^2}{(r^2+q_m^2)^{3/2}}\,,
\label{eq:metricbardeen}
\end{eqnarray}
with $q_m$ characterizing a specific hair and satisfying the condition $q_m \leq \sqrt{16/27} \approx 0.77$. We compute the corresponding shadow radius numerically, and show its evolution as a function of the magnetic charge $q_m$ in Fig.~\ref{fig:shadow_bardeen_bh}. As with the Reissner-Nordstr\"{o}m BH, we see that increasing $q_m$ decreases the shadow radius. Within $2\sigma$, the EHT observations are consistent with Sgr A$^*$ being a Bardeen BH for any (allowed) value of the magnetic charge, including the extremal value $q_m=\sqrt{16/27}M$.

A comment, relevant for several other space-times, is in order here. Since the Bardeen regular BH can emerge within a specific non-linear electrodynamics theory, there is potentially some ambiguity as to how one computes the corresponding shadow: one can compute it \textit{a)} using the ``standard'' formulae [Eqs.~(\ref{eq:photonspherecomplete}--\ref{eq:rshcomplete})], thus requiring only knowledge of the metric function Eq.~(\ref{eq:metricbardeen}), or \textit{b)} accounting for corrections to the effective geometry arising from the underlying non-linear electrodynamics theory. Approach \textit{a)} can be considered model-agnostic, in the sense that the metric is taken as a phenomenological input without further questioning its origin, whereas approach \textit{b)} assumes a specific microscopical origin for the metric. For the Bardeen BH, we have chosen to adopt approach \textit{a)} for three reasons:
\begin{itemize}
\item historically, Bardeen first proposed this specific metric on a phenomenological basis as a toy model, and only at a later stage was the associated non-linear electrodynamics theory obtained;
\item the corresponding underlying non-linear electrodynamics theory is arguably quite baroque, and was ``reconstructed'' to give the desired solution;
\item non-linear electrodynamics is not the only possible microscopical origin for the Bardeen BH.
\end{itemize}
Regarding the second point in particular, it is worth noting that it is almost always possible to reconstruct a possible non-linear electrodynamics source for a given (regular or non) BH metric. The corresponding Lagrangian is not always guaranteed to be particularly motivated, and in fact more often than not will be quite baroque. While this is somewhat a matter of taste, in our view a more aesthetically appealing approach is a ``top-down'' one starting from a specific well-motivated non-linear electrodynamics Lagrangian, and finding the corresponding solution. For this reason, if a specific metric can be supported by a non-linear electrodynamics source, but the latter is quite baroque or was reconstructed a posteriori, we choose to compute the shadow following the phenomenological/model-agnostic approach \textit{a)}. This is the case for the Bardeen BH, which is the reason why only one curve, obtained from the standard formulae Eqs.~(\ref{eq:photonspherecomplete}--\ref{eq:rshcomplete}), is present on Fig.~\ref{fig:shadow_bardeen_bh}, and similar considerations hold for other metrics we study later.

\subsection{Hayward regular black hole}
\label{subsec:hayward}

Another well-known regular BH solution has been proposed by Hayward~\cite{Hayward:2005gi}. This space-time replaces the $r=0$ GR singularity by a de Sitter core with effective cosmological constant $\Lambda=3\ell^2$. In doing so, one is assuming that an effective cosmological constant plays an important role at short distances, with Hubble length $\ell$. Such a behavior, while introduced at a phenomenological level for the Hayward BH, has been justified within the context of the equation of state of matter at high density~\cite{Sakharov:1966aja,1966JETP...22..378G}, or an upper limit on density or curvature~\cite{1982JETPL..36..265M,1987JETPL..46..431M,Mukhanov:1991zn}, the latter expected to emerge within a quantum theory of gravity~\cite{Addazi:2021xuf}.

\begin{figure}
\centering
\includegraphics[width=1.0\linewidth]{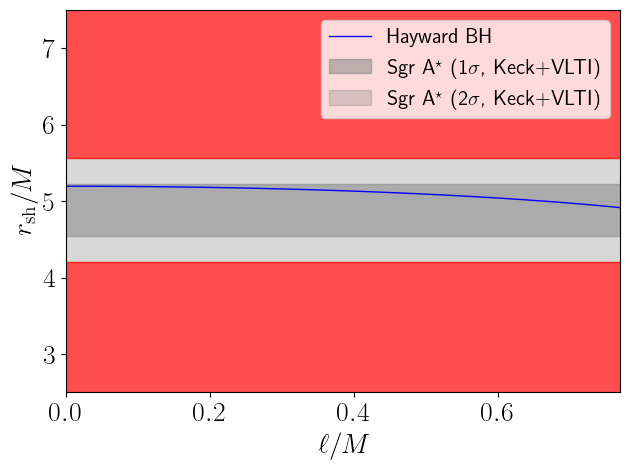}
\caption{Same as in Fig.~\ref{fig:shadow_reissner_nordstrom_bh_ns} for the Hayward regular BH with metric function given by Eq.~\eqref{eq:metrichayward}, as discussed in Sec.~\ref{subsec:hayward}.}
\label{fig:shadow_hayward_bh}
\end{figure}

The metric function for the Hayward regular BH is given by~\cite{Hayward:2005gi}:
\begin{eqnarray}
A(r) = 1-\frac{2Mr^2}{r^3+2\ell^2M} \,,
\label{eq:metrichayward}
\end{eqnarray}
with the Hubble length associated to the effective cosmological constant satisfying $\ell \leq \sqrt{16/27}M$ and, as originally envisaged by Hayward, characterizing an universal hair (although, since the space-time is introduced in a phenomenological way, nothing prevents the hair from being specific). We compute the corresponding shadow radius numerically, showing its evolution as a function of the Hubble length $\ell$ in Fig.~\ref{fig:shadow_hayward_bh}. Again, we see that increasing $\ell$ decreases the shadow radius. Within the present constraints, we see that the EHT observations are consistent with Sgr A$^*$ being a Hayward BH for any (allowed) value of the Hubble length associated to the effective cosmological constant, including the extremal value $\ell=\sqrt{16/27}M$.

\subsection{Frolov regular black hole}
\label{subsec:frolov}

A well-known generalization of the Hayward BH, with the same quantum gravity-inspired theoretical motivation and featuring an additional charge parameter, has been proposed by Frolov in Ref.~\cite{Frolov:2016pav}. The metric function for the Frolov BH is given by:
\begin{eqnarray}
A(r) = 1-\frac{(2Mr-q^2)r^2}{r^4+(2Mr+q^2)\ell^2} \,,
\label{eq:metricfrolov}
\end{eqnarray}
where the additional charge parameter $q$ characterizes a specific hair and satisfies $0<q \leq 1$, and the length below which quantum gravity effects become important satisfies $\ell \leq \sqrt{16/27}$ as with the Hayward BH. In principle, $q$ admits an interpretation in terms of electric charge measured by an observer situated at infinity, where the metric is asymptotically flat.

\begin{figure}
\centering
\includegraphics[width=1.0\linewidth]{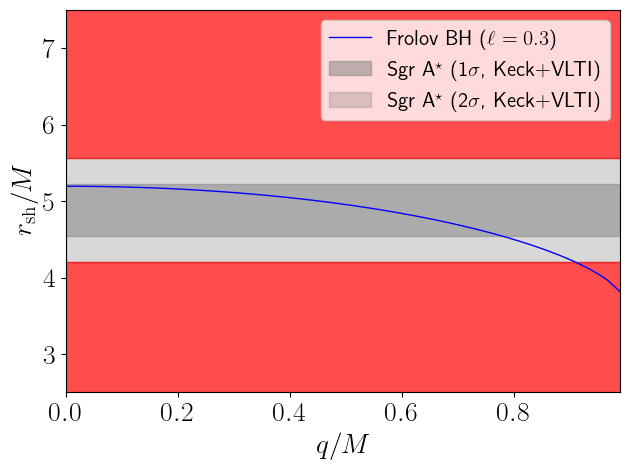}
\caption{Same as in Fig.~\ref{fig:shadow_reissner_nordstrom_bh_ns} for the Frolov regular BH with metric function given by Eq.~\eqref{eq:metricfrolov}, where we have fixed the Hubble length associated to the effective cosmological constant to $\ell=0.3$, as discussed in Sec.~\ref{subsec:frolov}.}
\label{fig:shadow_frolov_bh}
\end{figure}

Here, we focus on the charge $q$, while fixing $\ell=0.3$. This is similar to the choice reported in Ref.~\cite{EventHorizonTelescope:2022xqj}, which instead fixed $\ell=0.4$, and can be motivated by the fact that the shadow radius changes slowly with $\ell$, as is clear from Fig.~\ref{fig:shadow_hayward_bh}. Therefore, the limits on $q$ we will report can be interpreted as constraints along a particular slice of the full parameter space (see also the earlier work of Ref.~\cite{Kumar:2019pjp} by some of us). We compute the corresponding shadow radius numerically, showing its evolution as a function of the charge $q$ in Fig.~\ref{fig:shadow_frolov_bh}. We see that, for $\ell=0.3$, the EHT observations set the upper limits $q \lesssim 0.8M$ ($1\sigma$) and $q \lesssim 0.9M$ ($2\sigma$), qualitatively similar to the constraints on the electric charge of the RN metric $Q$ we obtained in Sec.~\ref{subsec:rn}. Recall once more that these constraints are valid for $\ell=0.3$. From the results reported in Sec.~\ref{subsec:hayward} and Fig.~\ref{fig:shadow_hayward_bh} for the Hayward BH, we can expect that fixing $\ell$ to a smaller/larger value would result in weaker/tighter constraints on $q$, as increasing $\ell$ leads to a slight decrease in the shadow size: in other words, we can expect a negative correlation/parameter degeneracy between $\ell$ and $q$.

\subsection{Magnetically charged Einstein-Bronnikov regular black hole}
\label{subsec:eb}

Non-linear electrodynamics (NLED) theories are well-motivated extensions of Maxwell's Lagrangian in the high-intensity regime~\cite{Stehle:1966wii}, and appear in the low-energy limit of various well-known theories, including Born-Infeld electrodynamics~\cite{Born:1934gh}, as well as several string or supersymmetric constructions~\cite{Fradkin:1985qd,Tseytlin:1986ti,Bern:1993tz,Dunne:2004nc,Jacobson:2018kso}. Another important class of regular BHs arises in the context of GR coupled to Bronnikov NLED, where the Lagrangian of the latter is given by~\cite{Bronnikov:2000vy}:
\begin{eqnarray}
\mathcal{L}(U) = F_{\mu\nu}F^{\mu\nu}\cosh^{-2}\left[ a \left( F_{\alpha\beta}F^{\alpha\beta}/2\right)^{1/4}\right]\,.
\label{eq:eb}
\end{eqnarray}
We recall that the relativistic invariants $U$ and $W$ are constructed from the electromagnetic field-strength tensor $F_{\alpha\beta}$ and its dual $F^{\star}_{\alpha\beta}$ as follows:
\begin{eqnarray}
U&=&F^{\alpha\beta}F_{\alpha\beta}\,, \quad W=F^{\alpha\beta}F_{\alpha\beta}^{\star}\,, \nonumber \\
F^{\star}_{ \alpha\beta}&=&\frac{1}{2}\epsilon_{\alpha\beta\mu\nu}F^{\mu\nu}\,, \quad F_{\mu \nu} = \partial_{\mu}A_{\nu}-\partial_{\nu}A_{\mu}\,,
\label{eq:uwf}
\end{eqnarray}
where $\star$, $\epsilon_{\alpha\beta\mu\nu}$, and $A_{\mu}$ denote respectively the Hodge dual operator, the Levi-Civita symbol, and the electromagnetic gauge field.

Einstein-Bronnikov gravity admits regular BHs only if these carry magnetic and not electric charge, i.e.\ choosing a purely magnetic configuration for the gauge field:
\begin{eqnarray}
A_{\mu} = q_m\cos\theta\delta_{\mu}^{\phi}\,,
\label{eq:amu}
\end{eqnarray}
where $q_m$ is the magnetic charge. In this case, the central singularity is removed if the parameter $a=q_m^{3/2}/M$ is different from zero. The metric function can be written explicitly as~\cite{Bronnikov:2000vy}:
\begin{eqnarray}
A(r) = 1-\frac{2M}{r} \left ( 1-\tanh \frac{q_m^2}{2Mr} \right ) \,,
\label{eq:metriceb}
\end{eqnarray}
where $q_m$ characterizes a specific hair. As anticipated earlier, however, in NLED theories photons do not move along null geodesics of the metric tensor, but of an effective geometry which depends on the details of the NLED theory. Therefore, knowledge of the metric function alone is in principle \textit{not} sufficient to compute the shadow cast by BHs within NLED theories.

To proceed, we follow Novello's method~\cite{Novello:1999pg,DeLorenci:2000yh} to compute the effective geometry. Since these can now be considered standard references in the field, here we very succinctly review the basic steps of Novello's method, and simply refer the reader to Refs.~\cite{Novello:1999pg,DeLorenci:2000yh} for further details. Novello's method is itself based on Hadamard's method to obtain the equations for the propagation of the electromagnetic field, where the discontinuity surface $\Sigma$ in the electromagnetic field is identified with the wavefronts. In short, one starts from the Bianchi identities involving the field-strength tensor $F_{\mu\nu}$ and its dual. It is assumed that the field-strength is continuous across $\Sigma$, while its first derivative is discontinuous, with the discontinuity parametrized by $f_{\alpha\beta}k_{\mu}$, where $k_{\mu}$ is the propagation vector. These conditions are then applied to the Bianchi identities and the resulting equation(s), taking the form of a cyclic identity (involving the tensor $f$ and the propagation vector) with indices, say, $_{\alpha\beta\mu}$, are contracted with $F^{\alpha\beta}k^{\mu}$ to obtain a scalar equation. These equations are evaluated on-shell and, after simple manipulations, can be expressed in the form $g_{\mu\nu}^{\rm eff}k^{\mu}k^{\nu}=0$, showing that photons move along null geodesics of $g_{\mu\nu}^{\rm eff}$, itself depending on the NLED Lagrangian.

\begin{figure}
\centering
\includegraphics[width=1.0\linewidth]{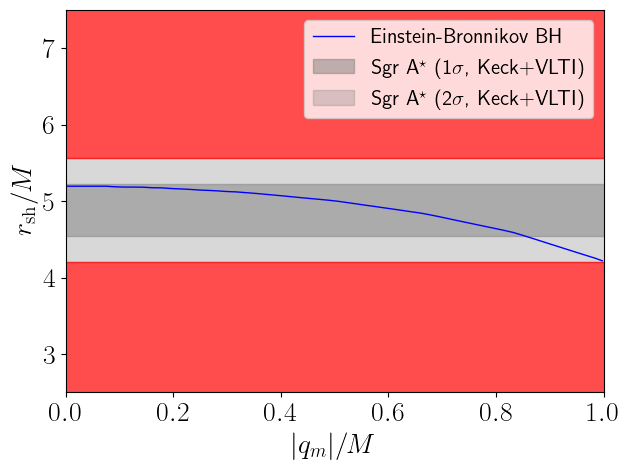}
\caption{Same as in Fig.~\ref{fig:shadow_reissner_nordstrom_bh_ns} for the Einstein-Bronnikov magnetically charged BH with metric function given by Eq.~\eqref{eq:eb}, as discussed in Sec.~\ref{subsec:eb}.}
\label{fig:shadow_bronnikov_bh}
\end{figure}

For the intermediate steps leading to the computation of the effective geometry, we refer the reader to Eqs.~(2.10--2.21) of Ref.~\cite{Allahyari:2019jqz} by some of us (see also Ref.~\cite{Atamurotov:2015xfa}), where these steps are explicitly spelled out. We compute the shadow radius numerically, following the procedure outlined in Ref.~\cite{Allahyari:2019jqz}. In Fig.~\ref{fig:shadow_bronnikov_bh}, we show the evolution of the shadow size as a function of the only free parameter, $\vert q_m \vert$. We see once more that increasing the magnetic charge $q_m$ acts to decrease the shadow radius. We also see that the EHT observations restrict the magnetic charge to $\vert q_m \vert \lesssim 0.8M$ ($1\sigma$) and $\vert q_m \vert \lesssim M$ ($2\sigma$), while noting that within Bronnikov NLED the magnetic charge is in principle allowed to take values $q_m>M$.

\subsection{Ghosh-Culetu-Simpson-Visser regular black hole}
\label{subsec:gcsv}

The Ghosh-Culetu-Simpson-Visser (GCSV) BH was introduced in Refs.~\cite{Ghosh:2014pba,Culetu:2014lca,Simpson:2019mud}, including work by one of us, as as toy model of a regular BH in the sense of Bardeen, where the core is asymptotically Minkowski rather than de Sitter:~\footnote{To the best of our knowledge, this specific metric does not carry a particular name, and here it is the first time it is being referred to as ``GCSV BH''.} stated differently, the associated energy density and pressure asymptote to zero rather than to a final value determined by the effective cosmological constant of the de Sitter case (see also Refs.~\cite{Kumar:2020yem,Banerjee:2022chn,Berry:2020tky,Berry:2020ntz,Ling:2021olm,Simpson:2021biv,Simpson:2021zfl,Ling:2022vrv,Guerrero:2022msp,Banerjee:2022len}). While the GCSV BH is introduced in a purely phenomenological way, Simpson and Visser argued in Ref.~\cite{Simpson:2019mud} that this space-time is mathematically interesting due to its tractability (the curvature tensors and invariants take forms which are much simpler than those of the Bardeen, Hayward, and Frolov BHs), and physically interesting because of its non-standard asymptotically Minkowski core. The metric function describing the GCSV space-time is given by~\cite{Ghosh:2014pba,Culetu:2014lca,Simpson:2019mud}:
\begin{eqnarray}
A(r) = 1-\frac{2Me^{-g^2/2Mr}}{r}\,,
\label{eq:metricgcsv}
\end{eqnarray}
where depending on the underlying origin for this space-time, $g$ can characterize either an universal or specific hair. As shown by some of us in Refs.~\cite{Kumar:2020ltt,Singh:2022xgi}, the space-time described by Eq.~(\ref{eq:metricgcsv}) can be obtained in the context of GR coupled to a suitable NLED source, with $g$ associated to the NLED charge.

\begin{figure}
\centering
\includegraphics[width=1.0\linewidth]{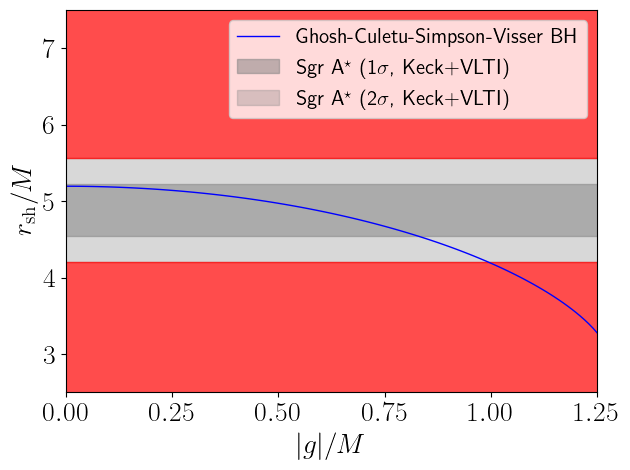}
\caption{Same as in Fig.~\ref{fig:shadow_reissner_nordstrom_bh_ns} for the Ghosh-Culetu-Simpson-Visser BH with metric function given by Eq.~\eqref{eq:metricgcsv}, as discussed in Sec.~\ref{subsec:gcsv}.}
\label{fig:shadow_gcsv_bh}
\end{figure}

We compute the shadow size numerically, and show its evolution against $\vert g \vert$ in Fig.~\ref{fig:shadow_gcsv_bh}. As for all solutions considered so far, we see that increasing the hair parameter decreases the shadow size, as also observed in the earlier work of Ref.~\cite{Kumar:2020yem} by some of us. The EHT observations set the limits $\vert g \vert \lesssim 0.8M$ ($1\sigma$) and $\vert g \vert \lesssim M$ ($2\sigma$). We are not aware of any a priori theoretical restriction on the value of $g$, given that the metric has been introduced as a toy model.

\subsection{Kazakov-Solodukhin regular black hole}
\label{subsec:ks}

As a further example of regular BH, we consider the Kazakov-Solodukhin (KS) BH, which arises within a string-inspired model where spherically symmetric quantum fluctuations of the metric and the matter fields are governed by the 2D dilaton-gravity action~\cite{Kazakov:1993ha}. The KS BH then provides a well-motivated example of quantum deformation of a Schwarzschild BH. The metric function for this space-time is given by:
\begin{eqnarray}
A(r) = -\frac{2M}{r}+\frac{\sqrt{r^2-\ell^2}}{r} \,,
\label{eq:metricks}
\end{eqnarray}
where the hair parameter $\ell>0$ sets the scale over which quantum deformations of the Schwarzschild BH shift the central singularity to a finite radius, and characterizes an universal hair. Physically speaking, what effectively occurs is that the singularity is smeared out over a two-dimensional sphere of area $4\pi\ell^2$. While $\ell$ is expected to be of the order of the Planck length $\ell_{\rm Pl}$, in a phenomenological approach it can in principle take any positive value, as done for instance in Ref.~\cite{Kocherlakota:2020kyu}.

\begin{figure}
\centering
\includegraphics[width=1.0\linewidth]{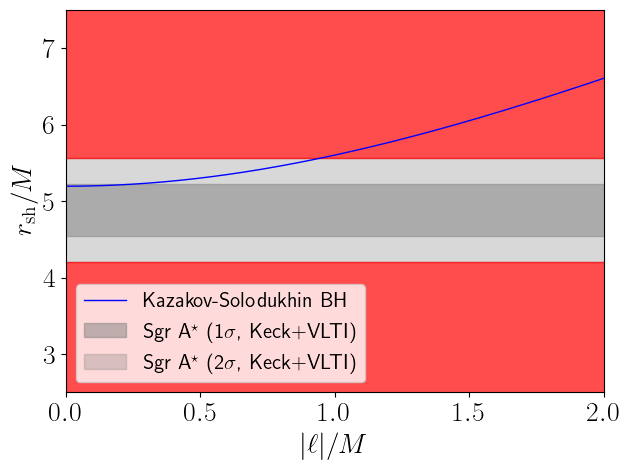}
\caption{Same as in Fig.~\ref{fig:shadow_reissner_nordstrom_bh_ns} for the Kazakov-Solodukhin BH with metric function given by Eq.~\eqref{eq:metricks}, as discussed in Sec.~\ref{subsec:ks}.}
\label{fig:shadow_ks_bh}
\end{figure}

We find that a closed-form expression for the shadow radius can be provided, and is given by:
\begin{eqnarray}
r_{\rm sh} &=& \frac{\sqrt[4]{ \left ( 9+3\ell^2+3\sqrt{9+2\ell^2} \right ) ^3}}{{\sqrt{-4\sqrt{2}+2\sqrt{9+\ell^2+3\sqrt{9+2\ell^2}}}}}\,,
\label{eq:rshks}
\end{eqnarray}
which correctly reduces to $3\sqrt{3}$ in the limit $\ell \to 0$. The evolution of the shadow size $r_{\rm sh}$ is shown against $\vert \ell \vert/M$ in Fig.~\ref{fig:shadow_ks_bh}. The KZ BH provides the first example, among those considered, of a space-time where the size of the BH shadow increases with the hair parameter. As per our previous discussion, it is easy to show that the location of the maximum of the effective potential for the KS space-time moves to the right with respect to the Schwarzschild case for any value of $\ell>0$. The EHT observations set the limits $\vert \ell \vert \lesssim 0.2M$ ($1\sigma$) and $\vert \ell \vert \lesssim M$ ($2\sigma$). The physical meaning of this limit is that the singularity cannot be spread out across a scale larger than the gravitational radius, and in any case has to be confined within the event horizon. This example also highlights how the EHT near-horizon image of Sgr A$^*$ places particularly tight limits on fundamental physics scenarios which lead to a larger shadow compared to the Schwarzschild BH.

\subsection{Ghosh-Kumar black hole}
\label{subsec:ghoshkumar}

Another interesting BH solution, which we shall refer to as the Ghosh-Kumar (GK) BH, was proposed by some of us in Ref.~\cite{Ghosh:2021clx}, and is described by the following metric function:
\begin{eqnarray}
A(r) = 1-\frac{2M}{\sqrt{r^2+k^2}} \,,
\label{eq:metricghoshkumar}
\end{eqnarray}
where depending on the underlying origin for this space-time, $k$ can characterize either an universal or specific hair. Although the above metric function is regular throughout the whole space-time, there is nonetheless a scalar polynomial singularity at $r=0$, which can be seen by computing the Kretschmann and Ricci scalars~\cite{Ghosh:2021clx}. In addition, the space-time exhibits an interesting reflection symmetry $r \to -r$. This can be viewed as a minimal deformation of the Schwarzschild BH, and is similar to the Simpson-Visser space-time we will consider later in Sec.~\ref{subsec:simpsonvisser}, albeit with a crucial difference in the angular metric function, which here is the standard $C(r)=r^2$.

\begin{figure}
\centering
\includegraphics[width=1.0\linewidth]{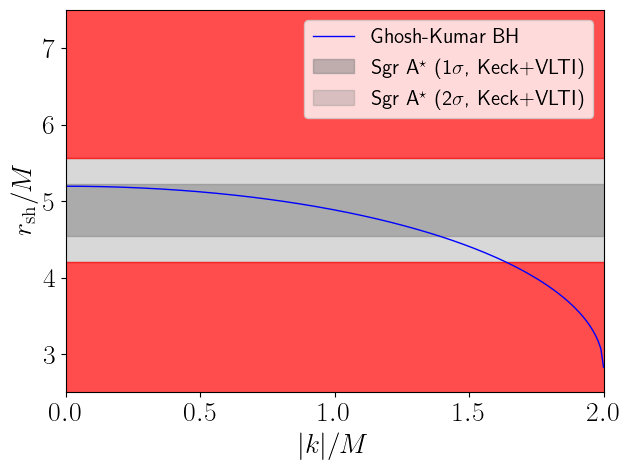}
\caption{Same as in Fig.~\ref{fig:shadow_reissner_nordstrom_bh_ns} for the Ghosh-Kumar BH with metric function given by Eq.~\eqref{eq:metricghoshkumar}, as discussed in Sec.~\ref{subsec:ghoshkumar}.}
\label{fig:shadow_ghosh_kumar_bh}
\end{figure}

We find that a closed-form expression for the shadow radius exists and is given by:
\begin{eqnarray}
r_{\rm sh} = \frac{\sqrt{{\cal X}-k^2 \left ( 1+\frac{2}{{\cal X}}\right ) +\frac{9}{{\cal X}}+3}}{\sqrt{1-\frac{2}{\sqrt{{\cal X}+\frac{9}{{\cal X}}-2\frac{k^2}{{\cal X}}+3}}}}\,,
\label{eq:rshghoshkumar}
\end{eqnarray}
where the quantity ${\cal X}$ is defined as:
\begin{eqnarray}
{\cal X} \equiv \sqrt[3]{\frac{k^4-18k^2+\sqrt{k^6 \left ( k^2-4 \right ) } +54}{2}}\,.
\label{eq:xghoshkumar}
\end{eqnarray}
It is easy to show that, with ${\cal X}$ defined as in Eq.~(\ref{eq:xghoshkumar}), $r_{\rm sh}$ correctly approaches $3\sqrt{3}$ in the limit $k \to 0$, where ${\cal X} \to 3$. We show the evolution of the shadow size against $\vert k \vert/M$ in Fig.~\ref{fig:shadow_ghosh_kumar_bh}. We see that increasing $\vert k \vert$ decreases the shadow size, as with most BH solutions studied so far. We find that the EHT observations set the upper limits on the NLED charge $\vert k \vert \lesssim 1.4M$ ($1\sigma$) and $\vert k \vert \lesssim 1.6M$ ($2\sigma$). To the best of our knowledge, these are the first ever constraints on this space-time.~\footnote{In Ref.~\cite{Ghosh:2021clx}, it was shown that this space-time can emerge within a specific NLED theory, where $k$ is related to the NLED charge. Following similar considerations as for the Bardeen BH in Sec.~\ref{subsec:bardeen}, here we have computed the shadow size following the phenomenological/model agnostic approach where we do not question the underlying origin of the GK BH.}

\subsection{Simpson-Visser regular black hole and traversable wormhole}
\label{subsec:simpsonvisser}

We now start moving away from regular BH metrics, considering a metric encompassing both a regular BH as well as a traversable wormhole (WH) regime. A WH is a (typically tunnel-like) space-time structure with a non-trivial topology, which can connect either two distant regions of the same Universe, or two different Universes (see e.g.\ Ref.~\cite{Bambi:2021qfo} for a review of astrophysical WHs). In most cases, the field content needed to sustain the WH geometry requires a violation of the null energy condition~\cite{Morris:1988tu,Visser:2003yf}, either in the form of exotic matter or a modification to Einstein's gravity.~\footnote{See for instance Refs.~\cite{Richarte:2007zz,Eiroa:2008hv,Lobo:2009ip,Richarte:2010bd,Kanti:2011jz,Boehmer:2012uyw,Bolokhov:2012kn,Capozziello:2012hr,Harko:2013yb,Richarte:2013fek,DiCriscienzo:2013ria,Shaikh:2015oha,Bahamonde:2016jqq,Ovgun:2016ujt,Moraes:2016akv,Ovgun:2017zao,Shaikh:2017zfl,Ovgun:2017jip,Jusufi:2017vta,Moraes:2017mir,Jusufi:2017mav,Moraes:2017dbs,Calza:2018ohl,Jusufi:2018kmk,Ovgun:2018xys,Amir:2018pcu,Shaikh:2018yku,Javed:2019qyg,Antoniou:2019awm,Singh:2020rai,Mishra:2021ato,DeFalco:2021btn,Benavides-Gallego:2021lqn,Karakasis:2021tqx,Mustafa:2021ykn,Sadeghi:2021pqg,Agrawal:2021gdq,Mishra:2021xfl,Moraes:2022myd,Kar:2022omn,Abdulxamidov:2022ofi,Hassan:2022ibc,Rosa:2022osy,Capozziello:2022zoz,Gao:2022cds,Cai:2023ite} for a diverse range of examples of WH solutions within modified theories of gravity which violate the null energy condition in a controlled way.}  The simplest WH configurations feature two ``mouths'' connected by a ``throat'', but more complex structures are possible.

In Ref.~\cite{Simpson:2018tsi}, Simpson and Visser proposed a one-parameter extension of the Schwarzschild space-time which interpolates between the latter and a traversable WH, while passing through a regular BH geometry. The Simpson-Visser (SV) space-time is described by the following metric functions:
\begin{eqnarray}
A(r) = 1-\frac{2M}{\sqrt{r^2+a^2}}\,,\quad C(r)=r^2+a^2 \,,
\label{eq:metricsimpsonvisser}
\end{eqnarray}
where the new parameter is required to satisfy $a>0$, and characterizes a specific hair. For $a=0$, Eq.~(\ref{eq:metricsimpsonvisser}) reduces to the Schwarzschild space-time. On the other hand, for $0<a<2M$, the metric describes the space-time of a regular BH with a one-way space-like throat. For $a=2M$, the space-time corresponds to that of a one-way WH with an extremal null throat. Finally, for $a>2M$, Eq.~(\ref{eq:metricsimpsonvisser}) describes a traversable WH with a two-way time-like throat. We also note that when $M=0$, Eq.~(\ref{eq:metricsimpsonvisser}) reduces to the metric of the Ellis-Bronnikov WH~\cite{Ellis:1973yv,Bronnikov:1973fh}. The SV space-time and extensions thereof have received significant attention in the literature, attracting several follow-up studies, see Refs.~\cite{Carballo-Rubio:2019nel,Huang:2019arj,Carballo-Rubio:2019fnb,Churilova:2019cyt,Bronnikov:2019sbx,Lobo:2020kxn,Nascimento:2020ime,Lima:2020auu,Lobo:2020ffi,Zhou:2020zys,Ayuso:2020vuu,Tsukamoto:2021fpp,Mazza:2021rgq,Bronnikov:2021liv,Cheng:2021hoc,Islam:2021ful,Franzin:2021vnj,Yang:2021diz,Guerrero:2021ues,Chakrabarti:2021gqa,Yang:2021cvh,Jiang:2021ajk,Stuchlik:2021tcn,Xu:2021lff,Bambhaniya:2021ugr,Ou:2021efv,Guo:2021wid,Tsukamoto:2022vkt,Yang:2022xxh,Calza:2022ioe} for examples of representative works.

As one sees from Eq.~(\ref{eq:metricsimpsonvisser}), the SV space-time is the first example of space-time we consider where $C(r) \neq r^2$. As a consequence, the coordinate $r$ is in general \textit{not} the areal radius, unless $a=0$. This is also the key difference between the SV space-time and the GK BH we studied previously, which are otherwise very similar, and underlies several peculiar features of the SV space-time we shall now discuss. The properties of the SV space-time are easy to compute analytically. To compute the location of the photon sphere we use Eq.~(\ref{eq:photonspherecomplete}), which upon inserting $A(r)$ and $C(r)$ as given by Eq.~(\ref{eq:metricsimpsonvisser}), and working in units where $M=1$, reduces to:
\begin{eqnarray}
&&2r \left ( 1-\frac{2}{\sqrt{r^2+a^2}} \right ) - (r^2+a^2) \frac{2r}{\sqrt{(r^2+a^2)^3}}\nonumber \\
&=&2r \left ( 1-\frac{3}{\sqrt{r^2+a^2}} \right ) = 0\,,
\label{eq:photonspheresimpsonvisser}
\end{eqnarray}
from which we trivially find the photon sphere location as being $r_{\rm ph}=\sqrt{9-a^2}$, as also obtained in Ref.~\cite{Simpson:2018tsi}. Note that these results are only valid for $a<3$: for $a \geq 3$, the space-time does not possess a photon sphere (although, we stress, this does not necessarily mean that it cannot cast a shadow).

\begin{figure}
\centering
\includegraphics[width=1.0\linewidth]{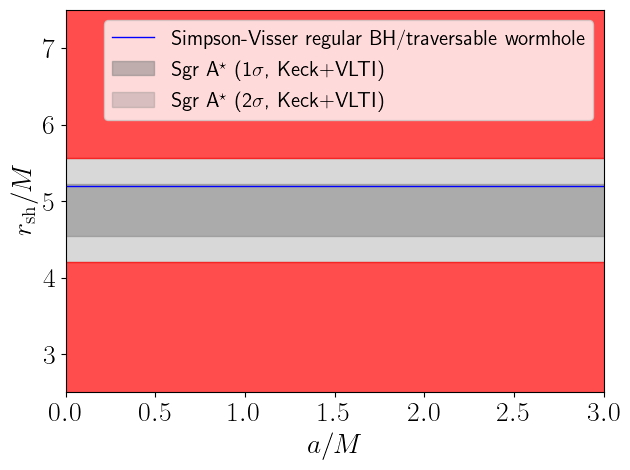}
\caption{Same as in Fig.~\ref{fig:shadow_reissner_nordstrom_bh_ns} for the Simpson-Visser regular BH and traversable wormhole with metric functions given by Eq.~\eqref{eq:metricsimpsonvisser}, as discussed in Sec.~\ref{subsec:simpsonvisser}.}
\label{fig:shadow_simpson_visser_bh_wh}
\end{figure}

To obtain the shadow size, we use the more complete expression given by Eq.~(\ref{eq:rshcomplete}), finding that $r_{\rm sh}=3\sqrt{3}$, with no dependence on the parameter $a$. Therefore, the shadow cast by the SV space-time is in perfect agreement with the EHT observations within all four the singular BH, regular BH, one-way WH, and traversable WH regimes, as shown in Fig.~\ref{fig:shadow_simpson_visser_bh_wh}. While phenomenological, this is therefore the first example (among those we have considered) of a BH mimicker space-time which could be a viable candidate in light of Sgr A$^*$'s shadow.~\footnote{While the SV space-time was introduced on purely phenomenological grounds in Ref.~\cite{Simpson:2018tsi}, it can emerge as a solution of GR coupled to a suitable NLED source and a minimally coupled phantom scalar field with an appropriate potential, as shown for the first time by one of us in Ref.~\cite{Bronnikov:2021uta}. Following similar considerations as for the Bardeen BH in Sec.~\ref{subsec:bardeen}, here we have computed the shadow size following the phenomenological/model agnostic approach where we do not question the underlying origin of the SV metric.} On the other hand, complementary X-ray reflection spectroscopy constraints~\cite{Reynolds:2013qqa} obtained by some of us set the very tight limit $a \lesssim 0.5M$ at $90\%$~C.L.~\cite{Riaz:2022rlx}, which is particularly tight since it is obtained from observations of the stellar-mass X-ray binary candidate BH EXO 1846-031, whose mass is 6 orders of magnitude lower than Sgr A$^*$.

The somewhat surprising result that the shadow cast by the SV space-time is identical in size to that of the Schwarzschild BH deserves further comments. The fact that the SV and Schwarzschild space-times are \textit{shadow-degenerate} had already been noted in Ref.~\cite{Lima:2021las}, and the easiest way of seeing this is to perform a coordinate transformation and express the metric described by Eq.~(\ref{eq:metricsimpsonvisser}) in terms of the areal radius $R=\sqrt{r^2+a^2}$. In these new coordinates, $g_{tt}$, $g_{\theta\theta}$, and $g_{\phi\phi}$ are exactly of the Schwarzschild form, whereas the only deviations from the Schwarzschild form appear in $g_{RR}$, which however does not enter in the computation of the shadow radius as is clear from Eqs.~(\ref{eq:hr}--\ref{eq:rsh}). More generally, Ref.~\cite{Lima:2021las} shows that the condition for a space-time to be shadow-degenerate with the Schwarzschild space-time is that the dominant photon sphere of both has the same impact parameter, after correcting for the potentially different redshift of comparable observers in the two space-times. Besides the SV space-time, Ref.~\cite{Lima:2021las} provides another example of a space-time which is shadow-degenerate with the Schwarzschild space-time, termed ``Class II''.

Finally, as anticipated earlier, we note that the absence of a photon sphere for $a>3$ does not necessarily imply that the SV space-time cannot cast a shadow in this regime. In fact, as noted in Ref.~\cite{Tsukamoto:2020bjm}, for $a>3$ the WH throat can act as photon sphere. For the specific case of $a=3$, two photon spheres and an anti-photon sphere at the throat degenerate into an effective marginally unstable photon sphere. This allows the SV space-time to cast a shadow. Given the significantly more complicated behavior of the system in the $a \geq 3$ regime, here we choose not to cover it further, and refer the reader to Ref.~\cite{Tsukamoto:2020bjm} for more detailed discussions on the subject.

\subsection{Morris-Thorne traversable wormhole}
\label{subsec:morristhorne}

We now consider what is arguably the simplest example of non-rotating, asymptotically flat, traversable wormhole, described by the well-known Morris-Thorne (MT) metric~\cite{Morris:1988cz}:
\begin{eqnarray}
{\rm d}s^2 = -e^{\Phi(r)}{\rm d}t^2+\frac{{\rm d}r^2}{1-\Psi(r)}+r^2{\rm d}\Omega^2\,,
\label{eq:metricmorristhorne}
\end{eqnarray}
where $\Phi(r)$ is referred to as the redshift function, and $\Psi(r)$ as the shape function (although the reader should note that different definitions and conventions for the redshift and shape functions exist in the WH literature). One of the simplest choices for these two in principle free functions consistent with a WH geometry is to set $\Phi=-r_0/r$ and $\Psi(r)=(r_0/r)^{\gamma}$~\cite{Morris:1988cz,Lemos:2003jb,Harko:2008vy,Bambi:2013jda,Bambi:2013nla}, where $r_0$ is the radial coordinate of the WH throat, and $\gamma$ is a constant which we shall henceforth fix to $\gamma=1$. For this choice of redshift and shape functions, it is easy to show that the Arnowitt-Deser-Misner (ADM) mass~\cite{Arnowitt:1959ah,Arnowitt:1962hi} is $M=r_0/2$, so the metric function can be rewritten as:
\begin{eqnarray}
A(r) = e^{-\frac{2M}{r}}\,.
\label{eq:metrictraversablewormhole}
\end{eqnarray}
Importantly, the size of the wormhole shadow in units of the ADM mass is then uniquely determined and does not depend on additional hair parameters, given that the only free parameter in the theory ($\gamma$) enters exclusively in $g_{rr}$, which does not affect the computation of the shadow.~\footnote{Note that in Eq.~(\ref{eq:metrictraversablewormhole}) the radial coordinate is only defined down to the WH throat, and therefore the limit $r \to 0$ (where formally $A(r) \to 0$) should not be taken. The location of the WH throat is controlled by the shape function, which does not in itself affect the computation of the shadow, if not implicitly through the parameter $\gamma$ which controls the ADM mass.}

\begin{figure}
\centering
\includegraphics[width=1.0\linewidth]{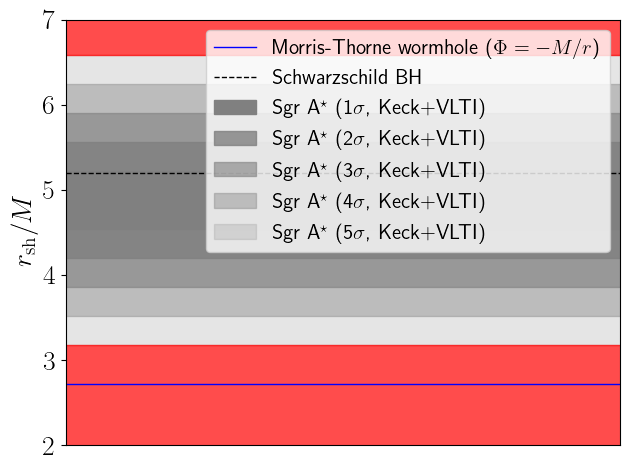}
\caption{Same as in Fig.~\ref{fig:shadow_reissner_nordstrom_bh_ns} for the Morris-Thorne traversable wormhole described by the metric in Eq.~\eqref{eq:metricmorristhorne}, whose shadow size is $r_{\rm sh}=eM$ and does not depend on any free parameters. Also shown are the $1\sigma$, $2\sigma$, $3\sigma$, $4\sigma$, and $5\sigma$ constraints derived from the EHT observations, given by different shades of gray. The red regions are excluded by the same observations at more than $5\sigma$. For comparison, the size of the shadow of a Schwarzschild BH with the same ADM mass is also shown (dashed black curve).}
\label{fig:shadow_morris_thorne_wh}
\end{figure}

For the choice of redshift function we have adopted, the equations governing the location of the photon sphere and the corresponding shadow size simplify considerably. The equation for $r_{\rm ph}$ takes the form $r_{\rm ph}(1-r_{\rm ph})=0$, for which the physical solution is $r_{\rm ph}=1$. Then, the shadow size is given by $r_{\rm sh}=1/\sqrt{e^{-2}}=e$. In other words, the size of the MT WH shadow in units of ADM mass is exactly equal to Euler's number, $r_{\rm sh}/M=e \approx 2.718$, i.e.\ almost a factor of 2 smaller than the size of the shadow of a Schwarzschild BH with the same ADM mass. By direct comparison, and as is visually clear in Fig.~\ref{fig:shadow_morris_thorne_wh}, the EHT measurements rule out at very high significance ($>5\sigma$) the possibility of Sgr A$^*$ being a MT traversable WH, for the particular well-motivated choice of redshift function $\Phi=-r_0/r$. Finally, we note that this result has been also been discussed by the EHT collaboration in Ref.~\cite{EventHorizonTelescope:2022xqj}.~\footnote{As per earlier discussions, one has to ensure that the size of the WH shadow is larger than that of the WH throat. For our choice of $\gamma=1$, it is easy to show that the throat is located at $r_0=2M$. Since the shadow size is $r_{\rm sh}=eM>2M$, the shadow size we have computed is indeed astrophysically relevant.}

\subsection{Damour-Solodukhin wormhole}
\label{subsec:damoursolodukhin}

Another interesting WH solution was proposed by Damour and Solodukhin in Ref.~\cite{Damour:2007ap} (see also Refs.~\cite{Nandi:2018mzm,Ovgun:2018fnk,Karimov:2019qfw,Tsukamoto:2020uay,Matyjasek:2020cmi} for further relevant work), with the goal of studying the extent to which globally static WHs can mimic observational features of BHs, including accretion of matter, no-hair properties, ringing of quasi-normal modes, and dissipative properties associated to the event horizon. The Damour-Solodukhin (DS) WH is described by the following metric~\cite{Damour:2007ap}:
\begin{eqnarray}
{\rm d}s^2 = - \left ( 1 - \frac{2M}{r} + \lambda^2 \right ) {\rm d}t^2+\frac{{\rm d}r^2}{1-\frac{2M}{r}}+r^2{\rm d}\Omega^2 \,,
\label{eq:metricdamoursolodukhin}
\end{eqnarray}
and differs from the Schwarzschild BH because of the presence of the dimensionless parameter $\lambda$, which characterizes an universal hair. When $\lambda \neq 0$, Eq.~(\ref{eq:metricdamoursolodukhin}), the would-be horizon at $r=2M$ actually corresponds to a throat joining two isometric regions. Note that the would-be Killing horizon where $g_{tt}=0$ would actually be located at $r=2M/(1+\lambda^2)<2M$, so behind the throat. As discussed in Ref.~\cite{Damour:2007ap}, this space-time provides an example of Lorentzian WH,~\footnote{That this space-time describes a WH and not a BH can explicitly be seen adopting the Kodama-Hayward invariant formalism (see e.g.\ Refs.~\cite{Kodama:1979vn,Hayward:2008jq,DiCriscienzo:2009kun,Vanzo:2011wq,Sebastiani:2022wbz}), and verifying whether the Kodama energy at the would-be horizon is $0$ (BH) or $>0$ (WH).} where the throat describes a brane-like structure (a high-tension distribution), with radial and tangential tensions proportional to $1/\lambda^2$ but with zero energy density, localized in a thin shell around the throat. We also note that a similar space-time was studied in 3D in Ref.~\cite{Solodukhin:2005qy}, with the goal of restoring Poincar\'{e} recurrence in BHs: in this specific case, $\lambda \sim e^{-4\pi GM^2}$ is exponentially small, so as to mimic the expected dependence of the Poincar\'{e} recurrence time on the BH entropy. In Ref.~\cite{Damour:2007ap}, it was argued that choosing exponentially small values of $\lambda$ allows the resulting WH space-time to mimic both classical and quantum properties of Schwarzschild BHs.

\begin{figure}
\centering
\includegraphics[width=1.0\linewidth]{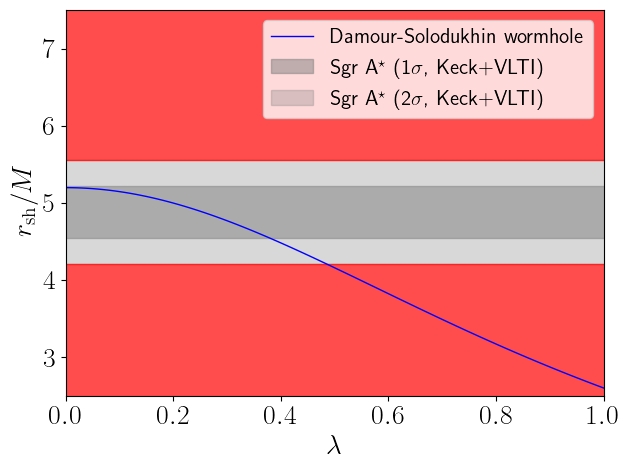}
\caption{Same as in Fig.~\ref{fig:shadow_reissner_nordstrom_bh_ns} for the Damour-Solodukhin wormhole with metric function given by Eq.~\eqref{eq:metricdamoursolodukhin}, as discussed in Sec.~\ref{subsec:damoursolodukhin}.}
\label{fig:shadow_damour_solodukhin_wh}
\end{figure}

In the following, we treat $\lambda$ as a free parameter and simply ask what values are compatible with the observed shadow of Sgr A$^*$, assuming that the latter is indeed described by a DS WH. Note that as soon as $\lambda \neq 0$ the space-time is no longer asymptotically flat, so in order to compute its shadow we adopt the more general computation given by Eqs.~(\ref{eq:alphashnotasymptoticallyflat},\ref{eq:rshnotasymptoticallyflat}). Following this procedure, we can easily obtain a closed-form expression for the shadow radius as follows:
\begin{eqnarray}
r_{\rm sh} \simeq \frac{3\sqrt{3}}{1+\lambda^2}\,.
\label{eq:rshdamoursolodukhin}
\end{eqnarray}
where the $\simeq$ sign is included as we have made the approximation $r_O \gg 1$ (in units $M=1$), valid for Sgr A$^*$. We plot the evolution of the shadow size against $\lambda$ in Fig.~\ref{fig:shadow_damour_solodukhin_wh}, where we see that increasing $\lambda$ decreases the size of the WH shadow. In particular, we find that the EHT observations set the upper limits $\lambda \lesssim 0.4$ ($1\sigma$) and $\lambda \lesssim 0.5$ ($2\sigma$). To the best of our knowledge, these are the first robust observational constraints on $\lambda$. We note of course that our limits are consistent with the exponentially small values of $\lambda$ suggested in Refs.~\cite{Solodukhin:2005qy,Damour:2007ap}, for which the impact on the shadow size is negligible as we could have expected.

\subsection{Janis-Newman-Winicour naked singularity} 
\label{subsec:jnw}

Having discussed a few interesting WH solutions, we now turn our attention to another fascinating class of BH mimickers: naked singularities (see Ref.~\cite{Joshi:2011rlc} for a review). The cosmic censorship conjecture, taken at face value, appears to forbid the existence of naked singularities~\cite{Penrose:1964wq}. In addition, the end result of gravitational collapse is commonly believe to be a BH, with the central singularity (if any) surrounded by an event horizon, a result which seemingly enjoys support from Birkhoff's theorem as well~\cite{Birkhoff:1923ghw}. The first gravitational collapse solution of a spherically symmetric and homogeneous dust cloud (the so-called OSD collapse) was derived in 1939 by Oppenheimer, Snyder, and Datt~\cite{Oppenheimer:1939ue,Datt:1938ghw}, and does indeed recover the Schwarzschild BH as final state. Within the OSD collapse scenario, trapped surfaces will form around the center before the formation of the central space-like singularity, which is therefore causally disconnected from all other points of space-time.

However, the OSD collapse model (as well as subsequent refinements thereof) is highly idealized in its assuming homogeneous density and zero pressure within the massive collapsing star: this is not expected to be the case for realistic astrophysical stars. Under physically more realistic situations, the introduction of inhomogeneities in the matter density profile and a non-zero pressure \textit{within GR} leads to the possibility of dynamical collapse resulting in a strong but non-space-like curvature singularity, not hidden behind an event horizon~\cite{Ori:1987hg,Lake:1991qrk,Joshi:1993zg,Harada:1998cq,Joshi:2001xi,Giambo:2003fd,Goswami:2006ph,Patil:2011yb,Chowdhury:2011aa,Shaikh:2018lcc}. Therefore, given suitable initial conditions, continuous gravitational collapse of an inhomogeneous matter cloud can lead to the formation of a class of exotic horizonless compact objects known as naked singularities (NSs), which have received significant attention, particularly in the recent literature (see e.g.\ Refs.~\cite{Virbhadra:1998kd,Ghosh:2001pv,Casadio:2001wh,Banerjee:2002sy,Debnath:2003fi,Miyamoto:2004ba,Torok:2005ct,Maeda:2007tk,Virbhadra:2007kw,Barausse:2010ka,Patil:2011ya,Sahu:2012er,Stuchlik:2014iia,Liu:2018bfx,Jusufi:2018gnz,Hioki:2019gnv,Paul:2020ufc,Tsukamoto:2021fsz,Tsukamoto:2021lpm,Atamurotov:2022srw}).

One of the simplest known NS solutions is the Janis-Newman-Winicour (JNW) NS. This is a solution of the Einstein-Klein-Gordon equations for a massless scalar field~\cite{Janis:1968zz} (see also Refs.~\cite{Fisher:1948yn,Bronnikov:1979uz,Wyman:1981bd,Virbhadra:1997ie}). Despite being a NS, it possesses a photon sphere and therefore casts a shadow. The JNW space-time is characterized by a scalar charge which one can continuously adjust to recover the Schwarzschild solution. Given its importance, this space-time has been the subject of a large number of studies, particularly regarding the possibility of observationally distinguishing it and other NSs from the Schwarzschild BH, see e.g.\ Refs.~\cite{Gyulchev:2008ff,Kovacs:2010xm,Chakraborty:2016ipk,Chakraborty:2016mhx,Rizwan:2018lht,Shaikh:2018lcc,Shaikh:2019hbm,Gyulchev:2019tvk,Capozziello:2020szy,Gyulchev:2020cvo,Chowdhury:2020rfj}.

The JNW space-time can be characterized by the following metric functions~\cite{Janis:1968zz}:
\begin{eqnarray}
A(r) = \left [ 1-\frac{2M}{r(1-\overline{\nu})} \right ] ^{1-\overline{\nu}}\,,\quad C(r)=r^2 \left [ 1-\frac{2M}{r(1-\overline{\nu})} \right ] ^{\overline{\nu}}\,, \nonumber \\
\label{eq:metricjnw}
\end{eqnarray}
Here, $\overline{\nu} = 1-\sqrt{1+q/M}$, $q$ is the scalar charge of the naked singularity, and the scalar field radial profile is given by the following:
\begin{eqnarray}
\phi(r) = \frac{q}{2\sqrt{M^2+q^2}} \left ( 1-\frac{2\sqrt{M^2+q^2}}{r} \right )\,.
\label{eq:scalarfieldprofile}
\end{eqnarray}
Overall, the parameter $\overline{\nu}$ characterizes a specific hair. Note that, since $C(r) \neq r^2$, the coordinate $r$ appearing in Eq.~(\ref{eq:metricjnw}) is not the areal radius, which can instead be obtained through an appropriate transcendental coordinate transformation. We also note that the JNW space-time can only cast a shadow for $0 \leq \overline{\nu} \leq 0.5$. 

\begin{figure}
\centering
\includegraphics[width=1.0\linewidth]{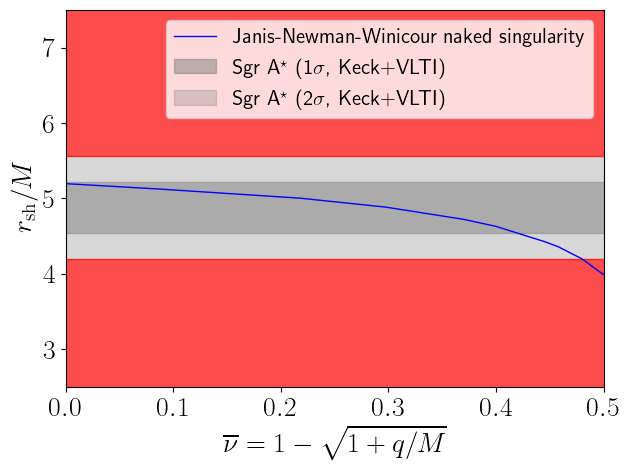}
\caption{Same as in Fig.~\ref{fig:shadow_reissner_nordstrom_bh_ns} for the JNW naked singularity with metric functions given by Eq.~\eqref{eq:metricjnw}, as discussed in Sec.~\ref{subsec:jnw}.}
\label{fig:shadow_jnw_ns}
\end{figure}

We compute the shadow radius numerically, and plot its evolution against the hair parameter $\overline{\nu}$ in Fig.~\ref{fig:shadow_jnw_ns}. There is a wide range of parameter space where Sgr A$^*$'s shadow is consistent with the corresponding compact object being a JNW NS. In fact, the EHT observations set rather weak limits on $\overline{\nu}$, with $\overline{\nu} \lesssim 0.4M$ ($1\sigma$) and $\overline{\nu} \lesssim 0.45M$ ($2\sigma$), barely excluding the extremal JNW NS, $\overline{\nu}=0.5$. These limits can of course be translated into rather weak limits on the strength of the scalar field from Eq.~(\ref{eq:scalarfieldprofile}). More importantly, among all the examples of BH mimickers we have considered so far, the JNW naked singularity is one of the few which is in excellent agreement with the EHT observations given the metric tests we have performed in this work.

\subsection{Joshi-Malafarina-Narayan naked singularity}
\label{subsec:jmn1}

Another well-known NS solution is the Joshi-Malafarina-Narayan (JMN) NS, developed by one of us in Ref.~\cite{Joshi:2011zm}. To distinguish it from another NS solution developed by the same authors in Ref.~\cite{Joshi:2013dva} we shall refer to the solution considered here as JMN singularity of the first type, or JMN-1 for short, as common practice in the literature. The JMN-1 NS can be formed as the end state of gravitational collapse given zero radial pressure but non-zero tangential pressure, i.e.\ in the presence of an anisotropic fluid. As shown in Ref.~\cite{Joshi:2011zm}, the non-zero tangential pressure can prevent trapped surfaces from forming around the high-density core region of the collapsing matter cloud, resulting in the formation of a central NS over the course of a large co-moving time (see also Refs.~\cite{Dey:2019fja,Pal:2022cxb}).

The JMN-1 space-time contains a compact region filled with an anisotropic fluid for $r<R_b$, which is then usually matched to an exterior space-time. The interior JMN-1 space-time is described by the following metric:
\begin{eqnarray}
{\rm d}s^2 = (\chi-1) \left ( \frac{r}{R_b} \right ) ^\frac{\chi}{(1- \chi)}{\rm d}t^2 + \frac{{\rm d}r^2}{(1 - \chi)} + r^2{\rm d}\Omega^2\,,
\label{eq:metricjmn1interior}
\end{eqnarray}
whose associated ADM mass is $M=\chi R_b/2$, with $\chi$ characterizing specific hair, and whose energy density $\rho$, radial pressure $p_r$, and tangential pressure $p_{\theta}$ are given by:
\begin{eqnarray}
\rho = \frac{\chi}{r^2}\,, \quad p_r=0\,, \quad p_{\theta}=\frac{\chi}{4(1-\chi)}\rho\,,
\label{eq:energydensitypressure}
\end{eqnarray}
satisfying all the energy conditions. It can be shown that the interior JMN-1 space-time can be matched smoothly at $r=R_b$ to an exterior Schwarzschild space-time with the same mass $M$, since the induced metrics and extrinsic curvatures of the two space-times can be matched on the matching hypersurface (see Ref.~\cite{Bambhaniya:2019pbr}).

\begin{figure}
\centering
\includegraphics[width=1.0\linewidth]{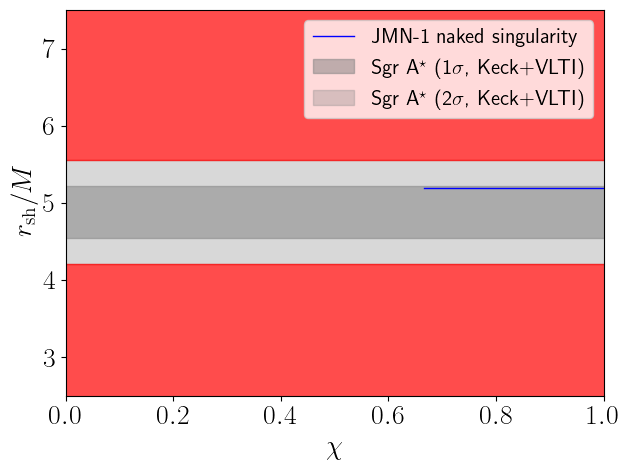}
\caption{Same as in Fig.~\ref{fig:shadow_reissner_nordstrom_bh_ns} for the JMN-1 naked singularity with interior metric given by Eq.~\eqref{eq:metricjmn1interior} and smoothly matched to an exterior Schwarzschild metric at $r=R_b$, as discussed in Sec.~\ref{subsec:jmn1}.}
\label{fig:shadow_jmn1_ns}
\end{figure}

The shadow cast by the JMN-1 space-time was studied in a number of works by some of us, including Refs.~\cite{Shaikh:2018lcc,Kaur:2021wgy,Saurabh:2022jjv}. It can be shown that the JMN-1 space-time does not support a photon sphere and cannot cast a shadow for $0<\chi<2/3$ (which can be re-written in the somewhat more illuminating form $R_b>3M$). On the other hand, for $2/3 \leq \chi<1$, the JMN-1 space-time casts a shadow of size $r_{\rm sh}=3\sqrt{3}\chi R_b/2=3\sqrt{3}M$, i.e.\ identical to that of a Schwarzschild BH with the same mass $M$. This is shown in Fig.~\ref{fig:shadow_jmn1_ns}. The physical reason for this is that ``shadow photons'' are actually traveling in the exterior Schwarzschild metric, since the interior metric does not support a photon sphere, and therefore the shadow size is identical to that of a Schwarzschild BH, thus trivially ensuring consistency with the findings of Ref.~\cite{Lima:2021las}. The photon sphere associated to the exterior Schwarzschild metric ($r_{\rm ph}=3M$) is astrophysically relevant only if it is located outside of the matching radius ($R_b$): when setting $M=\chi R_b/2$, the condition $3M \geq R_b$ becomes $\chi \geq 2/3$, recovering the previously discussed condition for the JMN-1 space-time to cast a shadow. Summarizing, based on the metric test considered in this work, the JMN-1 NS is completely indistinguishable from a Schwarzschild BH, and is in perfect agreement with the EHT observations, which therefore allow for the possibility of Sgr A$^*$ being a NS.

In closing, we note that one of us studied the effect of spherical accretion upon a JMN-1 NS, finding that the corresponding images remain indistinguishable from those of a Schwarzschild BH, even when considering the accretion flow spectra at $\approx 200\,{\rm GHz}$~\cite{Shaikh:2018lcc}. However, a subsequent study by some of us in Ref.~\cite{Kaur:2021wgy} showed that the two can in principle be distinguished by the intensity distribution of light inside the shadow region: although extremely challenging, this opens up the intriguing possibility of discriminating a JMN-1 NS from a Schwarzschild BH with the same mass, which is otherwise not possible with the test considered in this work. Finally, we also note that in the $0<\chi<2/3$ regime, for which no photon sphere exists, the central singularity can cast a peculiar full-moon-type image~\cite{Shaikh:2018lcc} which, although inconsistent with the EHT image of Sgr A$^*$, is certainly an interesting signature to look for in upcoming SMBH images.

\subsection{Naked singularity surrounded by a thin shell of matter}
\label{subsec:tsmns}

In all the space-times we have discussed so far, the existence of a shadow (if any) was tied to the existence of a photon sphere, which one could argue to be responsible for casting the shadow itself. However, as shown in a number of recent works, the existence of a photon sphere is a sufficient but not necessary condition for a space-time to cast a shadow. One particularly interesting example in this context is that where the existence of a thin-shell of matter is responsible for casting a shadow. Here, we consider a NS space-time surrounded by a thin shell of matter, studied by some of us in Ref.~\cite{Dey:2020haf}. Specifically, we glue two JMN-1 NS space-times (see Sec.~\ref{subsec:jmn1}), an interior one (JMN-1$_i$) characterized by the parameter $\chi_i$ and an exterior one (JMN-1$_e$) characterized by the parameter $\chi_e$, at $r=R_{b1}$. The JMN-1$_e$ space-time is then further glued to an external Schwarzschild space-time with mass $M=\chi_eR_{b2}/2$ at $r=R_{b2}$. Explicitly, the metric of the JMN-1$_i$ space-time is given by:
\begin{eqnarray}
{\rm d}s^2 &=& (\chi_e-1) \left ( \frac{R_{b1}}{R_{b2}} \right ) ^\frac{\chi_e}{(1- \chi_e)} \left ( \frac{r}{R_{b1}} \right ) ^\frac{\chi_i}{(1- \chi_i)}{\rm d}t^2 \nonumber \\
&&+ \frac{{\rm d}r^2}{(1 - \chi_i)} + r^2{\rm d}\Omega^2\,,
\label{eq:metricjmn1i}
\end{eqnarray}
whereas the metric of the JMN-1$_e$ space-time is given by:
\begin{eqnarray}
{\rm d}s^2 = (\chi_e-1) \left ( \frac{r}{R_{b2}} \right ) ^\frac{\chi_e}{(1- \chi_e)}{\rm d}t^2 + \frac{{\rm d}r^2}{(1 - \chi_e)} + r^2{\rm d}\Omega^2\,. \nonumber \\
\label{eq:metricjmn1e}
\end{eqnarray}
As shown by some of us in Ref.~\cite{Dey:2020haf} using Israel's junction conditions~\cite{Israel:1966rt}, the gluing procedure is possible even if $\chi_i \neq \chi_e$, provided a thin shell of matter exists at the time-like hypersurface $r=R_{b1}$. Note that $\chi_i$ and $\chi_e$ characterize specific hair.

If $\chi_e<2/3$, based on our earlier discussions in Sec.~\ref{subsec:jmn1} it is clear that the space-time does not support the existence of a photon sphere in the usual sense. Nonetheless, if $\chi_i>2/3$, the space-time can still cast a shadow, thanks to the presence of a cusp-like feature in the effective potential which leads to the existence of a critical impact parameter (much like in the case where a photon sphere is present), as shown in Ref.~\cite{Dey:2020haf}. One can then show that the shadow size is given by:
\begin{eqnarray}
r_{\rm sh} = \frac{R_{b1}}{\sqrt{1-\chi_e} \left ( \frac{R_{b1}}{R_{b2}} \right ) ^{\frac{\chi_e}{2(1-\chi_e)}}}\,,
\label{eq:shadowsizetsm}
\end{eqnarray}
which depends on the location of the thin shell of matter ($R_{b1}$), the radius where the JMN-1$_e$ space-time is matched to the external Schwarzschild space-time ($R_{b2}$), and the value of $\chi$ corresponding to the JMN-1$_e$ space-time ($\chi_e$), recalling that $M=\chi_eR_{b2}/2$, and subject to the conditions $R_{b2}>R_{b1}$ and $\chi_e<2/3$ (or equivalently $R_{b2}>3M$), with the latter required for the space-time to not possess a photon sphere, and therefore for the shadow to be cast exclusively by the thin shell of matter.

\begin{figure}
\centering
\includegraphics[width=1.0\linewidth]{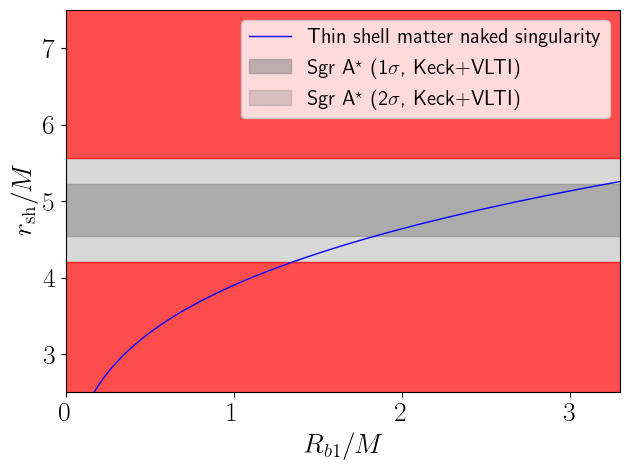}
\caption{Same as in Fig.~\ref{fig:shadow_reissner_nordstrom_bh_ns} for the naked singularity surrounded by a thin shell of matter with metric described by Eqs.~(\ref{eq:metricjmn1i},\ref{eq:metricjmn1e}), as discussed in Sec.~\ref{subsec:tsmns}.}
\label{fig:shadow_thin_shell_matter_ns}
\end{figure}

For illustrative purposes, we consider an example where we fix $\chi_e=0.6$ and $R_{b2} \approx 0.33$ (implying that $\chi_eR_{b2} \approx 2$ for consistency, as we are working in units of $M$), so that the shadow size only depends on the physically most interesting parameter $R_{b1}$, i.e.\ the location of the thin shell of matter. We show the evolution of the shadow size against $R_{b1}$ in Fig.~\ref{fig:shadow_thin_shell_matter_ns}. We immediately note that the space-time can cast a shadow only for non-zero values of $R_{b1} \gtrsim 1.5M$. This is consistent with our expectations: in fact, the limit $R_{b1} \to 0$ corresponds to the case where the JMN-1$_i$ space-time, and correspondingly the thin-shell of matter, is not present. This case reduces to that of a single JMN-1 space-time already studied in Sec.~\ref{subsec:jmn1}, for which we know that no shadow is cast for $\chi<2/3$ (see Fig.~\ref{fig:shadow_jmn1_ns}), which is clearly the case here since $\chi_e=0.6$. This result confirms that it is genuinely the thin shell of matter which is responsible for the space-time casting a shadow, even in the absence of a photon sphere. Of course, it is clear from Eq.~(\ref{eq:shadowsizetsm}) that the shadow size depends on $\chi_e$ and $R_{b2}$ separately, so that different choices of $\chi_e$ and $R_{b2}$ such that $\chi_eR_{b2}=2$ would lead to results which differ from those we have shown here. Nonetheless, the qualitative features we have shown and discussed would remain unchanged: this is sufficient for the purposes of the present discussion with the aim of showing that a thin shell of matter can cast a shadow.

\subsection{Null naked singularity}
\label{subsec:nullns}

In the absence of a photon sphere, a thin shell of matter is not the only possibility for casting a shadow. A relevant example in this sense was studied by some of us in Ref.~\cite{Joshi:2020tlq}, in the context of a novel NS space-time described by the following metric:
\begin{eqnarray}
{\rm d}s^2 = -\frac{{\rm d}t^2}{\left(1+\frac{M}{r}\right)^2}+\left(1+\frac{M}{r}\right)^2{\rm d}r^2 +  +r^2{\rm d}\Omega^2\,.
\label{eq:metricnullns}
\end{eqnarray}
The above metric describes a space-time of ADM mass $M$ which approaches the Schwarzschild space-time as $r \to \infty$, possesses a strong curvature singularity at $r=0$ (as can be verified by computing the Kretschmann and Ricci scalars), and is manifestly horizon-less. Note that this space-time is not characterized by additional hairs. Since the singularity is null-like, we shall refer to the space-time described by Eq.~(\ref{eq:metricnullns}) as ``null NS'' (NNS). We note that this particular NNS space-time was introduced on purely phenomenological grounds, as a toy model to explore the possibility of a shadow being cast in the absence of a photon sphere and a thin shell of matter. We leave open the question of whether gravitational collapse from suitable initial conditions can lead to the NNS space-time as end product. We note, however, that the formation of null- and even time-like singularities from the gravitational collapse of matter clouds is known to be possible and has been studied extensively in the literature (see e.g.\ Refs.~\cite{Ori:1989ps,Joshi:1993zg,Jhingan:2000pz,Kudoh:2000xs,Goswami:2004fx,Banik:2016qvf,Bhattacharya:2017chr,Dey:2019fja,Mosani:2020ena,Dey:2020bgo} for relevant works).

From Eq.~(\ref{eq:photonsphere}), it is clear that the NNS does not possess a photon sphere -- formally, the solution to Eq.~(\ref{eq:photonsphere}) would be $r_{\rm ph}=0$. In other words, the minimum turning point radius for a photon fired from infinity is $r=0$, where the effective potential for null geodesics has a finite value. Nonetheless, the impact parameter corresponding to the minimum turning point radius is non-zero, and this allows the NNS to cast a shadow of size $r_{\rm sh}=M$. In other words, despite the absence of an event horizon, null geodesics reaching the NNS from infinity with an impact parameter $b<M$ will be trapped closer and closer to the central singularity, and will not be able to return to the observer: this leads to the appearance of a central dark region in the observer's sky, which we associate to the NNS shadow. The situation is perhaps visually easier to grasp from Fig.~1f of Ref.~\cite{Joshi:2020tlq}, which shows examples of light trajectories around the NNS, making the physical origin of the shadow clear.

\begin{figure}
\centering
\includegraphics[width=1.0\linewidth]{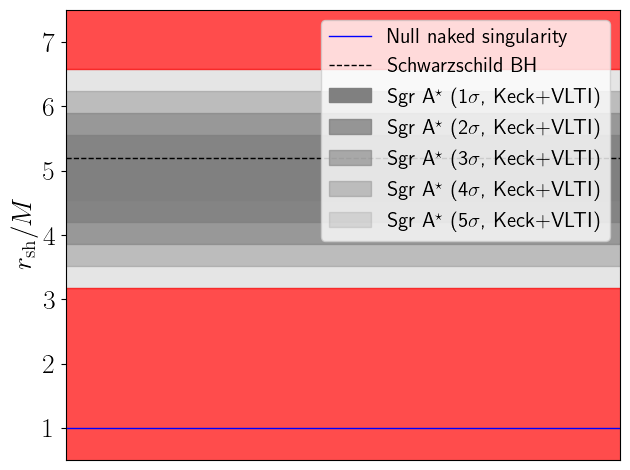}
\caption{Same as in Fig.~\ref{fig:shadow_morris_thorne_wh} for the null naked singularity with metric given by Eq.~\eqref{eq:metricnullns}, as discussed in Sec.~\ref{subsec:nullns}.}
\label{fig:shadow_null_ns}
\end{figure}

The shadow cast by the NNS is of course too small to be consistent with the EHT observations of Sgr A$^*$, as is clear from Fig.~\ref{fig:shadow_null_ns}. Therefore, much as the MT traversable WH, the EHT observations rule out the possibility of Sgr A$^*$ being a NNS at very high ($>5\sigma$) significance. Of course, as we have discussed earlier in Sec.~\ref{subsec:jnw}, Sec.~\ref{subsec:jmn1}, and Sec.~\ref{subsec:tsmns}, other NS space-times remain potentially viable descriptions of Sgr A$^*$, leaving open the intriguing possibility that the latter might not be a BH but an exotic horizonless object.

\subsection{Minimally coupled scalar field with a potential}
\label{subsec:minimallycoupledscalar}

So far we have discussed a few well-motivated regular BH solutions, as well as BH mimickers such as wormholes and naked singularities. To continue our journey into horizon-scale tests of alternatives to the Schwarzschild metric, we now move on to BH solutions which violate the no-hair theorem in a controlled way, in the presence of additional fields or exotic interactions. As is well known, one of the most straightforward (and well-motivated) ways of violating the NHT is through the presence of additional matter fields, such as a scalar field. The following subsections will therefore be devoted to the study of a number of ``hairy'' BH solutions arising broadly speaking in scenarios of ``new physics'', which include but are not limited to the presence of additional fields, modified theories of gravity (these two scenarios are not mutually exclusive), or new high-energy physics models.

\begin{figure}
\centering
\includegraphics[width=1.0\linewidth]{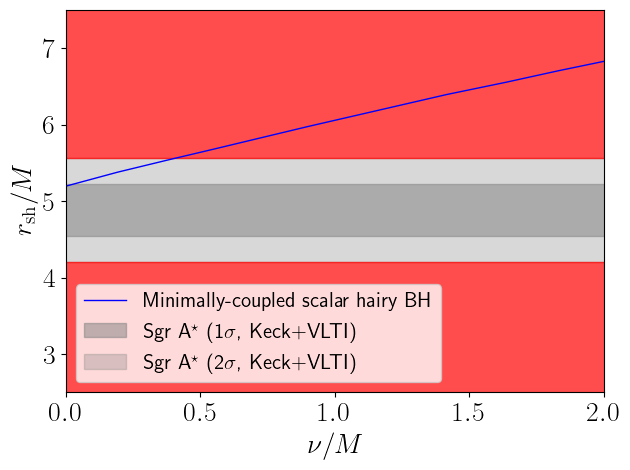}
\caption{Same as in Fig.~\ref{fig:shadow_reissner_nordstrom_bh_ns} for the BH with scalar hair, as discussed in Sec.~\ref{subsec:minimallycoupledscalar}.}
\label{fig:shadow_mcsh_bh}
\end{figure}

We begin by studying the simplest possible scenario where the no-hair theorem can be violated in a controlled way, considering a minimally coupled real scalar field $\phi$ featuring a potential. The dynamics of the field are thus governed by the following Lagrangian:
\begin{eqnarray}
\mathcal{L} \supset \frac{1}{2}\partial^{\mu}\phi\partial_{\mu}\phi-V(\phi)\,,
\label{eq:mc}
\end{eqnarray}
where here and in what follows the symbol $\supset$ indicates that the Lagrangian contains at least the terms on the right, plus potentially other terms which are irrelevant for the determination of the corresponding BH metric. Following Refs.~\cite{Gonzalez:2013aca,Gonzalez:2014tga,Khodadi:2020jij}, including work by some of us, we take the following form for the potential:
\begin{eqnarray}
V(\phi) = \frac{1}{2\nu^4}e^{-2\sqrt{2}\phi}g(\phi)\,,
\label{eq:potential}
\end{eqnarray}
where $g(\phi)$ is given explicitly in Ref.~\cite{Khodadi:2020jij}. The specific choice of potential is not particularly motivated from a high-energy theory perspective. However, in the context of BH physics, it is particularly interesting because it provides a toy example of a controlled violation of the NHT, which proceeds by explicitly introducing a new energy scale in the scalar sector. More specifically, the scalar field profile is given by~\cite{Gonzalez:2013aca,Gonzalez:2014tga,Khodadi:2020jij}:
\begin{eqnarray}
\phi(r) = \frac{1}{\sqrt{2}} \ln \left ( 1+\frac{\nu}{r} \right )
\label{eq:scalar}
\end{eqnarray}
where the scale $\nu>0$ governs the scalar field fall-off, and effectively acts as a hair parameter. The metric functions read (see e.g.\ Refs.~\cite{Gonzalez:2013aca,Gonzalez:2014tga,Khodadi:2020jij}):
\begin{eqnarray}
A(r) &=& -\frac{2r\nu+6M(2r+\nu)}{\nu^2}-\frac{2r(r+\nu)(6M+\nu)}{\nu^3}\ln\frac{r}{r+\nu}\,, \nonumber \\
C(r) &=& r(r+\nu)\,.
\label{eq:metricmch}
\end{eqnarray}
Although it is not straightforward to see from Eq.~(\ref{eq:metricmch}), the metric function $A(r)$ does indeed reduce to the Schwarzschild limit $1-2M/r$ as the hair parameter vanishes. This is best seen by Taylor-expanding the metric function in $\nu$, yielding:
\begin{eqnarray}
A(r) = 1-\frac{2M}{r}+\frac{3M-r}{3r^2}\nu+{\cal O}(\nu^2)\,,
\label{eq:metricmchtaylor}
\end{eqnarray}
from which one easily sees that the limit $\nu \to 0$ indeed recovers the Schwarzschild metric, and $\nu$ characterizes an universal hair. We compute the shadow radius of BHs within this theory numerically, as done earlier in Ref.~\cite{Khodadi:2020jij} by one of us. We show the evolution of the shadow size against the hair parameter $\nu$ in Fig.~\ref{fig:shadow_mcsh_bh}, finding that the shadow size increases quickly with increasing hair parameter. As in other similar examples we have already seen, the EHT observations set particularly tight constraints on this scenario, with limits on the hair parameter of order $\nu \lesssim 0.01M$ ($1\sigma$) and $\nu \lesssim 0.4M$ ($2\sigma$). This effectively constrains the energy scale at which violations of the NHT become important, within this particular theory.

\subsection{Black holes and wormholes with conformal scalar hair}
\label{subsec:conformallycoupledscalar}

We now study another scenario featuring a scalar field, which leads to a controlled violation of the no-hair theorem. We consider a real scalar field $\phi$, conformally coupled to gravity, through the following Lagrangian:
\begin{eqnarray}
\mathcal{L} \supset \partial^{\mu}\phi\partial_{\mu}\phi+\frac{1}{6}\phi^2R\,,
\label{eq:cc}
\end{eqnarray}
where $R$ is the Ricci scalar, and the $1/6$ factor ensures that the coupling is conformal (in 4 dimensions). BH solutions within this theory have been obtained by Astorino~\cite{Astorino:2013sfa} (see also Refs.~\cite{Martinez:2005di,Cisterna:2018hzf,Myung:2018jvi,Myung:2019oua,Myung:2019adj,Zou:2019ays,Zou:2020rlv,Caceres:2020myr} for BH and black brane solutions within related models), and contain a hair parameter $S$ related to the scalar field profile, and directly connected to the value of the scalar field at the horizon. The hair parameter, or scalar charge, emerges as an integration constant and characterizes primary hair, i.e.\ it cannot be expressed as a function of the other hair parameters. The metric function reads:
\begin{eqnarray}
A(r) = 1-\frac{2M}{r}+\frac{S}{r^2}\,,
\label{eq:metriccc}
\end{eqnarray}
where the scalar charge $S$ characterizes a specific hair and can take both positive and negative values, with $S<M^2$ required in the former regime for the solution to describe a BH. The space-time described by Eq.~(\ref{eq:metriccc}) is closely related to the well-known ``BBMB black hole''~\cite{Bocharova:1970skc,Bekenstein:1974sf}, and appears very similar to the Reissner-Nordstr\"{o}m metric [Eq.~(\ref{eq:metricrn})]. However, there is a crucial difference in that the scalar hair parameter can take negative values, which are instead effectively precluded within the RN metric as $Q^2>0$. The $S<0$ regime is in fact better known as \textit{mutated} Reissner-Nordstr\"{o}m BH, and effectively describes the geometry of a traversable wormhole~\cite{Chowdhury:2018pre} (see also Ref.~\cite{Khodadi:2020jij}). Formally, this corresponds to an analytic continuation of the RN metric for imaginary values of the electric charge (thus negative values of $Q^2$), although we discourage the reader from taking this interpretation too far. We also note that the RN-like form of the metric is no coincidence, and is directly related to the conformal symmetry enjoyed by the Lagrangian in Eq.~(\ref{eq:cc}), see in addition Ref.~\cite{Castro:2010fd} for earlier relevant results. Finally, it is worth noticing that a similar metric arises in the context of a conformal massive gravity model where Lorentz invariance is spontaneously broken (see Refs.~\cite{Bebronne:2009mz,Jusufi:2019caq,Jawad:2021hay}), although only for a very specific choice of the function ${\cal F}$ appearing in Eq.~(10) of Ref.~\cite{Bebronne:2009mz}.

\begin{figure}
\centering
\includegraphics[width=1.0\linewidth]{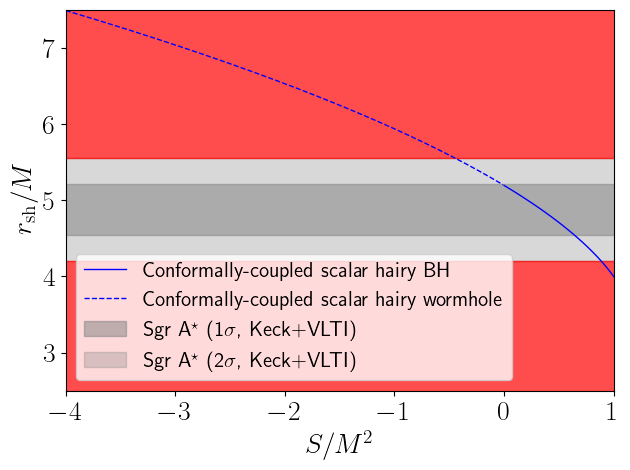}
\caption{Same as in Fig.~\ref{fig:shadow_reissner_nordstrom_bh_ns} for solution with conformally coupled scalar hair, leading to a BH ($S>0$, solid curve) or a wormhole ($S<0$, dashed curve) depending on the value of the hair parameter $S$, and with metric function given by Eq.~\eqref{eq:metriccc}, as discussed in Sec.~\ref{subsec:conformallycoupledscalar}.}
\label{fig:shadow_ccsh_bh_wh}
\end{figure}

Since the metric is RN-like, the shadow size takes the same functional form as Eq.~(\ref{eq:shadowsizern}):
\begin{eqnarray}
r_{\rm sh} = \frac{\sqrt{2} \left ( 3+\sqrt{9-8S/M^2} \right ) }{\sqrt{4+\frac{\sqrt{9-8S/M^2}-3}{S/M^2}}}\,,
\label{eq:shadowsizeconformallycoupledscalar}
\end{eqnarray}
valid within both the BH ($S>0$) and WH ($S<0$) regimes. We show the evolution of the shadow size against the scalar charge $S$ in Fig.~\ref{fig:shadow_ccsh_bh_wh}. We see that within the regime $S>0$ the shadow size decreases with increasing scalar charge, analogously to the case with the electric charge in the Reissner-Nordstr\"{o}m metric. In this regime, the EHT observations set the limits $S \lesssim 0.65M^2$ ($1\sigma$) and $S \lesssim 0.9M^2$ ($2\sigma$). In the mutated Reissner-Nordstr\"{o}m BH/WH regime $S<0$, the shadow size instead increases when increasing the absolute value of the scalar charge, quickly bringing the shadow size outside the range allowed by the EHT observations, which set the lower limits $S \gtrsim -0.04M^2$ (1$\sigma$) and $S \gtrsim -0.4M^2$ ($2\sigma$).

\subsection{Clifton-Barrow $f(R)$ gravity}
\label{subsec:cliftonbarrow}

Generalizations and extensions of Einstein's GR have a long and rich history, but undoubtedly one of the most studied theories in this sense is $f(R)$ gravity, where the Ricci scalar $R$ in the Einstein-Hilbert action is replaced by an arbitrary function $f(R)$. Several $f(R)$ models have been constructed in the literature with the aim of describing the late-time accelerated expansion of the Universe (and/or an early accelerated phase of inflation), which requires satisfying viability conditions including the positivity of effective gravitational couplings, stability of cosmological perturbations, asymptotic $\Lambda$CDM behavior in the high curvature regime, stability of the late-time de Sitter point, as well as observational constraints from the equivalence principle, Solar System, and cosmological tests. For an inevitably incomplete list of important works on a wide range of aspects of $f(R)$ gravity, with particular reference to BH solutions therein, we refer the reader to Refs.~\cite{Nojiri:2006gh,Amendola:2006we,Fay:2007uy,Hu:2007nk,Carloni:2007br,Starobinsky:2007hu,Cognola:2007zu,Brax:2008hh,delaCruz-Dombriz:2009pzc,Schmidt:2009am,Tsujikawa:2009ku,Faraoni:2010yi,Sebastiani:2010kv,Moon:2011hq,Olmo:2011uz,Elizalde:2011ds,Cembranos:2011sr,Jennings:2012pt,Li:2012by,delaCruz-Dombriz:2012bni,Basilakos:2013nfa,Cai:2014upa,Rinaldi:2014gua,Rinaldi:2014gha,Sebastiani:2015kfa,Odintsov:2015wwp,Khodadi:2015fav,Hu:2016zrh,Oikonomou:2016fxb,DeMartino:2016ltl,Peirone:2016wca,Burrage:2016bwy,Odintsov:2017hbk,Naik:2018mtx,Nashed:2018oaf,Kleidis:2018cdx,Banerjee:2019cjk,Odintsov:2020fxb,Nashed:2020tbp,Nashed:2021ffk,Nashed:2021lzq,Vagnozzi:2021quy,Farrugia:2021zwx,Oikonomou:2021msx,Odintsov:2021kup,Astashenok:2021btj,Odintsov:2021urx,Geng:2021hqc,Oikonomou:2022wuk,Nojiri:2022ski,Oikonomou:2022pdf,Oikonomou:2022tux,Oikonomou:2022ijs,Astashenok:2022kfj,Oikonomou:2022irx,Oikonomou:2023dgu}.

Clifton and Barrow considered a particularly interesting $f(R)$ model in Ref.~\cite{Clifton:2005aj}, whose action is given by:
\begin{eqnarray}
S = \int {\rm d}^4x\sqrt{-g}\,M_{\rm Pl}^{2-2\delta}R^{1+\delta}\,,
\label{eq:cliftonbarrow}
\end{eqnarray}
where $\delta$ is a dimensionless parameter, $M_{\rm Pl}$ is the Planck mass, and GR is recovered in the limit $\delta \to 0$. This theory was argued to possess several appealing properties, including its being a one-parameter extension of GR not introducing any additional scale besides the already present Planck mass, its providing a testing ground for new developments in particle production, holography and gravitational thermodynamics, as well as its being one of the few $f(R)$ models possessing simple exact solutions for FLRW-like cosmological models and exact static spherically symmetric solutions generalizing the Schwarzschild metric. Ref.~\cite{Clifton:2005aj} does indeed identify such a solution, whose metric function is given by:

\begin{eqnarray}
A(r)=r^{2\delta\frac{1+2\delta}{1-\delta}}-\frac{2M}{r^{\frac{1-4\delta}{1-\delta}}} \,,
\label{eq:metriccliftonbarrow}
\end{eqnarray}
where $\delta$ characterizes an universal hair. It is easy to see that the above metric is asymptotically flat only if $-1/2<\delta<0$ or $\delta>1$. We exclude the regime $\delta>1$, as we find a posteriori that within this range the shadow size is by far too small and completely excluded by observations, and therefore focus on the region $-1/2<\delta<0$.

\begin{figure}
\centering
\includegraphics[width=1.0\linewidth]{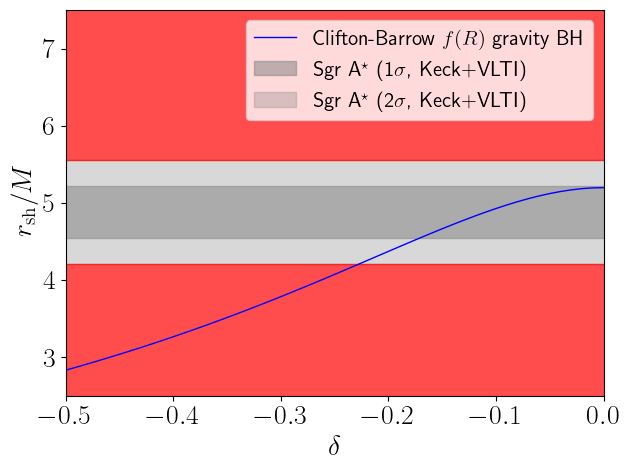}
\caption{Same as in Fig.~\ref{fig:shadow_reissner_nordstrom_bh_ns} for the BH in Clifton-Barrow $f(R)$ gravity with metric function given by Eq.~\eqref{eq:metriccliftonbarrow}, as discussed in Sec.~\ref{subsec:cliftonbarrow}.}
\label{fig:shadow_clifton_barrow_bh}
\end{figure}

We compute the shadow size numerically, and show its evolution against $\delta$ in Fig.~\ref{fig:shadow_clifton_barrow_bh}.~\footnote{We have found that a closed-form expression for the shadow radius exists, but is too cumbersome to report here. We have nonetheless verified that our numerical results perfectly reproduce those provided by this closed-form expression.} We see that the shadow size decreases with increasing $\vert \delta \vert$ (i.e.\ with $\delta$ becoming more negative), and that the EHT observations set the limits $\delta \gtrsim -0.15$ ($1\sigma$) and $\delta \gtrsim -0.20$ ($2\sigma$). In the $\delta<0$ regime, the existence of a stable matter-dominated period of evolution, which is itself a prerequisite for a viable cosmological model, requires $\delta<-1/4$~\cite{Clifton:2005aj}. This makes our limits particularly interesting, as they exclude this region, and therefore would appear to exclude this class of models, highlighting a very appealing complementarity between BH shadows and cosmological constraints.

\subsection{Horndeski gravity}
\label{subsec:horndeski}

We now move on to a well-known and much more general class of scalar-tensor theories: Horndeski gravity. On very general grounds, scalar-tensor theories are an appealing playground for theorists as they encapsulate the key characteristics of several theories of modified gravity, which typically possess a scalar-tensor limit. Back in 1974, Horndeski wrote down the most general scalar-tensor theory with equations of motion at most of second order~\cite{Horndeski:1974wa}, with the latter requirement sufficient to ensure that the theory is free of the Ostrogradsky ghost~\cite{Ostrogradsky:1850fid}.~\footnote{Note that having second order equations of motion is a sufficient but not necessary condition for the theory to be physically viable. As an example, the beyond Horndeski (GLPV) class of scalar-tensor theories has equations of motion of higher than second order, yet does not propagate unphysical degrees of freedom thanks to a degeneracy condition which allows to maintain the canonical number of degrees of freedom~\cite{Gleyzes:2014dya,Gleyzes:2014qga} (see for instance Refs.~\cite{Zumalacarregui:2013pma,Deffayet:2015qwa,Langlois:2015skt,Crisostomi:2016tcp,BenAchour:2016cay,Babichev:2016rlq,BenAchour:2016fzp,Langlois:2018jdg} for further relevant works).} After living almost ignored for over 30 years, the theory, which encompasses many well-known theories of gravity (including $f(R)$ gravity) as specific cases, was ``rediscovered'' in Refs.~\cite{Nicolis:2008in,Deffayet:2009wt}, making it a beautiful example of ``sleeping beauty'' theory which is now attracting a tremendous amount of attention (see e.g.\ Refs.~\cite{Bettoni:2013diz,Kase:2014yya,Kase:2014cwa,Tsujikawa:2014uza,Cisterna:2016vdx,Kobayashi:2016xpl,Rinaldi:2016oqp,Banerjee:2018svi,Bahamonde:2019shr,Oikonomou:2020sij,Bernardo:2021qhu,Bahamonde:2021dqn,Bernardo:2021izq,Bernardo:2021bsg,Oikonomou:2021hpc}) after decades of dormancy.

Here, we shall focus on the shift-symmetric subclass of Horndeski gravity, where the scalar field $\phi$ enjoys a shift symmetry $\phi \to \phi+\text{const}$. This sector is of particular interest since a well-known no-go theorem by Hui and Nicolis dictates that static, asymptotically flat BH solutions thereof are isometric to the Schwarzschild BH, with a trivial (constant) profile for the scalar field~\cite{Hui:2012qt}. Of course, this no-go theorem comes with several assumptions, which we shall here break (see e.g.\ Refs.~\cite{Rinaldi:2012vy,Anabalon:2013oea,Minamitsuji:2013ura,Sotiriou:2014pfa,Maselli:2015yva,Babichev:2017guv}) in order to play with hairy BH solutions. The shift-symmetric sector of Horndeski gravity we shall consider is described by the following action:
\begin{eqnarray}
S =  \int {\rm d}^4x\sqrt{-g}\, \left ( {\cal L}_2+{\cal L}_3+{\cal L}_4+{\cal L}_5 \right ) \,,
\label{eq:actionhorndeski}
\end{eqnarray}
where the four Lagrangian terms ${\cal L}_2$, ${\cal L}_3$, ${\cal L}_4$, and ${\cal L}_5$ are given by the following~\cite{Horndeski:1974wa}:
\begin{eqnarray}
\label{eq:l2}
{\cal L}_2 &=& G_2(X)\,,\\
\label{eq:l3}
{\cal L}_3 &=& -G_3(X) \Box \phi\,,\\
\label{eq:l4}
{\cal L}_4 &=& G_4(X) R + G_{4,X} \left [ (\Box \phi)^2 -(\nabla_\mu\nabla_\nu\phi)^2 \right ] \,,\\
\label{eq:l5}
{\cal L}_5 &=& G_5(X) G_{\mu\nu}\nabla^\mu \nabla^\nu \phi - \frac{1}{6} G_{5,X} \big [ (\Box \phi)^3 - 3\Box \phi(\nabla_\mu\nabla_\nu\phi)^2 \nonumber \\
&~&~~ + 2(\nabla_\mu\nabla_\nu\phi)^3 \big ]\,,
\end{eqnarray}
which involve four arbitrary functions ($G_2$, $G_3$, $G_4$, and $G_5$) of the kinetic term $X\equiv-\partial^\mu\phi\partial_\mu\phi/2$, $(\nabla_\mu\nabla_\nu\phi)^2 \equiv \nabla_\mu\nabla_\nu\phi \nabla^\nu\nabla^\mu\phi$, and $(\nabla_\mu\nabla_\nu\phi)^3 \equiv \nabla_\mu\nabla_\nu\phi \nabla^\nu\nabla^\rho\phi \nabla_\rho\nabla^\mu\phi$, and where $G_{4,X}$ and $G_{5,X}$ are respectively the derivatives of the functions $G_4$ and $G_5$ with respect to $X$.

As discussed in Ref.~\cite{Babichev:2017guv}, the no-go theorem of Ref.~\cite{Hui:2012qt} essentially rests upon three assumptions:
\begin{enumerate}
\item the derivatives with respect to $X$ of the $G_i$s only contain positive or zero powers of $X$ in the limits $X \to 0$, $r \to \infty$;
\item the kinetic term $X$ is canonical;
\item the norm of the Noether current associated to the shift symmetry is finite down to the horizon.
\end{enumerate}
Clearly, then, in order to find hairy BH solutions within Horndeski gravity, one must at the very least break one of the above assumptions -- the solutions which we shall consider are obtained by breaking the first one. As shown in Ref.~\cite{Babichev:2017guv}, doing so is possible at least for the following particular choices of the $G_i$ functions:
\begin{eqnarray}
G_2 &\supseteq& \sqrt{-X}\,,\quad G_3 \supseteq \ln \vert X \vert \,, \nonumber \\
\quad G_4 &\supseteq& \sqrt{-X}\,,\quad G_5 \supseteq \ln \vert X \vert \,.
\label{eq:hairgifunction}
\end{eqnarray}
A choice of the $G_i$ functions satisfying Eqs.~(\ref{eq:hairgifunction}) has the potential to admit non-Schwarzschild-isometric, asymptotically flat BH solutions, with a non-trivial scalar field profile, and with a regular Noether current. In the following, we shall consider two specific solutions obtained in Refs.~\cite{Babichev:2017guv,Bergliaffa:2021diw} for a particular choice of the $G_i$ functions satisfying Eqs.~(\ref{eq:hairgifunction}), which we shall overall refer to as ``sqrt quartic Horndeski gravity'',~\footnote{We are obviously using ``sqrt'' as a shorthand for ``square root''.} while simply referring to the solutions of Ref.~\cite{Babichev:2017guv} and Ref.~\cite{Bergliaffa:2021diw} as ``Case 1'' and ``Case 2'' respectively.

\subsubsection{Case 1}
\label{subsubsec:horndeskicase1}

The first BH solution in the shift-symmetric sector of Horndeski gravity we consider was studied in Ref.~\cite{Babichev:2017guv}. In the notation of Eqs.~(\ref{eq:l2}--\ref{eq:l5}), and within the system of units we have adopted, the particular choice of $G_i$ functions is given by:
\begin{eqnarray}
G_2=\eta X\,,\quad G_4=1/16\pi+\beta\sqrt{-X}\,,\quad G_3=G_5=0\,, \nonumber \\
\label{eq:ghorndeskicase1}
\end{eqnarray}
Clearly, $G_2$ corresponds to a canonical kinetic term, whereas it is the particular $X$-dependence of $G_4$ (the $1/16\pi$ term simply leads to an Einstein-Hilbert term in the action, since it is multiplied by $R$) which allows for hairy BH solutions, by satisfying the specific set of conditions laid out in Eqs.~(\ref{eq:hairgifunction}).

Hairy BH solutions within this theory are described by the following metric function (again within the system of units we have adopted)~\cite{Babichev:2017guv}:
\begin{eqnarray}
A(r) = 1-\frac{2M}{r}-\frac{8\pi p_{\rm eff}}{r^2}\,,
\label{eq:metrichorndeskicase1peff}
\end{eqnarray}
which is clearly asymptotically flat, and where we have defined the hair parameter $p_{\rm eff} \equiv \beta^2/\eta$ which characterizes an universal hair. Note that, depending on the sign of $\eta$ (which, in principle, can be either positive or negative), $p_{\rm eff}$ can carry either sign: in particular, positive $\eta$ corresponds to positive $p_{\rm eff}$, and vice-versa.~\footnote{As explicitly discussed in Ref.~\cite{Babichev:2017guv}, although the kinetic term is phantom for $\eta<0$, this is in itself not a guarantee of the instability of the theory, as the latter depends sensibly on the quartic term and hence on the value of $\beta$, see e.g.\ Refs.~\cite{Deffayet:2010qz,Babichev:2012re}. To the best of our knowledge, an analysis of the stability of this specific Horndeski theory has not been carried out in the literature, and doing so would go well beyond the scope of our work. Nonetheless, we note that the BH solution is equally valid for the $\eta>0$ and $\eta<0$ regimes, provided $p_{\rm eff}$ satisfies $p_{\rm eff}>-M^2/8\pi$, as noted in Ref.~\cite{Babichev:2017guv} (albeit with a different notation compared to us), and as we later impose.}

Note that this BH solution is of the RN-like form. In Ref.~\cite{Babichev:2017guv}, this was argued to be possibly related to the remnant of global conformal invariance enjoyed by the action defined by Eq.~(\ref{eq:ghorndeskicase1}), as the extremal RN space-time itself is known to possess a discrete conformal isometry related to spatial inversion.~\footnote{This symmetry is best known as Couch-Torrence inversion symmetry, and maps $(t,r) \to (t,M+M^2/(r-M))$~\cite{Couch:1984ghw} (see also Refs.~\cite{Bizon:2012we,Fernandes:2020jto}). Under this symmetry, the metric of an extremal RN BH transforms as $g_{\mu\nu} \to \omega^2 g_{\mu\nu}$, with $\omega=M/(r-M)$, making this a (discrete) conformal symmetry. Moreover, it can be shown that the symmetry interchanges the future event horizon ${\cal H}^+$ with future null infinity $\mathscr{I}^+$, and similarly the past event horizon ${\cal H}^-$ with past null infinity $\mathscr{I}^-$, allowing for an interpretation as spatial inversion symmetry.} Within the negative $p_{\rm eff}$ regime, $p_{\rm eff}>-M^2/8\pi$ is required (else the solution describes a naked singularity), whereas no such restrictions are present in the positive $p_{\rm eff}$ regime. Finally, we note that the positive $p_{\rm eff}$ regime is expected to described a mutated RN WH, as the $S<0$ regime of the BH/WH with conformal scalar hair considered in Sec.~\ref{subsec:conformallycoupledscalar} (although we note that such an interpretation was not considered in Ref.~\cite{Babichev:2017guv}).

\begin{figure}
\centering
\includegraphics[width=1.0\linewidth]{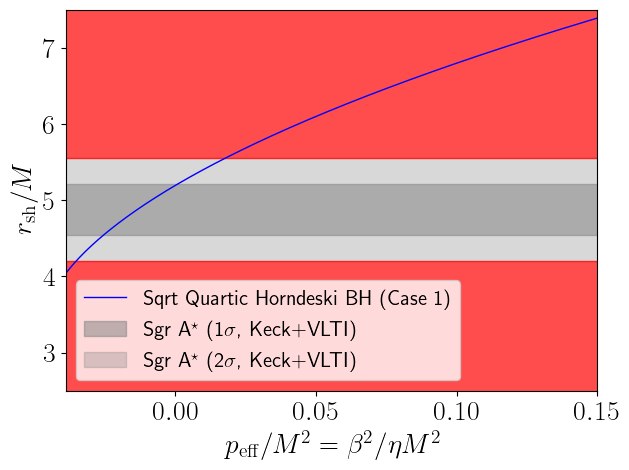}
\caption{Same as in Fig.~\ref{fig:shadow_reissner_nordstrom_bh_ns} for the BH in the Horndeski gravity sector described by Eq.~(\ref{eq:ghorndeskicase1}) and with metric function given by Eq.~\eqref{eq:metrichorndeskicase1peff}, as discussed in Sec.~\ref{subsubsec:horndeskicase1}.}
\label{fig:shadow_horndeski_case_1_bh}
\end{figure}

By virtue of the RN-like form of the metric, we can immediately borrow upon the previous results for RN-like space-times (Sec.~\ref{subsec:rn} and Sec.~\ref{subsec:conformallycoupledscalar}) to obtain the shadow radius as a function of $p_{\rm eff}$ [see e.g.\ Eq.~(\ref{eq:shadowsizeconformallycoupledscalar})]:
\begin{eqnarray}
r_{\rm sh} = \frac{\sqrt{2} \left ( 3+\sqrt{9+64\pi p_{\rm eff}} \right ) }{\sqrt{4+\frac{3-\sqrt{9+64\pi p_{\rm eff}}}{8\pi p_{\rm eff}}}}\,,
\label{eq:shadowsizehorndeskicase1}
\end{eqnarray}
which, we recall, is valid within both the $p_{\rm eff}<0$ and $p_{\rm eff}>0$ regimes. We show the evolution of the shadow size against the scalar charge $p_{\rm eff}=\beta^2/\eta$ in Fig.~\ref{fig:shadow_horndeski_case_1_bh}, which of course is qualitatively identical to Fig.~\ref{fig:shadow_ccsh_bh_wh}, albeit with the $x$ axis inverted and rescaled. We see that the EHT observations set very stringent constraints on $p_{\rm eff}$. In the $p_{\rm eff}>0$ regime, we find the limits $p_{\rm eff} \lesssim 0.002M^2$ ($1\sigma$) and $p_{\rm eff} \lesssim 0.01M^2$ ($2\sigma$). On the other hand, in the $p_{\rm eff}<0$ regime, we find the limits $p_{\rm eff} \gtrsim -0.005M^2$ ($1\sigma$) and $p_{\rm eff} \gtrsim -0.01M^2$ ($2\sigma$). These constraints limit the combination $\beta^2/\eta$, with $\beta$ governing the strength of the $\sqrt{-X}$ term in $G_4$ and $\eta$ governing the strength of the kinetic term in $G_2$. A detailed study of the theoretical and model-building implications of these constraints as well as the complementarity with cosmological, astrophysical, and GW constraints (which is the subject of a rich and rapidly growing body of literature, see e.g.\ Refs.~\cite{Bellini:2015wfa,Bellini:2015xja,Myrzakulov:2015ysa,Sakstein:2016ggl,Zumalacarregui:2016pph,Pogosian:2016pwr,Perenon:2016blf,Renk:2017rzu,Arroja:2017msd,Creminelli:2017sry,Sakstein:2017xjx,Ezquiaga:2017ekz,Baker:2017hug,Kreisch:2017uet,Dima:2017pwp,Kase:2018iwp,Casalino:2018tcd,Kase:2018aps,Frusciante:2018jzw,Casalino:2018wnc,Noller:2018wyv,Duniya:2019mpr,Traykova:2019oyx,Perenon:2019qmd,Pace:2019uow,Peirone:2019yjs,Bahamonde:2019ipm,Traykova:2021hbr,Reyes:2021owe,Petronikolou:2021shp} for examples of these constraints and observational signatures of Horndeski gravity) would be very interesting, but goes significantly the scope of the paper, and is therefore left to future work.

\subsubsection{Case 2}
\label{subsubsec:horndeskicase2}

We now consider a different BH solution in the shift-symmetric sector of Horndeski gravity, studied in  Ref.~\cite{Bergliaffa:2021diw}, and whose shadow was first computed in Ref.~\cite{Afrin:2021wlj} by some of us. Using the same notation of Eqs.~(\ref{eq:l2}--\ref{eq:l5}), the authors begin from the following particular choice of $G_i$ functions~\cite{Bergliaffa:2021diw}:
\begin{eqnarray}
G_2&=&\alpha_{21}X+\alpha_{22}(-X)^{\omega_2}\,,\quad G_3=-\alpha_{31}(-X)^{\omega_3}\,,\nonumber \\
G_4&=&1/16\pi+\alpha_{42}(-X)^{\omega_4}\,,\quad G_5=0\,,
\label{eq:ghorndeskicase2}
\end{eqnarray}
which clearly generalizes Eqs.~(\ref{eq:ghorndeskicase1}), with the latter recovered for $\alpha_{22}=\alpha_{31}=0$ and $\omega_4=1/2$. Note also that the theory described by Eq.~(\ref{eq:ghorndeskicase2}) is known to be weakly hyperbolic for $\alpha_{21} \neq 0$~\cite{Papallo:2017ddx}, but its stability has not been examined for $\alpha_{21}=0$. The authors of Ref.~\cite{Bergliaffa:2021diw} at a later stage focus on a particular choice of the $\alpha$ parameters, which still preserves the regularity of the Noether current associated to the shift symmetry, and the finiteness of the energy. More specifically, the following choice of $G_i$ functions is studied~\cite{Bergliaffa:2021diw}:
\begin{eqnarray}
G_2&=&\alpha_{22}(-X)^{\frac{3}{2}}\,,\quad G_4=1/16\pi+\alpha_{42}\sqrt{-X}\,,\nonumber \\
G_3&=&G_5=0\,.
\label{eq:ghorndeskicase2restrict}
\end{eqnarray}
The above is clearly not a generalization of Eq.~(\ref{eq:ghorndeskicase1}): while the $G_4$ functions are identical, there is a crucial difference in the $G_2$ functions. Therefore, we should not expect the BH solution derived within this sector to (necessarily) reduce to the one studied in Sec.~\ref{subsubsec:horndeskicase1}.

The corresponding BH solution is indeed qualitatively very different from the one considered in Sec.~\ref{subsubsec:horndeskicase1}, and is described by the following metric function:
\begin{eqnarray}
A(r) = 1-\frac{2M}{r}+\frac{2Mh}{r} \ln \left ( \frac{r}{2M} \right ) \,,
\label{eq:metrichorndeskicase2}
\end{eqnarray}
where the hair parameter $h$ characterizes an universal hair and is related to the fundamental Lagrangian parameters $\alpha_{22}$ and $\alpha_{42}$ appearing in Eq.~(\ref{eq:ghorndeskicase2restrict}) through the following:
\begin{eqnarray}
h \equiv \frac{2\sqrt{2}}{3\sqrt{3}}\alpha_{42}\sqrt{-\frac{\alpha_{42}}{\alpha_{22}}}\,.
\label{eq:hairhorndeski}
\end{eqnarray}
We see that the above space-time is asymptotically flat, while computing the Kretschmann and Ricci scalars signals the presence of a scalar polynomial singularity at $r=0$. In the $h>0$ regime, which for consistency requires $\alpha_{22}<0$ and $\alpha_{42}>0$, the space-time only contains one event horizon, analogously to the Schwarzschild metric. In the $h<0$ regime (and more precisely for $h \in [-2;0]$), which for consistency requires $\alpha_{22}>0$ and $\alpha_{42}<0$, the space-time contains two horizons: an outer event horizon, and an inner Cauchy horizon.

\begin{figure}
\centering
\includegraphics[width=1.0\linewidth]{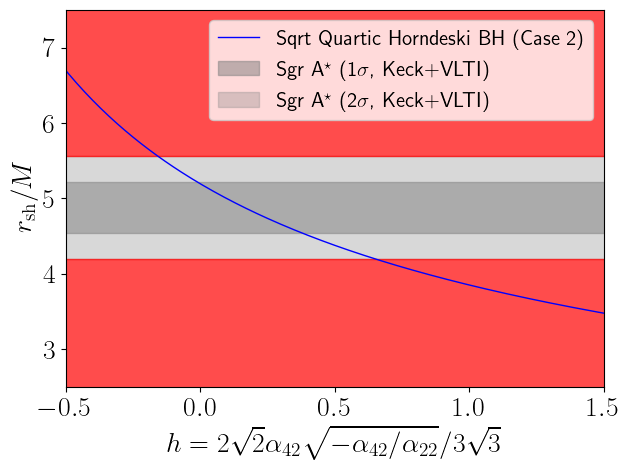}
\caption{Same as in Fig.~\ref{fig:shadow_reissner_nordstrom_bh_ns} for the BH in the Horndeski gravity sector described by Eq.~(\ref{eq:ghorndeskicase2restrict}) and with metric function given by Eq.~\eqref{eq:metrichorndeskicase2}, as discussed in Sec.~\ref{subsubsec:horndeskicase2}.}
\label{fig:shadow_horndeski_case_2_bh}
\end{figure}

Unlike the earlier Ref.~\cite{Afrin:2021wlj} by some of us, where the analysis was restricted to the $h<0$ regime, here we do not set any prior restriction on the sign of the hair parameter $h$. While slightly cumbersome, a closed-form expression for the shadow radius exists and is given by:
\begin{eqnarray}
r_{\rm sh} = \frac{3h{\cal X}}{\sqrt{1+\frac{2 \left ( -1+h\ln{\cal X} \right )}{3h{\cal X}}}}\,,
\label{eq:rshhorndeskicase2}
\end{eqnarray}
where the quantity ${\cal X}$ is defined as:
\begin{eqnarray}
{\cal X} \equiv W \left [ \frac{2e^{\frac{1}{3}+\frac{1}{h}}}{3h} \right ]\,,
\label{eq:xhorndeski}
\end{eqnarray}
with $W(z)$ denoting the principal branch of the Lambert $W$ function, also known as omega function or product logarithm function. In Fig.~\ref{fig:shadow_horndeski_case_2_bh} we show the evolution of the shadow size against $h$. We see that the shadow radius quickly increases with increasing $\vert h \vert$ in the $h<0$ regime, while decreasing more slowly with increasing $h$ in the $h>0$ regime. As a result, negative values of the hair parameter are very tightly constrained: we find that consistency with the EHT observations requires $-0.01 \lesssim h \lesssim 0.40$ ($1\sigma$) and $-0.15 \lesssim h \lesssim 0.60$ ($2\sigma$). As with the BH in Horndeski gravity studied in Sec.~\ref{subsubsec:horndeskicase1}, these constraints limit a particular combination of two fundamental Lagrangian parameters (in this case $\alpha_{42}\sqrt{-\alpha_{42}/\alpha_{22}}$), and a detailed study of the theoretical and model-building implications of these constraints, as well as the complementarity with cosmological and astrophysical constraints, while interesting, is left to future work.

\subsection{MOdified Gravity (scalar-tensor-vector gravity)}
\label{subsec:modifiedgravity}

A particularly interesting modified theory of gravity was put forward by Moffat in Ref.~\cite{Moffat:2005si}, and typically goes by the name of scalar-tensor-vector gravity, but is also often referred to as MOG (short for MOdified Gravity), as we shall do here. MOG is a covariant scalar-tensor-vector theory of gravity which introduces an additional massive scalar field and an additional massive vector field, while allowing the coupling strength of the latter, as well as the gravitational constant, to vary in space and time. The theory effectively adds a repulsive Yukawa force to Newton's acceleration law. For the full action of MOG, we refer the reader to Ref.~\cite{Moffat:2005si}. We also note that MOG has a long history of successfully explaining several astrophysical and cosmological observations: for an inevitably incomplete list of important works on MOG, see for example Refs.~\cite{Brownstein:2005zz,Brownstein:2005dr,Brownstein:2007sr,Moffat:2007yg,Moffat:2013sja,Moffat:2013uaa,Moffat:2014pia,Moffat:2015kva,Frusciante:2015maa,Hussain:2015cga,LopezArmengol:2016irf,DeMartino:2017ztt,Perez:2017spz,Oman:2017vkl,Moffat:2019uxp,Haghi:2019hjn,Liu:2019cxm,Moffat:2020mug,Moffat:2020jic,Cai:2020kue,Cai:2020igv,Moffat:2021log,DellaMonica:2021xcf,Davari:2021mge,Moffat:2021tfs}.

A particular BH solution within MOG was derived in Ref.~\cite{Moffat:2014aja}, assuming that the gravitational coupling $G$ is constant (although not necessarily taking its Newtonian value), i.e.\ $\partial_{\mu}G=0$ (see also Refs.~\cite{Guo:2018kis,Wang:2018prk,Qin:2022kaf}). In this configuration, the metric function of the resulting BH solution is given by~\cite{Moffat:2014aja}:
\begin{eqnarray}
A(r)=1-\frac{2M(1+\alpha)}{r}+\frac{M^2\alpha(1+\alpha)}{r^2} \,,
\label{eq:metricmodifiedgravity}
\end{eqnarray}
where $\alpha$ is a parameter controlling the strength of the effective gravitational coupling $G=G_N(1+\alpha)$, and characterizes an universal hair. We immediately notice that the space-time is of the RN form, with effective charge proportional to the BH mass. Clearly, one of the effects of MOG is modify the BH ADM mass as $M \to M(1+\alpha)$, as already noticed in many works~\cite{Moffat:2014aja,Moffat:2015kva,Moffat:2019uxp}. In the following, however, we shall opt for the same strategy adopted by many earlier works, and report the shadow size in units of the ``bare''/source/Newtonian mass $M$ rather than the ADM mass. The rationale is that this is the ADM mass quantifies the energy measured by an observer at infinity, but dynamical measurements of the BH mass sufficiently close to the horizon will actually probe a quantity which is much closer to the Newtonian mass. This is indeed the case for the S-stars which underlie our mass-to-distance ratio prior, as explicitly discussed in Ref.~\cite{Moffat:2019uxp}. In addition, one can envisage constraining $\alpha$ through independent tests of the strength of Newton's constant, allowing to disentangle the Newtonian mass from the ADM mass.

\begin{figure}
\centering
\includegraphics[width=1.0\linewidth]{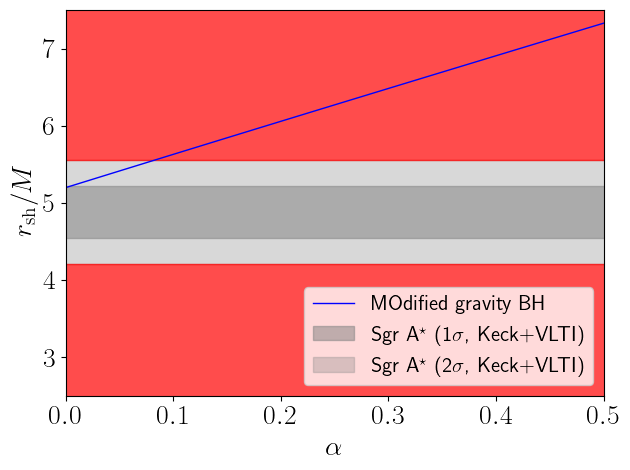}
\caption{Same as in Fig.~\ref{fig:shadow_reissner_nordstrom_bh_ns} for the MOG BH with metric function given by Eq.~\eqref{eq:metricmodifiedgravity}, as discussed in Sec.~\ref{subsec:modifiedgravity}.}
\label{fig:shadow_mog_bh}
\end{figure}

Having made these considerations (see Sec.~\ref{subsubsec:gup} for further considerations of this sort in the context of another model), we compute the shadow size in units of $M$ as done in various earlier works (we refer especially to Ref.~\cite{Moffat:2019uxp}), and show its evolution against the MOG parameter $\alpha$ in Fig.~\ref{fig:shadow_mog_bh}. Moreover, we note that a closed-form expression for the shadow radius can be found, and is given by the following:
\begin{eqnarray}
r_{\rm sh} = \sqrt{2\alpha}\frac{3+3\alpha+\sqrt{(1+\alpha)(9+\alpha)}}{\sqrt{-3+\alpha+\sqrt{(1+\alpha)(9+\alpha)}}}\,,
\label{eq:rshmodifiedgravity}
\end{eqnarray}
which, although not obvious at first glance, does indeed reduce to $3\sqrt{3}$ as $\alpha \to 0$, as can be seen by Taylor expanding Eq.~(\ref{eq:rshmodifiedgravity}) in $\alpha$. We see that the effect of increasing $\alpha$ is that of very quickly increasing the shadow size. Despite the RN-like form of the space-time, this is a result of our choice of reporting the shadow size in units of $M$ rather than $M(1+\alpha)$, for the reasons discussed above.

We note that the EHT observations set extremely tight upper limits on $\alpha$, of the order of $\alpha \lesssim 0.01$ ($1\sigma$) and $\alpha \lesssim 0.1$ ($2\sigma$). These limits are much tighter than the limit $\alpha \lesssim 0.662$ obtained from the motion of the S2-star in Ref.~\cite{DellaMonica:2021xcf}, and safely exclude the range of values of $\alpha$ required for MOG to explain the rotation curves of dwarf galaxies, $2.5 \lesssim \alpha \lesssim 11.2$~\cite{Brownstein:2005zz}. On the other hand, they are competitive with the upper limits of $\alpha \lesssim 0.1$ arising from neutron stars~\cite{LopezArmengol:2016irf}, and $\alpha \lesssim 0.03$ coming from globular clusters~\cite{Moffat:2007yg}. Overall, the limits we have provided on the MOG parameter of order $\alpha \lesssim 0.01$ are therefore among the tightest in the literature.

\subsection{Brane-world black holes}
\label{subsec:rsii}

We now consider one of the most well-motivated high-energy physics models: the Randall-Sundrum II (RSII) brane-world model, which has been tested over a wide range of scales and using an enormous variety of probes (see e.g.\ Refs.~\cite{Antoniadis:1998ig,Johannsen:2008aa,Salumbides:2015qwa,Visinelli:2017bny,Pardo:2018ipy,Chakraborty:2016lxo,Chakravarti:2019aup,Banerjee:2019sae,Yu:2019jlb,Dey:2020lhq,Trivedi:2020wxf,Trivedi:2021ivk,Chakraborty:2021gdf,Banerjee:2021aln,Mishra:2021waw}). This is an AdS$_5$ brane-world model where the extra dimension has an infinite size and negative bulk cosmological constant~\cite{Randall:1999vf}. When projecting the full 5-dimensional BH solution of the theory onto our 4D space-time, BHs effectively inherit a tidal charge $q$ which carries projected information about 5D bulk (and in particular the AdS$_5$ curvature radius $\ell$), while characterizing the effect of the bulk on the brane. More specifically, the tidal charge controls the strength of projected Weyl tensor ${\cal E}_{\mu\nu}$ which transmits tidal charge stresses from the bulk to the brane, and takes the form~\cite{Dadhich:2000am}:
\begin{eqnarray}
{\cal E}_{\mu\nu} = - \left ( \frac{q}{\widetilde{M}_{\rm Pl}^2} \right ) \frac{1}{r^4} \left [ u_{\mu}u_{\nu}-2r_{\mu}r_{\nu}+h_{\mu\nu} \right ]\,,
\label{eq:weylrandallsundrum}
\end{eqnarray}
where $\widetilde{M}_{\rm Pl}$ is the 5-dimensional Planck mass, $u_{\mu}$ is the velocity 4-vector, $r_{\mu}$ is an unit radial vector, and $h_{\mu\nu} = g_{\mu\nu}+u_{\mu}u_{\nu}$. Physically speaking, a negative tidal charge amplifies gravitational effects of the bulk on the brane.

BH solutions within the RSII model have been studied in a large number of works (see e.g.\ Refs.~\cite{Modgil:2001hm,Aliev:2005bi,Schee:2008kz,Aliev:2009cg,Amarilla:2011fx,Aliev:2012rj,Neves:2012it,Eiroa:2017uuq,Abdujabbarov:2017pfw,Stuchlik:2018qyz,Banerjee:2019cjk}, as well as Refs.~\cite{Papnoi:2014aaa,Abdujabbarov:2015rqa}). The metric function for spherically symmetric solutions takes the following form:
\begin{eqnarray}
A(r) = 1-\frac{2M}{r}+\frac{4M^2q}{r^2} \,.
\label{eq:metricrsii}
\end{eqnarray}
where $q$ characterizes a specific hair. We note that the metric function is again RN-like as with a few other metrics discussed earlier. Therefore, we can immediately borrow from the previous results, and write the shadow radius as a function of the tidal charge in a closed form:
\begin{eqnarray}
r_{\rm sh} = \frac{\sqrt{2} \left ( 3+\sqrt{9-32q} \right ) }{\sqrt{4+\frac{\sqrt{9-32q}-3}{4q}}}\,,
\label{eq:shadowsizersii}
\end{eqnarray}

\begin{figure}
\centering
\includegraphics[width=1.0\linewidth]{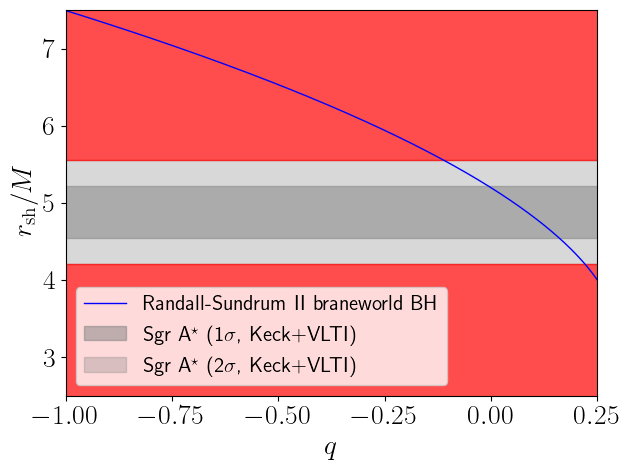}
\caption{Same as in Fig.~\ref{fig:shadow_reissner_nordstrom_bh_ns} for the Randall-Sundrum II brane-world BH with metric function given by Eq.~\eqref{eq:metricrsii}, as discussed in Sec.~\ref{subsec:rsii}.}
\label{fig:shadow_randallsundrumii_bh}
\end{figure}

In Fig.~\ref{fig:shadow_randallsundrumii_bh} we plot the size of the shadow radius as a function of the tidal charge: the plot is of course analogous to Fig.~\ref{fig:shadow_ccsh_bh_wh} and Fig.~\ref{fig:shadow_horndeski_case_1_bh}. It is clear that the EHT observations place very tight limits on the tidal charge, $q \lesssim 0.15$ ($1\sigma$) and $q \lesssim 0.2$ ($2\sigma$). The same observations also place very tight limits on the negative tidal charge regime, $q \gtrsim -0.01$ ($1\sigma$) and $q \gtrsim -0.1$ ($2\sigma$). This is in contrast with the shadow of M87$^*$, which instead exhibited a very slight preference for negative tidal charge, as noted earlier in Refs.~\cite{Banerjee:2019nnj,Neves:2020doc}. This is due to the fact that M87$^*$'s shadow size is slightly larger than what would be expected given its mass-to-distance ratio, or conversely one could argue that its mass-to-distance ratio is biased low. At any rate, this fact again highlights how Sgr A$^*$'s shadow places particularly tight constraints on scenarios which tend to enlarge the shadow radius compared to the Schwarzschild BH. We note that the negative tidal charge regime of the RSII BH would in principle appear to admit an interpretation in terms of mutated Reissner-Nordstr\"{o}m BH/wormhole (just as the space-time discussed in Sec.~\ref{subsec:conformallycoupledscalar}), though to the best of our knowledge we are not aware of this possibility having being discussed in the literature on brane-world BHs.

\subsection{Magnetically charged Einstein-Euler-Heisenberg black hole}
\label{subsec:eeh}

Earlier in Sec.~\ref{subsec:eb} we discussed regular magnetically charged BHs arising from Einstein-Bronnikov NLED. Another well-motivated NLED theory is Einstein-Euler-Heisenberg NLED~\cite{Heisenberg:1936nmg}, which appears in the low-energy limit of Born-Infeld electrodynamics~\cite{Born:1934gh}, and is described by the following Lagrangian:
\begin{eqnarray}
\mathcal{L}(U,W)=-\frac{1}{4}U+\frac{\mu}{4}\left(U^2+\frac{7}{4}W^2\right)\,,
\label{eq:eeh}
\end{eqnarray}
Choosing the same gauge field configuration as in Eq.~(\ref{eq:amu}), and carefully accounting for the NLED-induced effective geometry, BH shadows within the Einstein-Euler-Heisenberg model were studied by some of us in Ref.~\cite{Allahyari:2019jqz}, whose procedure we follow. For completeness, we report the metric function for magnetically charged BH solutions within this theory, given by:
\begin{eqnarray}
A(r) = 1-\frac{2M}{r}+\frac{q_m^2}{r^2}-\frac{2\mu}{5}\frac{q_m^4}{r^6}\,,
\label{eq:metriceeh}
\end{eqnarray}
where $q_m$ is the BH magnetic charge and characterizes a specific hair, and $\mu$ is the NLED coupling appearing in Eq.~(\ref{eq:eeh}). We note that as long as $\mu \neq 0$, values of the magnetic charge $q_m>1$ are allowed.

We compute the shadow radius numerically, following Ref.~\cite{Allahyari:2019jqz} by some of us and once more carefully taking into account the effective geometry induced by the NLED coupling by adopting Novello's method~\cite{Novello:1999pg,DeLorenci:2000yh} (again, as in Sec.~\ref{subsec:eb}), for the intermediate steps of this calculation we refer the reader to Ref.~\cite{Allahyari:2019jqz}. Note that we now have 2 free parameters: $q_m$ and $\mu$. However, for simplicity, we fix the Lagrangian coupling to $\mu=0.3$. The reason is two-fold: \textit{a)} as noted in Ref.~\cite{Allahyari:2019jqz}, most of the effect on the shadow size come from the magnetic charge $q_m$ rather than the NLED coupling $\mu$, and \textit{b)} $\mu \sim 0.3$ is approximately the largest allowed coupling before the perturbative approach of the theory around the Maxwell Lagrangian for $\mu \to 0$ ceases to be meaningful.

\begin{figure}
\centering
\includegraphics[width=1.0\linewidth]{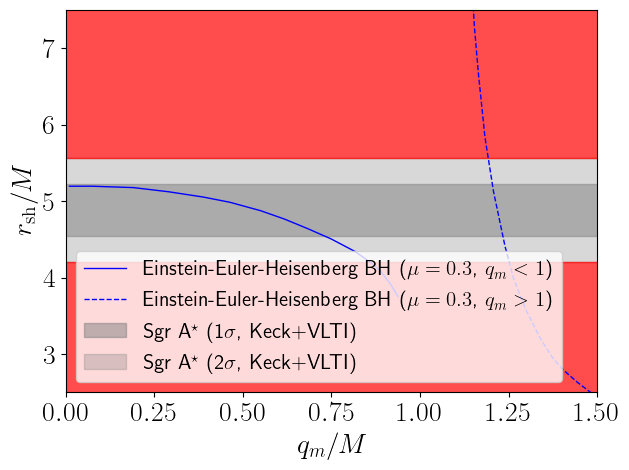}
\caption{Same as in Fig.~\ref{fig:shadow_reissner_nordstrom_bh_ns} for the Einstein-Euler-Heisenberg magnetically charged BH with metric function given by Eq.~\eqref{eq:eeh}, where we have fixed the NLED coupling to $\mu=0.3$, as discussed in Sec.~\ref{subsec:eeh}.}
\label{fig:shadow_euler_heisenberg_bh}
\end{figure}

In Fig.~\ref{fig:shadow_euler_heisenberg_bh} we plot the evolution of the shadow radius against $q_m$, at fixed $\mu=0.3$. We immediately note that the behavior of the shadow radius, and therefore the constraints from the EHT observations, depend on whether we are in the $q_m<1$ or $q_m>1$ regime. For $q_m<1$, we find the upper limits $q_m \lesssim 0.7M$ ($1\sigma$) and $q_m \lesssim 0.8M$ ($2\sigma$). On the other hand, for $q_m>1$, we observe a sharp discontinuity in the shadow radius, with the shadow size apparently diverging. This feature is due to the appearance of a singularity in the NLED-induced effective geometry: for more details see Ref.~\cite{Allahyari:2019jqz}.

Leaving aside the issue of whether the $q_m>1$ regime is physical in first place, given this singularity, in this regime we find that the shadow size varies drastically as $q_m$ changes. For $\mu=0.3$, consistency with the EHT observations requires $q_m \sim 1.25M$. This number depends to some extent on the value to which $\mu$ is fixed (see Ref.~\cite{Allahyari:2019jqz} for further discussions), although for values of $\mu \lesssim 0.3$, following the results of Ref.~\cite{Allahyari:2019jqz}, we can expect consistency with the EHT observations for values of the magnetic charge $1.1M \lesssim q_m \lesssim 1.5M$. We stress once again, however, that it is unclear whether we can consider this range of parameter space physical.

\subsection{Sen black hole} 
\label{subsec:sen}

The Sen BH is a non-regular space-time appearing in the low-energy limit of heterotic string theory~\cite{Gross:1984dd}, whose field content includes the electromagnetic field $A^{\mu}$ with field-strength tensor $F^{\mu\nu}$, the second-rank antisymmetric tensor gauge field $B_{\mu\nu}$ (also known as Kalb-Ramond field~\cite{Kalb:1974yc}), and the dilaton field $\Phi$,~\footnote{In heterotic string theory the dilaton, the Kalb-Ramond field, and the metric tensor appear as a set of massless excitations of a closed string.} as described by the following action in the string frame~\cite{Sen:1992ua} (see also Refs.~\cite{Gibbons:1987ps,Garfinkle:1990qj,Horowitz:1991cd,Shapere:1991ta} for other examples of BH and black \textit{p}-brane solutions in the low-energy limit of string theory):
\begin{eqnarray}
S = \int {\rm d}^4x\sqrt{-g}\,e^{-\phi}\,{\cal L} \,,
\label{eq:actionsen}
\end{eqnarray}
where the Lagrangian ${\cal L}$ is given by:
\begin{eqnarray}
{\cal L} = R-\frac{1}{8}F^{\mu\nu}F_{\mu\nu}+\partial^{\mu}\Phi\partial_{\mu}\Phi-\frac{1}{12}H_{\kappa\lambda\mu}H^{\kappa\lambda\mu}\,,
\label{eq:lagrangiansen}
\end{eqnarray}
with $H_{\kappa\lambda\mu}$ the third-rank tensor field (Kalb-Ramond 3-form) defined in terms of the tensor gauge field and the field-strength tensor of the electromagnetic field as:
\begin{eqnarray}
H_{\kappa\mu\nu} &\equiv& \partial_{\kappa}B_{\mu\nu}+\partial_{\nu}B_{\kappa\mu}+\partial_{\mu}B_{\nu\kappa} \nonumber \\
&&-\frac{1}{4} \left ( A_{\kappa}F_{\mu\nu}+A_{\nu}F_{\kappa\mu}+A_{\mu}F_{\nu\kappa} \right )\,.
\label{eq:hklm}
\end{eqnarray}
The metric function derived in this configuration was found by Sen in Ref.~\cite{Sen:1992ua} and reads:
\begin{eqnarray}
A(r) = 1-\frac{2M}{r+\frac{q_m^2}{M}} \,,
\label{eq:metricsen}
\end{eqnarray}
where $q_m$ is an effective charge (including both the electric charge and an effective charge associated with the dilaton field) measured by a static observer at infinity, which characterizes specific hair, and is bound by the requirement $q_m \leq \sqrt{2}$. Various studies have been devoted to the Sen BH and its rotating generalization (the Kerr-Sen BH) in the literature, see for instance Refs.~\cite{Gyulchev:2009dx,Dastan:2016bfy,Lin:2017oag,Guo:2019lur,Xavier:2020egv,Zhang:2020tfz,Zhang:2020pay,Belhaj:2020idr,Wu:2021pgf,Carleo:2022qlv}.

\begin{figure}
\centering
\includegraphics[width=1.0\linewidth]{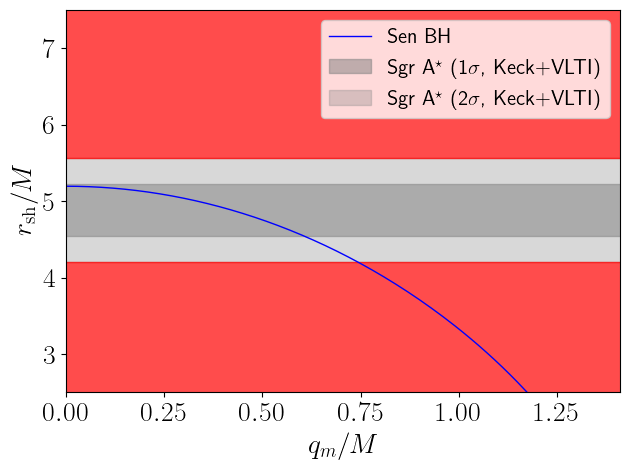}
\caption{Same as in Fig.~\ref{fig:shadow_reissner_nordstrom_bh_ns} for the Sen BH with metric function given by Eq.~\eqref{eq:metricsen}, as discussed in Sec.~\ref{subsec:sen}.}
\label{fig:shadow_sen_bh}
\end{figure}

A closed-form expression exists for the shadow radius, and is given as follows:
\begin{eqnarray}
r_{\rm sh} = \frac{3-2q_m^2+\sqrt{9-4q_m^2}}{2\sqrt{\frac{\sqrt{9-4q_m^2}+q_m^2-3}{q_m^2}}}\,.
\label{eq:rshsen}
\end{eqnarray}
Although not obvious, Eq.~(\ref{eq:rshsen}) reduces to $3\sqrt{3}$ as $q_m \to 0$, as can be seen by Taylor expanding it in $q_m$. We plot the evolution of the shadow radius against $q_m$ in Fig.~\ref{fig:shadow_sen_bh}, from which we see that the shadow radius quickly decreases with increasing charge. We also see that the EHT observations set the upper limits $q_m \lesssim 0.6M$ ($1\sigma$) and $q_m \lesssim 0.75M$ ($2\sigma$), ruling out the possibility of Sgr A$^*$ being an extremal Sen BH with $q_m=\sqrt{2}M$. We further note that complementary X-ray reflection spectroscopy constraints obtained by one of us set a much tighter limit, $q_m \lesssim 0.6M$ at $90\%$~C.L.~\cite{Tripathi:2021rwb}, where as in Sec.~\ref{subsec:simpsonvisser} $M$ denotes the mass of EXO 1846-031.

\subsection{Kalb-Ramond black hole}
\label{subsec:kalbramond}

Remaining in string-inspired scenarios, another interesting BH solution has been derived in a scenario where the Kalb-Ramond field $B_{\mu\nu}$ plays a key role in driving spontaneous breaking of Lorentz symmetry~\cite{Kalb:1974yc}. Specifically, the scenario considered by Ref.~\cite{Lessa:2019bgi} envisages a field content including the Kalb-Ramond field $B_{\mu\nu}$ and the Kalb-Ramond 3-form $H_{\kappa\mu\nu}$.~\footnote{Since the electromagnetic field does not appear within the field content considered by Ref.~\cite{Lessa:2019bgi}, the contributions related to $A_{\mu}$ in Eq.~(\ref{eq:hklm}) vanish.} Inspired by the Standard Model Extension (an effective field theory containing all possible Lorentz breaking operators in addition to the Standard Model and GR Lagrangian terms~\cite{Kostelecky:1988zi,Colladay:1998fq,Kostelecky:2003fs}) and by heterotic string theory, Ref.~\cite{Lessa:2019bgi} considered a scenario where the Kalb-Ramond field acquires a non-vanishing vacuum expectation value (VEV) $b_{\mu\nu}$ (with $b^2=b_{\mu\nu}b^{\mu\nu}$), as well as a non-minimal coupling to the Ricci tensor with strength $\xi_2$ (for the full action of the theory, see Ref.~\cite{Lessa:2019bgi}). The non-zero VEV spontaneously breaks Lorentz invariance, leading to a host of interesting observational signatures~\cite{Cruz:2013zka,Maluf:2018jwc,Maluf:2021eyu}.

BHs within this configuration have been studied by Ref.~\cite{Lessa:2019bgi}, and found to be described by the following metric function:
\begin{eqnarray}
A(r)=1-\frac{2M}{r}+\frac{\Gamma}{r^{\frac{2}{k}}} \,,
\label{eq:metrickalbramond}
\end{eqnarray}
where $\Gamma$ is an integration constant, and $k$ is related to the Kalb-Ramond VEV and non-minimal coupling strength through $k=\vert b \vert^2\xi_2$ and therefore characterizes an universal hair. For illustrative purposes, here we choose to focus our attention on the parameter $k$ which is of more direct interest (since it enjoys a direct physical interpretation) and is expected to be small. To this end, we fix $\Gamma=0.5$: although there is no deep motivation for this value, this allows us to also compare our results to the earlier work of Ref.~\cite{Kumar:2020hgm} by some of us.

\begin{figure}
\centering
\includegraphics[width=1.0\linewidth]{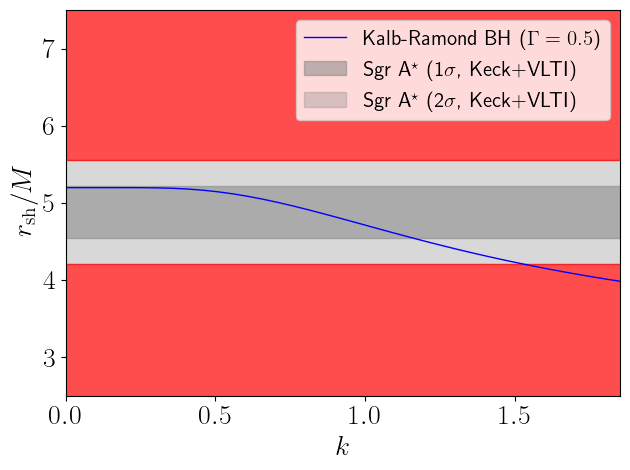}
\caption{Same as in Fig.~\ref{fig:shadow_reissner_nordstrom_bh_ns} for the Kalb-Ramond BH with metric function given by Eq.~\eqref{eq:metrickalbramond}, where we have fixed $\Gamma=0.5$, as discussed in Sec.~\ref{subsec:kalbramond}.}
\label{fig:shadow_kalb_ramond_bh}
\end{figure}

We compute the size of the Kalb-Ramond BH numerically, and show its evolution against $k$ in Fig.~\ref{fig:shadow_kalb_ramond_bh}. As already found earlier in Ref.~\cite{Kumar:2020hgm}, we see that increasing $k$ decreases the size of the BH shadow (see also Ref.~\cite{Atamurotov:2022wsr}. For this particular choice of $\Gamma$, we find that the EHT observations set the upper limits $k \lesssim 1.2$ ($1\sigma$) and $k \lesssim 1.5$ ($2\sigma$). If independent limits on the Kalb-Ramond VEV are available, these limits can be translated to limits on the Kalb-Ramond non-minimal coupling strength (and vice-versa). However, we stress that these results only correspond to a slice of the full $\Gamma$-$k$ parameter space, and that smaller (larger) values of $\Gamma$ would lead to weaker (tighter) constraints on $k$ (for more details see Ref.~\cite{Kumar:2020hgm}).

In closing, it is worth noting that the space-time described by Eq.~(\ref{eq:metrickalbramond}) is closely related to the 4D string-corrected BH solution found by one of us in Ref.~\cite{Jusufi:2022uhk}, which generalizes the Callan-Myers-Perry BH~\cite{Callan:1988hs} through a dimensional regularization-like procedure akin to the one considered in 4D Einstein-Gauss-Bonnet gravity, which we will discuss later in Sec.~\ref{subsec:4degb}. The solution derived in Ref.~\cite{Jusufi:2022uhk} corresponds to Eq.~(\ref{eq:metrickalbramond}) with $k=1/2$ (thus leading to a $1/r^4$ correction to the Schwarzschild BH), and requires $\Gamma \leq 27/16M^4 \sim 1.69M^4$. However, it is easy to show (as we have explicitly checked) that within this particular space-time, and given the above upper limit on $\Gamma$, the EHT observations are not able to set any meaningful limit on $\Gamma$: thus, we will not discuss this space-time further.~\footnote{Note that for this particular space-time we cannot translate the Reissner-Nordstr\"{o}m constraints on $Q$ to constraints on $\Gamma$, as the latter space-time corresponds to $k=1$ (and $\Gamma=Q^2$), whereas here we have $k=1/2$.}

\subsection{Einstein-Maxwell-dilaton gravity} 
\label{subsec:emd1}

Still remaining in the string-inspired realm, one of the models in this context which enjoys particularly strong theory motivation is Einstein-Maxwell dilaton gravity (EMDG), which arises as an effective theory of the heterotic string at low energies once the additional six dimensions of the ten-dimensional manifold are compactified onto a torus~\cite{Sen:1992ua}, see also Sec.~\ref{subsec:sen}. In EMDG, a dynamical dilaton field $\Phi$ evolving under a potential $V(\Phi)$ is included in the GR Lagrangian as:
\begin{eqnarray}
{\cal L} \supset \frac{1}{16\pi} \left ( {\cal R}+\frac{1}{2}(\nabla^\mu\Phi)^2+V(\Phi) \right ) +\frac{\alpha'}{8}\Phi F^{\mu\nu}F_{\mu\nu}\,, \nonumber \\
\end{eqnarray}
where $\alpha'$ regulates the coupling of the dilaton to photons. Given a BH of dilaton charge $Q$, the corresponding metric function of the so-called Einstein-Maxwell-dilaton-1 BH is given by~\cite{Gibbons:1987ps,Garfinkle:1990qj,Garcia:1995qz}:
\begin{eqnarray}
A(r) = 1-\frac{2M}{r}\left(\sqrt{1 + \frac{q^4}{4M^2r^2}}-\frac{q^2}{2Mr}\right) \,,
\label{eq:metricemd1}
\end{eqnarray}
with the requirement that the parameter $q = Q\sqrt{\alpha'/8}$ associated to the dilaton charge, which characterizes a specific hair, satisfies $q \leq \sqrt{2}$.

\begin{figure}
\centering
\includegraphics[width=1.0\linewidth]{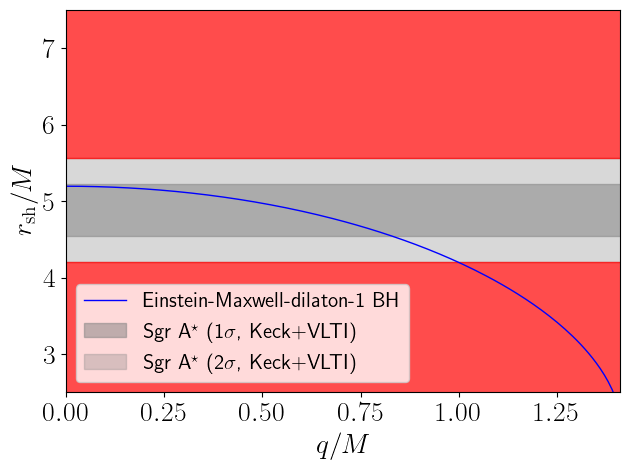}
\caption{Same as in Fig.~\ref{fig:shadow_reissner_nordstrom_bh_ns} for the Einstein-Maxwell-dilaton 1 BH with metric function given by Eq.~\eqref{eq:metricemd1}, as discussed in Sec.~\ref{subsec:emd1}.}
\label{fig:shadow_einstein_maxwell_dilaton_1_bh}
\end{figure}

A closed-form expression exists for the shadow radius, and is given as follows:
\begin{eqnarray}
r_{\rm sh} = \frac{\sqrt{6 \left ( \Tilde{q}+6 \right ) -q^2 \left ( q^2+\sqrt{q^4-20 q^2+36}+16 \right ) }}{2 \sqrt{2-\frac{16}{2 q^2+\sqrt{2} \sqrt{q^4- \left ( \Tilde{q}+16 \right ) q^2+6 \left ( \Tilde{q}+6 \right ) }}}}\,, \nonumber \\
\label{eq:rshemd1}
\end{eqnarray}
where the quantity $\Tilde{q}$ is defined as:
\begin{eqnarray}
\Tilde{q} \equiv \sqrt{q^4-20 q^2+36}\,.
\label{eq:qtilde}
\end{eqnarray}
Although not obvious, Eq.~(\ref{eq:rshemd1}) reduces to $3\sqrt{3}$ as $q \to 0$, as can be seen by Taylor expanding it in $q$. We show the evolution of the shadow radius as a function of the charge parameter $q$ in Fig.~\ref{fig:shadow_einstein_maxwell_dilaton_1_bh}. We see that the EHT observations set the upper limits $q \lesssim 0.8M$ ($1\sigma$) and $q \lesssim M$ ($2\sigma$), ruling out the extremal Einstein-Maxwell-dilaton-1 BH ($q = \sqrt{2}M$).

\subsection{Einstein-\ae ther gravity} 
\label{subsec:einsteinaether}

Einstein-\ae ther gravity is a theory of gravity that dynamically violates Lorentz symmetry by means of a unit norm time-like vector field (the ``\ae ther''), which defines a preferred time-like direction at each point of space-time. The Einstein-\ae ther gravity Lagrangian is governed by 4 free parameters $c_i$ with $i=1,2,3,4$, and is given by the following~\cite{Jacobson:2000xp,Zlosnik:2006zu,Jacobson:2007veq} (see also Refs.~\cite{Jacobson:2004ts,Foster:2005dk,Konoplya:2006rv,Jacobson:2013xta,Yagi:2013ava,Ding:2015kba,Gong:2018cgj,Leon:2019jnu,Paliathanasis:2020bgs,Zhu:2021tgb} for other important works):
\begin{eqnarray}
{\cal L} \supset -h^{\alpha\beta}_{\mu\nu} \left ( \nabla_{\alpha}u^{\mu} \right ) \left ( \nabla_{\beta}u^{\nu} \right ) +\lambda \left ( u^2+1 \right )\,,
\label{eq:einsteinaether}
\end{eqnarray}
where $u^{\mu}$ is the \ae ther field, $\lambda$ is a Lagrange multiplier enforcing the condition $u^{\mu}u_{\mu}=-1$, and the tensor $h^{\alpha\beta}_{\mu\nu}$ is given by:
\begin{eqnarray}
h^{\alpha\beta}_{\mu\nu} = c_1g^{\alpha\beta}g_{\mu\nu}+c_2\delta^{\alpha}_{\mu}\delta^{\beta}_{\nu}+c_3\delta^{\alpha}_{\nu}\delta^{\beta}_{\mu}-c_4u^{\alpha}u^{\beta}g_{\mu\nu}\,.
\label{eq:halphabetamunu}
\end{eqnarray}
In what follows, we will adopt the widely used shorthand notation $c_{ij}=c_i+c_j$, $c_{ijk}=c_i+c_j+c_k$.

\begin{figure}
\centering
\includegraphics[width=1.0\linewidth]{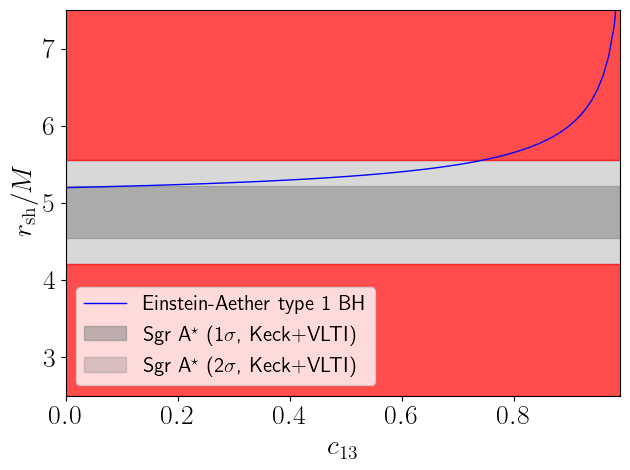}
\caption{Same as in Fig.~\ref{fig:shadow_reissner_nordstrom_bh_ns} for the Einstein-\ae ther-1 BH with metric function given by Eq.~\eqref{eq:metriceinsteinaether1}, as discussed in Sec.~\ref{subsec:einsteinaether}.}
\label{fig:shadow_einstein_aether_1_bh}
\end{figure}

Two exact static spherically symmetric BH solutions within Einstein-\ae ther gravity are known~\cite{Eling:2006ec,Barausse:2011pu}. The first solution, commonly referred to as Einstein-\ae ther type 1, holds for $c_{14}=0$ and $c_{123} \neq 0$. The metric function for this space-time is given by:
\begin{eqnarray}
A(r) = 1-\frac{2M}{r}-\frac{27c_{13}M^4}{16(1-c_{13})r^4} \,,
\label{eq:metriceinsteinaether1}
\end{eqnarray}
with the requirement $0<c_{13}<1$, where $c_{13}$ obviously characterizes an universal hair. We compute the shadow radius numerically, plotting it as a function of the Lagrangian parameter $c_{13}$ in Fig.~\ref{fig:shadow_einstein_aether_1_bh}. We see yet again an example of shadow radius increasing with increasing fundamental parameter, which leads to this scenario being tightly constrained by the ETH observations, which set the upper limits $c_{13} \lesssim 0.1$ ($1\sigma$) and $c_{13} \lesssim 0.75$ ($2\sigma$) -- note that the $c_i$s are dimensionless. While this is an independent constraint on $c_{13}$ obtained in the strong-field regime, we note that it is still significantly weaker than constraints arising from GWs. For instance, the multi-messenger GW event GW170817 sets the very stringent limit $\vert c_{13} \vert \lesssim 10^{-15}$~\cite{Oost:2018tcv}.

\begin{figure}
\centering
\includegraphics[width=1.0\linewidth]{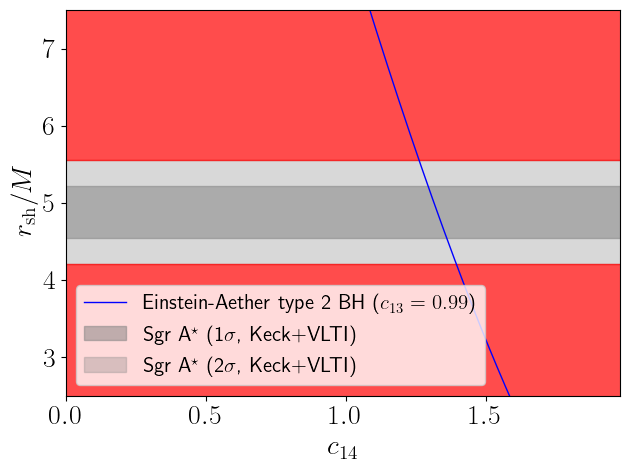}
\caption{Same as in Fig.~\ref{fig:shadow_reissner_nordstrom_bh_ns} for the Einstein-\ae ther-2 BH with metric function given by Eq.~\eqref{eq:metriceinsteinaether2}, where we have fixed the Lagrangian parameter $c_{13}$ to $c_{13}=0$, as discussed in Sec.~\ref{subsec:einsteinaether}.}
\label{fig:shadow_einstein_aether_2_bh}
\end{figure}

The second exact BH solution in this theory holds for $c_{14} \neq 0$ and $c_{123}=0$, and is commonly referred to as Einstein-\ae ther type 2. The metric function for this space-time is given by:
\begin{eqnarray}
A(r) = 1-\frac{(2-c_{14})M}{r}-\frac{(2-c_{13}-c_{14})(2-c_{14})^2M^2}{8(1-c_{13})r^2} \,, \nonumber \\
\label{eq:metriceinsteinaether2}
\end{eqnarray}
with the requirement $0 \leq c_{14} \leq 2c_{13}<2$, where $c_{13}$ and $c_{14}$ obviously characterize universal hairs. In this case, the space-time is described by 2 free parameters. For illustrative purposes, we fix $c_{13}=0.99$, close to the maximum value allowed, which as argued earlier is completely excluded in the case of the Einstein-\ae ther type 1 BH. The purpose is simply to illustrate how a non-zero value of $c_{14}$ can in principle help bring a previously excluded range of $c_{13}$ values into agreement with the EHT observations. We compute the shadow radius numerically (see also Ref.~\cite{Atamurotov:2013dpa}), plotting it as a function of the Lagrangian parameter $c_{13}$ in Fig.~\ref{fig:shadow_einstein_aether_2_bh}. In this case we see that in principle a value of $1.3 \lesssim c_{14} \lesssim 1.4$ is allowed by the EHT observations, when fixing $c_{13}=0.99$. However, we note that these values of $c_{14}$ are excluded by GW observations, with the GW170817 event setting the very stringent limit $c_{14} \lesssim 2.5 \times 10^{-5}$~\cite{Oost:2018tcv}.

In closing, we discuss important relations between Einstein-\ae ther gravity and other well-known theories of gravity, to which our results could therefore carry over. We first note that Einstein-\ae ther gravity is closely related to Ho\v{r}ava-Lifshitz (HL) gravity~\cite{Horava:2009uw,Mukohyama:2010xz}, a well-known and well-studied candidate theory of quantum gravity which achieves power-counting renormalizability by adding higher order spatial derivatives but not time derivatives, thus taming the UV behavior of the graviton propagator, at the cost of explicitly breaking diffeomorphism invariance.~\footnote{This explicit diffeomorphism invariance breaking has been argued to lead to several pathologies in the IR, see e.g.\ Refs.~\cite{Charmousis:2009tc,Nojiri:2009th,Blas:2009yd,Blas:2009qj,Papazoglou:2009fj,Henneaux:2009zb,Blas:2009ck,Nojiri:2010tv,Blas:2010hb,Cognola:2016gjy} for discussions thereof, as well as possible ways out.} Einstein-\ae ther gravity is known to appear in the IR limit of HL gravity (when operators higher than second order are neglected), provided one requires the \ae ther field to be hypersurface orthogonal at the level of the action~\cite{Jacobson:2010mx}. In fact, the Einstein-\ae ther BH solutions we have studied are also solutions of the IR limit of HL gravity, as discussed in Ref.~\cite{Barausse:2011pu} (see also Ref.~\cite{Cai:2009pe} for other BH solutions in HL gravity). Another well-known modified theory of gravity appearing in the IR limit of (projectable) HL gravity is mimetic gravity~\cite{Lim:2010yk,Chamseddine:2013kea,Chamseddine:2014vna}: see Ref.~\cite{Ramazanov:2016xhp} for a full proof of this equivalence, and Refs.~\cite{Myrzakulov:2015nqa,Chamseddine:2019gjh} for discussions on relations to Einstein-\ae ther and HL gravity. Mimetic gravity is related to GR by a singular (non-invertible) disformal transformation, and has gained significant attention recently (see e.g.\ Refs.~\cite{Momeni:2014qta,Nojiri:2014zqa,Leon:2014yua,Arroja:2015wpa,Rabochaya:2015haa,Odintsov:2015ocy,Arroja:2015yvd,Nojiri:2016ppu,Nojiri:2016vhu,Firouzjahi:2017txv,Zheng:2017qfs,Takahashi:2017pje,Paston:2017das,Dutta:2017fjw,Gorji:2018okn,Ganz:2018mqi,Gorji:2019ttx,Khalifeh:2019zfi,Ganz:2019vre,Sur:2020lzd,Izaurieta:2020kuy,deCesare:2020got,Benisty:2021cin,Mansoori:2021fjd,Nashed:2022yfc,Jirousek:2022rym}, and Ref.~\cite{Sebastiani:2016ras} for a review). Spherically symmetric solutions in mimetic gravity have been studied in Refs.~\cite{Deruelle:2014zza,Myrzakulov:2015sea,Myrzakulov:2015kda,Chen:2017ify,Nashed:2018qag,Sheykhi:2019gvk,Gorji:2020ten,Sheykhi:2020fqf,Casalino:2020vhl,Nashed:2021ctg,Nashed:2021hgn,Nojiri:2022cah}, though the issue remains open as to whether mimetic gravity can actually support non-pathological BH solutions. A full investigation of the relation between Einstein-\ae ther and mimetic gravity as far as BH solutions are concerned is therefore warranted, and is left to future work, where we also plan to study the shadows cast by spherically symmetric space-times (BHs, WHs, or NSs) in mimetic gravity.

\subsection{4D Einstein-Gauss-Bonnet gravity}
\label{subsec:4degb}

The Gauss-Bonnet invariant ${\cal G}$ is a very important quantity, which in more than 4 dimensions plays a role in determining the allowed Lagrangian terms (Lovelock invariants) of diffeomorphism-invariant, metric theories of gravity with equations of motion of second order~\cite{Lovelock:1971yv}, and is given by:
\begin{eqnarray}
{\cal G} = R^{\mu\nu}_{\rho\sigma}R^{\rho\sigma}_{\mu\nu}-4R^{\mu}_{\nu}R^{\nu}_{\mu}+R^2\,.
\label{eq:gaussbonnet}
\end{eqnarray}
Crucially, in 4 dimensions ${\cal G}$ amounts to a total derivative, and therefore cannot contribute to the gravitational dynamics. However, in Ref.~\cite{Glavan:2019inb} Glavan and Lin proposed what effectively amounts to a dimensional regularization procedure applied to the following action:
\begin{eqnarray}
S = \int {\rm d}^Dx\sqrt{-g}\,\alpha{\cal G}\,,
\label{eq:action4degb}
\end{eqnarray}
by rescaling the coupling constant as $\alpha \to \alpha/(D-4)$ (see also the generalization of Ref.~\cite{Casalino:2020kbt}). This procedure is shown to lead to non-trivial contributions even in $D=4$, and we refer to the resulting theory as 4D Einstein-Gauss-Bonnet gravity (4DEGB). See e.g.\ Ref.~\cite{Fernandes:2022zrq} for a recent review of the theory, Refs.~\cite{Lu:2020iav,Kobayashi:2020wqy,Ai:2020peo,Arrechea:2020evj,Gurses:2020ofy,Fernandes:2020nbq,Hennigar:2020lsl,Mahapatra:2020rds,Aoki:2020lig,Gurses:2020rxb} for important discussions on possible pathologies associated to the rescaling procedure, and Refs.~\cite{Doneva:2020ped,EslamPanah:2020hoj,Aragon:2020qdc,Malafarina:2020pvl,Yang:2020czk,Shu:2020cjw,Jusufi:2020yus,Yang:2020jno,Chakraborty:2020ifg,Banerjee:2020dad,Wei:2020rcd,Hohmann:2020cor,Jafarzade:2020ilt,Aoki:2020ila,Wang:2021kuw,Atamurotov:2021imh,Ovgun:2021ttv,Gyulchev:2021dvt,Nashed:2022zxm,Donmez:2022dze} for other important works on the theory.

With all the caveats discussed in Refs.~\cite{Lu:2020iav,Ai:2020peo,Gurses:2020ofy,Fernandes:2020nbq,Hennigar:2020lsl,Mahapatra:2020rds,Aoki:2020lig} in mind, static spherically symmetric solutions within 4DEGB gravity have been studied, and the metric function is given by the following (see e.g.\ Refs.~\cite{Glavan:2019inb,Konoplya:2020bxa,Wei:2020ght,Konoplya:2020qqh,Kumar:2020owy,Kumar:2020uyz,HosseiniMansoori:2020yfj,Islam:2020xmy,Kumar:2020xvu,Kumar:2020sag,ElMoumni:2020wrf,Zahid:2021vdy,Donmez:2021fbk,Narzilloev:2021jtg,Papnoi:2021rvw}):
\begin{eqnarray}
A(r) = 1+\frac{r^2}{32\pi\alpha} \left ( 1\pm\sqrt{1+\frac{128\pi\alpha M}{r^3}} \right ) \,,
\label{eq:metric4degb}
\end{eqnarray}
which, although $\alpha$ characterizes universal hair, requires $\alpha \leq M^2/16\pi \sim 0.02M^2$, i.e.\ $\alpha \leq {\cal O}(10^{18})\,{\rm m^2}$ for Sgr A$^*$. We need to take the negative branch of Eq.~(\ref{eq:metric4degb}), as it is the only one which recovers the Schwarzschild metric function $A(r)=1-2M/r$ in the limit $r \to \infty$.

\begin{figure}
\centering
\includegraphics[width=1.0\linewidth]{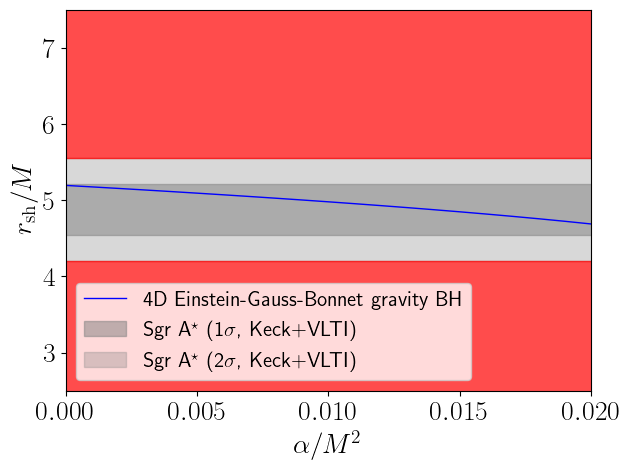}
\caption{Same as in Fig.~\ref{fig:shadow_reissner_nordstrom_bh_ns} for the BH in 4D Einstein-Gauss-Bonnet gravity with metric function given by Eq.~\eqref{eq:metric4degb}, as discussed in Sec.~\ref{subsec:4degb}.}
\label{fig:shadow_4degb_bh}
\end{figure}

We compute the shadow radius numerically, and show its evolution as a function of $\alpha$ in Fig.~\ref{fig:shadow_4degb_bh}.~\footnote{We have found that a closed-form expression for the shadow radius exists, but is too cumbersome to report here. We have nonetheless verified that our numerical results perfectly reproduce those provided by this closed-form expression.} We see that the shadow radius decreases with increasing $\alpha$. However, for all values of $\alpha \leq 1/16\pi$, the shadow radius is always consistent with the EHT observations within better than $1\sigma$. Therefore, the size of Sgr A$^*$'s shadow does not place meaningful constraints on the 4DEGB coupling $\alpha$, which is instead constrained by alternative astrophysical probes. For instance, the tightest constraint on $\alpha$ comes from binary BH systems, which limit $\alpha \lesssim 10^8m^2$~\cite{Clifton:2020xhc}. For comparison, it is useful to recast the previous limit in units of Sgr A$^*$'s mass: noting that Sgr A$^*$'s gravitational radius is of order $r_g \approx {\cal O}(10^7)\,{\rm km}$, the previous constraint translates to the extremely stringent limit $\alpha \lesssim 10^{-28}M^2$, much tighter than any constraint realistically achievable from BH shadows.

\subsection{Asymptotically safe gravity}
\label{subsec:asg}

The issue of the non-renormalizability of GR, currently one of the biggest obstacles towards constructing a full quantum theory of gravity, has prompted a number of interesting ideas, many of which conjecture the existence of novel symmetries. Asymptotically safe (AS) gravity broadly falls within this category, in that it provides a quantum realization of scale symmetry. While, classically, the absence of dimensionful couplings is enough to ensure the realization of scale symmetry, this is no longer true at the quantum level, due to quantum fluctuations in the vacuum. In AS gravity, quantum scale symmetry is realized thanks to the presence of an interacting fixed point of the renormalization group (RG)~\cite{Weinberg:1980gg,Reuter:1996cp}: beyond a certain energy scale, a transition to a quantum scale-invariant regime occurs, and the strength of the gravitational constant flows to a constant but non-zero value (the case where this constant is zero would essentially correspond to a realization of asymptotic freedom). Therefore, within AS gravity quantum scale symmetry is achieved at fixed points of the RG, enabling an UV extension of GR (interpreted as an effective field theory) and restoring the predictivity of the latter in the UV. Currently, AS gravity is considered a very promising approach towards a theory of quantum gravity (see Refs.~\cite{Koch:2014cqa,Bonanno:2017pkg,Platania:2020lqb,Bonanno:2020bil} for reviews), and has been the subject of study of an enormous body of literature, see e.g.\ Refs.~\cite{Shaposhnikov:2009pv,Eichhorn:2010tb,Bonanno:2010bt,Bonanno:2010mk,Bonanno:2012jy,Eichhorn:2012va,Litim:2014uca,Litim:2015iea,Bonanno:2015fga,Meibohm:2015twa,Dona:2015tnf,Biemans:2016rvp,Mann:2017wzh,Eichhorn:2017sok,Eichhorn:2018vah,Bonanno:2018gck,Eichhorn:2018whv,Adeifeoba:2018ydh,deAlwis:2019aud,Draper:2020knh,Platania:2020knd,Basile:2021krr,Eichhorn:2021tsx,Fehre:2021eob,deBrito:2020afe,Trivedi:2022svr,Lambiase:2022xde,Scardigli:2022jtt,Knorr:2022lzn,Platania:2022gtt,Fraaije:2022uhg,Platania:2023srt} for important examples.

RG-improved BH space-times within the framework of AS gravity were first studied in detail by Bonanno and Reuter in Ref.~\cite{Bonanno:2000ep}, considering the following expression for the running of the gravitational constant as a function of momentum $k$:
\begin{eqnarray}
G(k) = \frac{G(k_0)}{1+\omega G(k_0)(k-k_0)^2}\,,
\label{eq:grunningmomentum}
\end{eqnarray}
where $k_0$ is a reference scale later set to $k_0 \to 0$, so that $G(k_0)$ effectively corresponds to Newton's constant $G_N$ as measured by low-energy observations. Close to the RG fixed point, i.e.\ for $k^2 \gg G_N^{-1}$, the gravitational constant runs as $G(k)=1/\omega k^2$, where $\omega$ is the inverse dimensionless fixed point value: thus, the gravitational constant ``forgets'' its IR value when entering the quantum scale-invariant regime. To consider the impact of Eq.~(\ref{eq:grunningmomentum}) on BH space-times, one needs a prescription for converting $k$ to a position-dependent IR cutoff. Various well-motivated prescriptions were studied in Ref.~\cite{Bonanno:2000ep}. Here, we simply quote the final expression, leaving the reader to Ref.~\cite{Bonanno:2000ep} for further details:
\begin{eqnarray}
G(r) = \frac{G_Nr^3}{r^3+\widetilde{\omega}G_N \left ( r+\gamma G_NM \right )}\,,
\label{eq:grunningposition}
\end{eqnarray}
where $M$ is the BH mass, and $\widetilde{\omega}$ is an in principle free parameter, related to the inverse dimensionless fixed point value $\omega$, and expected to be of order unity from first principles considerations.~\footnote{It is worth noting that the arguments presented in Refs.~\cite{Donoghue:1993eb} lead to the specific prediction of $\widetilde{\omega}=118/15\pi \approx 2.5$.} To obtain the RG-improved AS-driven corrections to the Schwarzschild space-time, one therefore simply replaces $G$ in the latter with Eq.~(\ref{eq:grunningposition}), after identifying $r$ with the usual radial coordinate (see also Refs.~\cite{Reuter:2010xb,Becker:2012js,Falls:2012nd,Koch:2013rwa,Saueressig:2015xua,Gonzalez:2015upa,Bonanno:2016dyv,Bonanno:2017zen,Platania:2019kyx,Held:2019xde,Rincon:2020iwy,Eichhorn:2021etc,Eichhorn:2021iwq,Knorr:2022kqp,Delaporte:2022acp,Eichhorn:2022oma,Eichhorn:2022bbn} for other relevant works on BHs in AS gravity). As argued explicitly in Ref.~\cite{Bonanno:2000ep}, $\gamma$ plays a very minor role in the phenomenology of the model, especially at the scales relevant for BH observations. Therefore, in the following, we shall set $\gamma=0$, which effectively corresponds to considering the IR limit ($r \gg \sqrt{\gamma}$) of Eq.~(\ref{eq:grunningposition}). Explicitly, the the metric function we consider is therefore given by (where we have obviously set $G_N=1$ for convenience):
\begin{eqnarray}
A(r) = 1-\frac{2Mr}{r^2+\widetilde{\omega}} \,,
\label{eq:metricasg}
\end{eqnarray}
where $\widetilde{\omega}$ characterizes an universal hair. One can easily show that the Kretschmann and Ricci scalars associated to the above metric are finite throughout the entire space-time, which therefore describes a globally regular BH. Because the curvature scalars are finite everywhere, including $r=0$, the metric describes a globally regular BH space-time. The impact of this and related space-times on BH observables such as shadows and gravitational deflection of light have been recently studied in Refs.~\cite{Kumar:2019ohr,Held:2019xde}.

\begin{figure}
\centering
\includegraphics[width=1.0\linewidth]{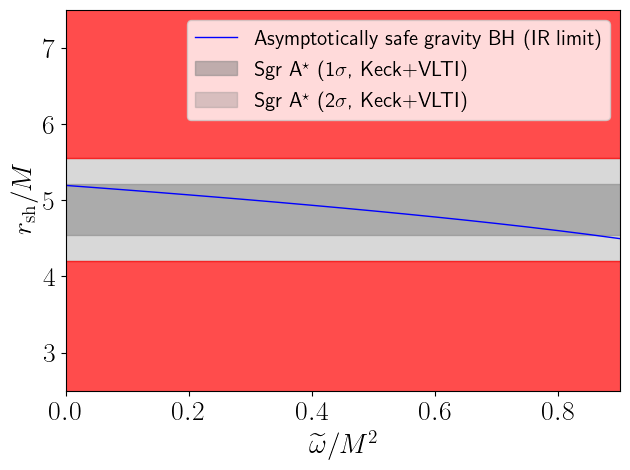}
\caption{Same as in Fig.~\ref{fig:shadow_reissner_nordstrom_bh_ns} for the BH in asymptotically safe gravity (in the IR limit) with metric function given by Eq.~\eqref{eq:metricasg}, as discussed in Sec.~\ref{subsec:asg}.}
\label{fig:shadow_asymptotically_safe_gravity_bh}
\end{figure}

Here, we shall focus on the region $\widetilde{\omega}<M^2$, for which the BH possesses two horizons. We compute the shadow size numerically, and show its evolution against $\widetilde{\omega}$ in Fig.~\ref{fig:shadow_asymptotically_safe_gravity_bh}. We see that the EHT observations set very weak limits on $\widetilde{\omega}$, with the upper limit $\widetilde{\omega} \lesssim 0.9M^2$ at $1\sigma$. Converting this to the dimensionless $\widetilde{\omega}$, we find that this corresponds to a very weak limit, and we are left to conclude that based on this metric test the EHT observations are unable to set strong constraints on asymptotically safe gravity (though the situation may be different with other types of tests, see e.g.\ Refs~\cite{Eichhorn:2021etc,Eichhorn:2021iwq,Delaporte:2022acp,Eichhorn:2022oma,Eichhorn:2022bbn}). On the other hand, complementary X-ray reflection spectroscopy constraints obtained by one of us set a much tighter limit, $\widetilde{\omega} \lesssim 0.05$ at $1\sigma$~\cite{Zhou:2020eth} from observations of GRS 1915+105, an X-ray binary system with mass of order ${\cal O}(10)M_{\odot}$.

\subsection{Rastall gravity}
\label{subsec:rastall}

Rastall's theory of gravity is perhaps one of the best known and studied non-conservative theories of gravity, i.e.\ theories where the stress-energy tensor is not covariantly conserved, $\nabla_{\mu}T^{\mu\nu} \neq 0$. The theory was introduced by Rastall in Ref.~\cite{Rastall:1972swe} with the motivation that in the strong curvature regime the covariant derivative of the stress-energy tensor may not vanish but could depend on the space-time curvature. Based on phenomenological arguments, Rastall proposed that $\nabla_{\mu}T^{\mu\nu} \propto \lambda g^{\mu\nu}\nabla_{\mu}R$, with Einstein's equations being modified to:
\begin{eqnarray}
R_{\mu\nu} + \left ( \kappa-\frac{1}{2} \right ) g_{\mu\nu}R = T_{\mu\nu}\,,
\label{eq:rastall}
\end{eqnarray}
where we refer to $\kappa$ as the Rastall coupling parameter, and Einstein's equations are recovered in the limit $\kappa \to 0$. Rastall gravity has been at the center of a surge of interest in recent years, with applications to a number of interesting scenarios from thermodynamics to cosmology (see e.g.\ Refs.~\cite{Fabris:2011wz,Batista:2011nu,Moradpour:2016ubd,Moradpour:2016fur,Moradpour:2017shy,Xu:2017vse,Darabi:2017coc,Hansraj:2018zwl,Cruz:2019jiq,Khyllep:2019odd,Pourhassan:2019tti,Akarsu:2020yqa,Lobo:2020jfl,Ghosh:2021byh,Gogoi:2021dkr,Shahidi:2021lxt,Narzilloev:2021ifl,Guo:2021bwr,Narzilloev:2022bbs,Zhong:2022wlw}), although we note that concerns have been raised about its possibly being equivalent to GR in the gravity sector, modulo an artificially isolated part of the physical conserved stress-energy tensor in the matter sector~\cite{Visser:2017gpz}.

By construction, while all vacuum solutions of GR are also solutions of Rastall gravity, the non-vacuum solutions depend on the Rastall coupling parameter and can be significantly different from their GR counterparts. BH solutions in Rastall gravity have been studied in several works~\cite{Heydarzade:2016zof,Heydarzade:2017wxu,Kumar:2017qws}. Here, we shall consider the solution obtained by Heydarzade and Darabi in Ref.~\cite{Heydarzade:2017wxu}, describing BHs in Rastall gravity surrounded by a perfect fluid with equation of state $\omega$. In this case, the metric function is given by:
\begin{eqnarray}
A(r)=1-\frac{2M}{r}-N_sr^{\frac{6\kappa(1+\omega)-1-3\omega}{1-3\kappa(1+\omega)}} \,,
\label{eq:metricrastall}
\end{eqnarray}
where $N_s$ is the field structure parameter, which is related to the energy density of the surrounding fluid and therefore characterizes a specific hair, as does $\omega$, whereas $\kappa$ characterizes an universal hair. The reason why it is interesting to consider a surrounding fluid is that, as is clear from Eq.~(\ref{eq:metricrastall}), depending on the value of $\kappa$ the surrounding field acts as if it had a different effective equation of state. In the following, we aim to focus on the Rastall coupling $\kappa$, while also exploring the role of the equation of state of the surrounding fluid. In our baseline cases, we shall therefore fix $N_s=0.005$ (so that it has little impact on the resulting shadow size), considering the cases where the BH is surrounded by dust ($\omega=0$), which we later extend to a stiff fluid ($\omega=1$, which can be generated during a period of kination, where the energy is dominated by the kinetic energy of a scalar field), radiation ($\omega=1/3$), and a fluid with equation of state $\omega=-1/3$ (which can be associated to cosmic strings, and is the effective equation of state corresponding to spatial curvature).

\begin{figure}
\centering
\includegraphics[width=1.0\linewidth]{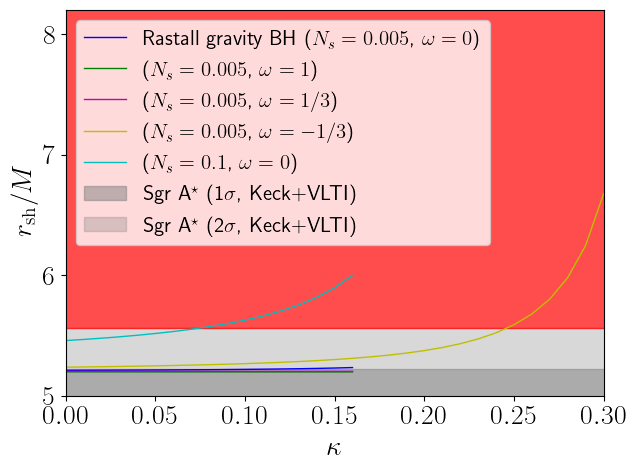}
\caption{Same as in Fig.~\ref{fig:shadow_reissner_nordstrom_bh_ns} for the BH in Rastall gravity with metric function given by Eq.~\eqref{eq:metricrastall}, with the field structure parameter $N_s$ and equation of state of the surrounding fluid $\omega$ fixed to different values (our baseline case corresponds to $N_s=0.005$ and $\omega=0$), as discussed in Sec.~\ref{subsec:rastall}.}
\label{fig:shadow_rastall_bh}
\end{figure}

We compute the resulting shadow size numerically, and show its evolution as a function of the Rastall coupling $\kappa$ in Fig.~\ref{fig:shadow_rastall_bh}. For the cases where the equation of state of the surrounding fluid is $\omega \geq 0$, we set the upper limit $\kappa \leq 1/6$~\cite{Kumar:2017tdw}, ensuring that the energy conditions are not violated (by construction this cannot be done for the $\omega=-1/3$ fluid) and the space-time is asymptotically flat. In these cases, and in particular for our baseline case with $N_s=0.005$ and $\omega=0$, we see that the shadow size grows extremely slowly with increasing $\kappa$, in a way which is hardly visible by eye. We find that the largest correction to the shadow size relative to the Schwarzschild counterpart within the whole $\kappa \leq 1/6$ regime is at most ${\cal O}(10^{-2})$, and as a result the EHT observations are unable to set meaningful constraints on the Rastall coupling parameter $\kappa$. On the other hand, for $\omega=-1/3$ the shadow size starts increasing quickly once $\kappa \gtrsim 0.25$, and we find the upper limits $\kappa \lesssim 0.01$ ($1\sigma$) and $\kappa \lesssim 0.25$ ($2\sigma$). To further show the role of $N_s$ instead, we return to our baseline case with $\omega=0$, but setting $N_s=0.1$. In this case we see that the shadow size increases more markedly, and is excluded by the EHT observations within $1\sigma$, while at $2\sigma$ we find the upper limit $\kappa \lesssim 0.07$.

\subsection{Loop quantum gravity}
\label{subsec:lqg}

Loop quantum gravity (LQG) is arguably one of the leading approaches towards constructing a quantum theory of gravity. One of the main hurdles towards constructing such a theory is the difficulty in constructing a quantum Riemannian geometry, going beyond the classical (fixed) reference space-time. The first steps towards this task were completed by reformulating gravity without reference to any background field (not even a metric), and subsequently introducing non-perturbative techniques from gauge theories, once more without reference to any background (see e.g.\ Refs.~\cite{Ashtekar:2004eh,Rovelli:2004tv,Thiemann:2007pyv,Rovelli:2014ssa,Ashtekar:2021kfp}). A theory with no background fields only has access to an underlying manifold and is therefore inevitably diffeomorphism-covariant: this, together with the non-perturbative methods mentioned previously, leads to a fundamentally discrete geometry, whereas the familiar continuum space-time of GR results as an emergent phenomenon upon a coarse grained description of the fundamental quantum Riemannian discrete structures. Summarizing the main features of LQG in a few lines is close to impossible: loosely speaking, we can say that LQG leads to a granular space-time structure, with a minimum length of order the Planck length, and ``woven'' out of Planck-scale loops whose network and evolution thereof are referred to as spin network and spin foam respectively. A significant amount of work has been devoted to LQG in recent years, and we refer the reader to Refs.~\cite{Bojowald:2001xe,Bojowald:2005epg,Bojowald:2006gr,Modesto:2006qh,Engle:2007wy,Rovelli:2008aa,Modesto:2008jz,Bianchi:2010gc,Henderson:2010qd,Singh:2010qa,Bojowald:2011aa,Cailleteau:2012fy,Agullo:2013ai,Rovelli:2013osa,Haggard:2014rza,Odintsov:2014gea,Odintsov:2015uca,Haro:2015oqa,Odintsov:2016apy,Bianchi:2018mml,DeHaro:2018hia,deHaro:2018sqw,Odintsov:2018awm,Benetti:2019kgw,Casalino:2019tho,Bouhmadi-Lopez:2020oia,Graef:2020qwe,Brahma:2020eos,Barboza:2022hng,SVicente:2022ebm} for examples thereof (including cosmological applications), as well as Ref.~\cite{Rovelli:1997yv} for a comprehensive review of the status of LQG.

BHs in LQG have been explored in several works~\cite{Modesto:2005zm,Modesto:2006mx,Modesto:2008jz,Caravelli:2010ff,Brown:2011tv}. Here we shall consider the LQG BH constructed in Ref.~\cite{Modesto:2008im}, whose shadow was studied by one of us in Ref.~\cite{Liu:2020ola}, and described by the following metric functions:
\begin{eqnarray}
A(r)=\frac{(r-r_+)(r-r_-)(r+r_*)^2}{r^4+a_0^2}\,,\quad C(r)=r^2+\frac{a_0^2}{r^2}\,, \nonumber \\
\label{eq:metriclqg}
\end{eqnarray}
where $r_+=2 M/(1+P)^2$ and $r_{-} = 2MP^2/(1+P)^2$ are the BH's two horizons, while $r_*=\sqrt{r_+r_-}=2 MP/(1+P)^2$. On the other hand $P=(\sqrt{1+\epsilon^2}-1)/(\sqrt{1+\epsilon^2}+1)$ is the polymeric function, where $\epsilon$ denotes the product of the Immirzi parameter $\gamma$ and the polymeric parameter $\delta$ (where typically $\epsilon=\gamma\delta \ll 1$): the former is an area parameter~\cite{Immirzi:1996di,Immirzi:1996dr,Perez:2005pm}, whereas the latter was introduced to define the Hamiltonian constraint in terms of holonomies~\cite{Ashtekar:2002vh}, and was conservatively assumed to be constant in the approach of Ref.~\cite{Modesto:2008im}. Finally, $a_0=A_{\min}/8\pi$ is related to the minimum area gap of LQG, i.e.\ the smallest non-zero eigenvalue of the kinematical area operator. Note that all three $a_0$, $\gamma$, and $\delta$ (and therefore $\epsilon$ and $P$) characterize universal hairs.

\begin{figure}
\centering
\includegraphics[width=1.0\linewidth]{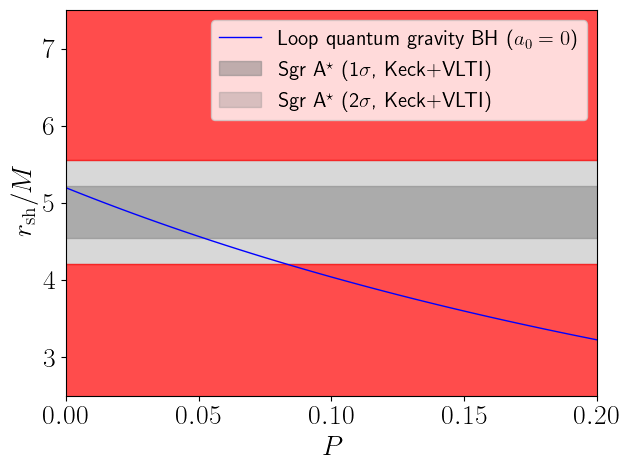}
\caption{Same as in Fig.~\ref{fig:shadow_reissner_nordstrom_bh_ns} for the Loop Quantum Gravity BH with metric described by Eq.~\eqref{eq:metriclqg}, as discussed in Sec.~\ref{subsec:lqg}.}
\label{fig:shadow_loop_quantum_gravity_bh}
\end{figure}

The above space-time depends upon the two parameters $P$ (or equivalently $\epsilon$) and $a_0$, and reduces to the Schwarzschild space-time in the limit $P \to 0$, $a_0 \to 0$. However, since $a_0$ is expected to be extremely small in LQG, we do not expect it to play a relevant role as far as astrophysical BHs are concerned. Indeed, we have explicitly verified that $a_0$ plays a minor role in determining the shadow size. Therefore, for the purposes of this work, we set $a_0=0$, so that the only free parameter is $P$. We compute the shadow size numerically, and show its evolution as a function of $P$ in Fig.~\ref{fig:shadow_loop_quantum_gravity_bh}.~\footnote{We have found that a closed-form expression for the shadow radius exists, but is far too cumbersome to report here, to the extent that it would easily occupy an entire page, but would lead to little physical insight. We have nonetheless verified that our numerical results perfectly reproduce those provided by this closed-form expression.} We see that increasing the polymeric function decreases the shadow size, and the EHT observations set the upper limits $P \lesssim 0.05$ ($1\sigma$) and $P \lesssim 0.08$ ($2\sigma$). These can be translated into the upper limits $\epsilon \lesssim 0.5$ ($1\sigma$) and $\epsilon \lesssim 0.6$ ($2\sigma$), which are rather loose as on theoretical grounds we expect $\epsilon \ll 1$. However, this remains a valuable constraint on an interesting theoretical scenario, particularly given that observational constraints on LQG are not easy to obtain.

\subsection{Kottler black hole}
\label{subsec:kottler}

The Kottler space-time~\cite{Kottler:1918ghw}, also known as Schwarzschild-de Sitter space-time, is the unique spherically symmetric solution to Einstein's equations sourced by a positive cosmological constant $\Lambda$. It is a particular case of the McVittie metric, which describes models interpolating between Schwarzschild and FLRW space-times in a continuous manner, and is part of a larger class of space-times which generically describe BHs embedded in an expanding Universe (see e.g.\ Refs.~\cite{McVittie:1933zz,Einstein:1945id,Einstein:1946zz,Lake:1977ui,Stuchlik:1983ghw,Nolan:1998xs,Nolan:1999kk,Nolan:1999wf,Stuchlik:1999qk,Gibbons:2009dr,Nandra:2011ug,Nandra:2011ui,daSilva:2012nh,Piattella:2015xga,Piattella:2016nzt,Perlick:2018iye,Bisnovatyi-Kogan:2018vxl,Firouzjaee:2019aij,Tsupko:2019pzg,Tsupko:2019mfo,Roy:2020dyy,Belhaj:2020kwv,Chang:2021ngy,Renzi:2022fmw,Odintsov:2022umu} for further relevant works). The Kottler BH is described by the following metric function~\cite{Kottler:1918ghw}:
\begin{eqnarray}
A(r)=1-\frac{2M}{r}-\frac{\Lambda}{3}r^2 \,,
\label{eq:metrickottler}
\end{eqnarray}
where $\Lambda$ characterizes an universal hair. The Kottler space-time possesses two horizons for $0<\Lambda<\sqrt{3M}$, with the vector field $\partial_t$ only being time-like between these two horizons. We therefore restrict our discussion to static observers whose radial coordinate is bounded above by the coordinate of the outer horizon, which is anyhow always guaranteed to be the case since the latter corresponds to our de Sitter cosmological horizon in the physically interesting case where $\Lambda$ is the cosmological constant responsible for cosmic acceleration.

As discussed at length in Ref.~\cite{Perlick:2021aok}, the Kottler space-time is an excellent example of a BH whose shadow size depends explicitly on the radial coordinate of the observer, due to the metric described by Eq.~(\ref{eq:metrickottler}) being manifestly non-asymptotically flat. Considering the importance of the Kottler space-time, we shall derive the exact expression for the shadow size, following the discussion of Sec.~\ref{subsec:bhshadowradius}. Explicitly, we find:
\begin{eqnarray}
r_{\rm sh} = \frac{3\sqrt{3}M}{\sqrt{1-9\Lambda M^2}}\sqrt{1-\frac{2M}{r_O}-\frac{\Lambda}{3}r_O^2}\,,
\label{eq:shadowsizekottler}
\end{eqnarray}
which clearly reduces to $3\sqrt{3}M$ in the limit $\Lambda \to 0$ and $r_O \gg M$, the latter being clearly the case for the EHT observations of Sgr A$^*$. We further note that the requirement of the observer lying within the cosmological horizon ensures that the term in the square root to the right of Eq.~(\ref{eq:shadowsizekottler}) does not turn negative. As per our discussion in Sec.~\ref{subsec:bhshadowradius}, for the purposes of our discussion it suffices to consider a static observer, whereas considering an observer comoving with the expansion of the Universe would complicate the discussion considerably.

\begin{figure}
\centering
\includegraphics[width=1.0\linewidth]{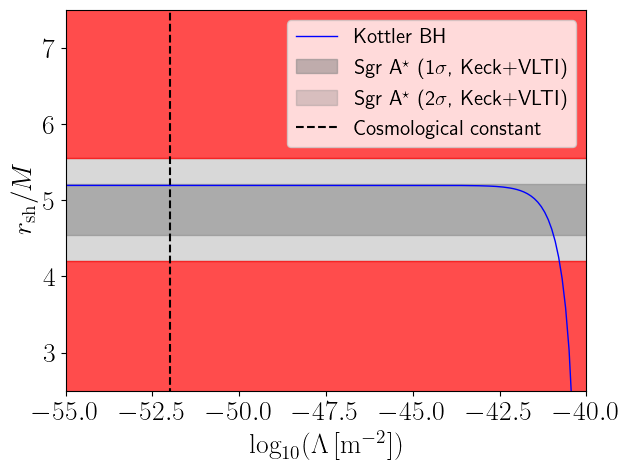}
\caption{Same as in Fig.~\ref{fig:shadow_reissner_nordstrom_bh_ns} for the Kottler BH with metric function given by Eq.~\eqref{eq:metrickottler}, as discussed in Sec.~\ref{subsec:kottler}. The dashed black vertical line corresponds to the value of the cosmological constant indicated by observations, $\Lambda \sim 1.1 \times 10^{-52}\,{\rm m^2}$.}
\label{fig:shadow_kottler_bh}
\end{figure}

We show the evolution of the shadow size as a function of the cosmological constant $\Lambda$ in Eq.~(\ref{fig:shadow_kottler_bh}). As a note of caution, we plot the shadow radius in units of the Newtonian mass $M$: since the Kottler space-time is not asymptotically flat, the ADM mass is not well-defined, however for the purposes of our discussion considering $M$ will be sufficient. Fig.~\ref{fig:shadow_kottler_bh} displays the features we could have expected from a qualitative assessment. For $\Lambda \lesssim 10^{-41}\,{\rm m}^2$, the product $\Lambda r_O^2 \ll 1$ considering $r_O \sim 8\,{\rm kpc} \simeq 2.5 \times 10^{20}\,{\rm m}$, and therefore the size of the Kottler BH is essentially that of a Schwarzschild BH of the same mass. Clearly, this includes the value of $\Lambda$ indicated by cosmological observations, $\Lambda \sim {\cal O}(10^{-52})\,{\rm m}^2$ (corresponding to the dashed vertical line in Fig.~\ref{fig:shadow_kottler_bh}), for which the product $\Lambda r_O^2 \sim {\cal O}(10^{-12})$, clearly too small to be appreciated by current EHT measurements. Only for larger values of $\Lambda$, when $\Lambda r^2$ becomes of order unity, does the effect of the cosmological constant become visible, acting to quickly decrease the shadow size. We therefore conclude that within a Kottler space-time where $\Lambda$ is responsible for cosmic acceleration, there is no detectable imprint on the size of the BH shadow, as we could have expected from back-of-the-envelope calculations, but reassuringly confirmed by the full computation (see also Ref.~\cite{Afrin:2021ggx} by two of us for different perspectives on the subject, as well as Refs.~\cite{Adler:2022pqw,Adler:2022qtb}).

\subsection{Electromagnetic-Weyl coupling}
\label{subsec:weylcorrected}

One of the simplest extensions of the Einstein-Maxwell equations of GR and electromagnetism involves a direct coupling between the electromagnetic and Weyl tensors. The Weyl tensor $C_{\mu\nu\rho\sigma}$, invariant under conformal transformations of the metric, is a measure of the space-time shear, and is given by the following~\cite{Weyl:1918pdp}:
\begin{eqnarray}
C_{\mu\nu\rho\sigma} = R_{\mu\nu\rho\sigma}- \left ( g_{\mu[\rho}R_{\sigma]\nu} - g_{\nu[\rho}R_{\sigma]\mu} \right )+\frac{1}{3}Rg_{\mu[\rho}g_{\sigma]\nu}\,,\nonumber \\
\label{eq:weyltensor}
\end{eqnarray}
where the square brackets indicate the antisymmetric part. A direct coupling between the electromagnetic and Weyl tensors is allowed, in which case the electromagnetic sector of the Lagrangian is given by~\cite{Drummond:1979pp,Ritz:2008kh}:
\begin{eqnarray}
{\cal L} = -\frac{1}{4}F_{\mu\nu}F^{\mu\nu}+\alpha C^{\mu\nu\rho\sigma}F_{\mu\nu}F_{\rho\sigma}\,,
\label{eq:weylcoupling}
\end{eqnarray}
where $\alpha$ is a constant with dimensions of length squared. Assuming that the gravitational sector is still described by GR, Maxwell's equations are modified as follows:
\begin{eqnarray}
\nabla_{\mu} \left ( F^{\mu\nu}-4\alpha C^{\mu\nu\rho\sigma}F_{\rho\sigma} \right ) = 0\,.
\label{eq:maxwellweyl}
\end{eqnarray}
Clearly, the coupling to the Weyl tensor modifies the propagation of photons across the space-time, leading to a series of interesting phenomena including apparent superluminal propagation and birefringence~\cite{Cho:1997vg,Cai:1998ij,Zhang:2021hit}. This model is perhaps not a model of modified gravity in a strict sense, but is perhaps best interpreted as a model of modified electrodynamics, with such modifications actually expected when considering the effects of one-loop vacuum polarization on the photon effective action in QED as well as the effects of tidal gravitational forces (see Ref.~\cite{Drummond:1979pp} for early discussions).

As in most modified electrodynamics models, here as well photons travel along null geodesics of an effective geometry. We follow Ref.~\cite{Chen:2015cpa} who computed the effective geometry experienced by photons traveling in a Schwarzschild space-time in the presence of electromagnetic-Weyl coupling. As discussed in Ref.~\cite{Chen:2015cpa}, only two of the three possible polarizations turn out to be physical (likely as a consequence of gauge invariance and the Ward-Takahashi identities), which the authors refer to as $\ell$-polarization and $m$-polarization, or PPL and PPM respectively: we refer the reader to Ref.~\cite{Chen:2015cpa} for further details on the derivation of the modified light-cone conditions and of the physical polarizations. The Weyl-corrected effective geometry corresponding to an underlying Schwarzschild space-time is then described by the following metric functions~\cite{Chen:2015cpa} (see also Refs.~\cite{Lu:2016gsf,Huang:2016qnl,Chen:2016hil,Zhang:2017pxr,Jing:2017cmg,Ovgun:2018ran,Zhang:2018yzr,Zhang:2019hob,Bergliaffa:2020ivp,Lu:2020kpo} for further relevant works):
\begin{eqnarray}
A(r)&=&1-\frac{2M}{r} \,,\nonumber \\
C(r)&=& r^2 \left ( \frac{r^3-8\alpha M}{r^3+16\alpha M} \right ) ^{-1}\quad \text{(PPL polarization)}\,,\nonumber \\
C(r)&=& r^2 \left ( \frac{r^3+16\alpha M}{r^3-8\alpha M} \right ) ^{-1}\quad \text{(PPM polarization)}\,,\nonumber \\
\label{eq:metricweylcorrected}
\end{eqnarray}
where $\alpha$ characterizes an universal hair. The fact that only the angular metric function $C(r)$ is modified is a direct reflection of the modifications to the light-cone structure due to the direct coupling between photons and the Weyl tensor.

\begin{figure}
\centering
\includegraphics[width=1.0\linewidth]{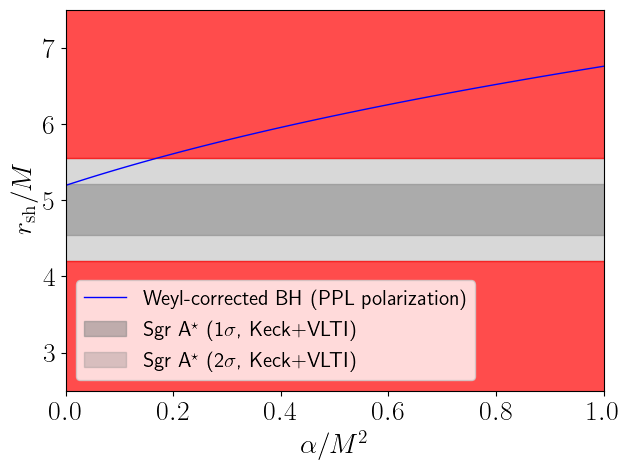}
\caption{Same as in Fig.~\ref{fig:shadow_reissner_nordstrom_bh_ns} for the Weyl-corrected (PPL polarization) BH with metric described by Eq.~\eqref{eq:metricweylcorrected}, as discussed in Sec.~\ref{subsec:weylcorrected}.}
\label{fig:shadow_weyl_corrected_ppl_bh}
\end{figure}

\begin{figure}
\centering
\includegraphics[width=1.0\linewidth]{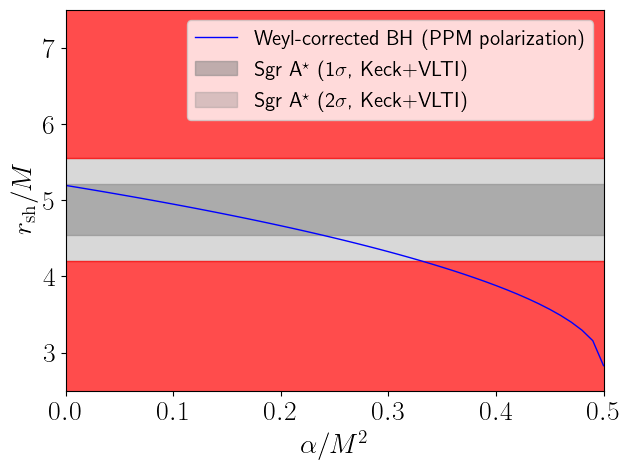}
\caption{Same as in Fig.~\ref{fig:shadow_reissner_nordstrom_bh_ns} for the Weyl-corrected (PPM polarization) BH with metric described by Eq.~\eqref{eq:metricweylcorrected}, as discussed in Sec.~\ref{subsec:weylcorrected}.}
\label{fig:shadow_weyl_corrected_ppm_bh}
\end{figure}

We compute the shadow size numerically for both PPL- and PPM-polarized photons, and show its evolution as a function of the Weyl coupling $\alpha$ in Fig.~\ref{fig:shadow_weyl_corrected_ppl_bh} and Fig.~\ref{fig:shadow_weyl_corrected_ppm_bh} for PPL- and PPM-polarized photons respectively. We see that in the former case increasing $\alpha$ increases the shadow size, whereas the reverse happens in the latter case: this is not surprising, as the respective angular metric functions are the inverse of one another. For PPL-polarized photons, we find that the EHT observations set the upper limits $\alpha \lesssim 0.01M^2$ ($1\sigma$) and $\alpha \lesssim 0.2M^2$ ($2\sigma$). For PPM-polarized photons, the limits are instead looser and read $\alpha \lesssim 0.25M^2$ ($1\sigma$) and $\alpha \lesssim 0.3M^2$ ($2\sigma$). Taking $\alpha \lesssim {\cal O}(0.1)M^2$ as a representative limit, in SI units this translates to $\alpha \lesssim {\cal O}(10^{17})\,{\rm m}^2$, significantly weaker than some of the best constraints on this parameter arising from gravitational time delay/advancement, light deflection, and frequency shifts, with Cassini tracking setting the upper limit $\alpha \lesssim {\cal O}(10^{12})\,{\rm m}^2$~\cite{Li:2017zsb,Li:2018jjk}.

\subsection{DST black hole}
\label{subsec:dst}

In Ref.~\cite{Deser:2007za}, Deser, Sarioglu, and Tekin (DST hereafter) introduced a class of gravitational models whose modifications to GR forbid Ricci-flat geometries, while preserving as many features of GR as possible, including the derivative order of the Einstein equations in Schwarzschild gauge. This class of models was introduced as a tool to better appreciate the nature of Ricci-flat geometries, and relies on the addition to the Einstein-Hilbert Lagrangian of terms which are non-polynomial in the Weyl tensor, as follows:
\begin{eqnarray}
\Delta{\cal L} = \sqrt{-g}\,\beta_n \left \vert \text{tr}\,{\cal C}^n \right \vert ^{\frac{1}{n}}\,,
\label{eq:lagrangiandst}
\end{eqnarray}
where ${\cal C}$ is the Weyl tensor already encountered earlier in Eq.~(\ref{eq:weyltensor}), the index $n$ labels the various models belonging to this class, and the trace operator acts upon the (mixed) Weyl tensor as follows:
\begin{eqnarray}
\text{tr}\,{\cal C}^n \equiv C_{ab}^{\,\,\,\,\,cd}C_{cd}^{\,\,\,\,\,ef}...C_{..}^{\,\,\,\,\,yz}C_{yz}^{\,\,\,\,\,ab}\,,
\label{eq:traceweyltensordst}
\end{eqnarray}
for $n$ copies of the Weyl tensor, where $n$ can be any arbitrary positive integer. To the best of our knowledge, Ref.~\cite{Deser:2007za} was the first to introduce the non-polynomial gravity approach. Models of these type have later been considered in various works (particularly in the search for regular BHs), see for instance Refs.~\cite{Bellini:2010ar,Cognola:2011nj,Myrzakulov:2013hca,Colleaux:2015yta,Chinaglia:2017wim,Sebastiani:2017rxr,Colleaux:2017ibe,Chinaglia:2018gvf,Vetsov:2018dte,Gonzalez:2018zuu}.

Without loss of generality, we shall follow Ref.~\cite{Deser:2007za} and take $n=2$, while defining $\sigma=\beta_2/\sqrt{3}$, so that Eq.~(\ref{eq:lagrangiandst}) reduces to:
\begin{eqnarray}
\Delta{\cal L}_{n=2} = \sqrt{-g}\,\sigma \sqrt{3 \left \vert C_{ab}^{\,\,\,\,\,cd}C_{cd}^{\,\,\,\,\,ef} \right \vert }\,.
\label{eq:lagrangiandst2}
\end{eqnarray}
This particular Lagrangian admits an explicit static spherically symmetric solution, the so-called DST BH, with metric function given by~\cite{Deser:2007za}:
\begin{eqnarray}
A(r)= \left [ \frac{1-\sigma}{1-4\sigma} - \left ( \frac{2M}{r} \right ) ^{\frac{1-4\sigma}{1-\sigma}} \right ] \left ( \frac{r}{2M} \right ) ^{\frac{6\sigma}{\sigma-1}} \,,
\label{eq:metricdst}
\end{eqnarray}
where $\sigma$ characterizes an universal hair. It is easy to show that the above space-time is asymptotically flat for $0<\sigma<1$. In our subsequent analysis we have imposed $\sigma>0$, whereas the upper limit guaranteeing asymptotic flatness $\sigma<1$ turns out a posteriori to be satisfied given the very stringent limits on $\sigma$ imposed by the EHT observations.

\begin{figure}
\centering
\includegraphics[width=1.0\linewidth]{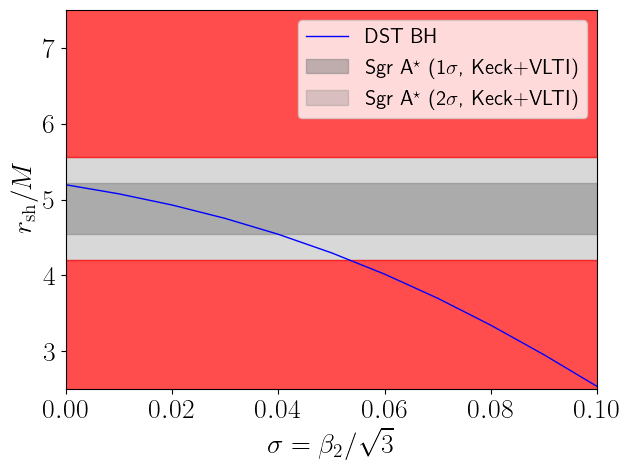}
\caption{Same as in Fig.~\ref{fig:shadow_reissner_nordstrom_bh_ns} for the DST BH with metric function given by Eq.~\eqref{eq:metricdst}, as discussed in Sec.~\ref{subsec:dst}.}
\label{fig:shadow_dst_bh}
\end{figure}

While slightly cumbersome, there exists a closed-form expression for the shadow radius, given by:
\begin{eqnarray}
r_{\rm sh}=\frac{2^{\frac{3\sigma}{\sigma-1}}3^{\frac{\sigma-1}{4\sigma-1}}{\cal X}}{\sqrt{ \left ( 3^{\frac{\sigma-1}{4\sigma-1}}{\cal X} \right ) ^{\frac{6\sigma}{\sigma-1}} \left ( \frac{\sigma-1}{4\sigma-1}-2^{\frac{4\sigma-1}{\sigma-1}} \left ( 3^{\frac{1-\sigma}{4\sigma-1}}{\cal X} \right ) ^{\frac{4\sigma-1}{\sigma-1}} \right ) }}\,, \nonumber \\
\label{eq:rshdst}
\end{eqnarray}
where the quantity ${\cal X}$ is defined as:
\begin{eqnarray}
{\cal X} \equiv \frac{8^{\frac{\sigma}{\sigma-1}}(4\sigma-1)}{(\sigma-1)(2\sigma+1)}\,,
\label{eq:xdst}
\end{eqnarray}
which clearly reduces to ${\cal X} \to 1$ when $\sigma \to 0$, from which it is relatively straightforward to see that the limit $r_{\rm sh}=3\sqrt{3}$ is indeed recovered when $\sigma \to 0$ in Eq.~(\ref{eq:rshdst}). We show the evolution of the shadow size as a function of $\sigma$ in Fig.~\ref{fig:shadow_dst_bh}, where we see that increasing $\sigma$ quickly decreases the shadow size. We note that the EHT observations set the very stringent limits $\sigma \lesssim 0.04$ ($1\sigma$) and $\sigma \lesssim 0.05$ ($2\sigma$), which directly translate into upper limits on the Lagrangian coupling $\beta_2$ in Eq.~(\ref{eq:lagrangiandst2}) of $\beta_2 \lesssim 0.07$ ($1\sigma$) and $\beta_2 \lesssim 0.09$ ($2\sigma$). To the best of our knowledge, we are not aware of any observational constraints on $\sigma$: a study on collisions of spinning particles near DST BHs was conducted by one of us in Ref.~\cite{Gonzalez:2018zuu}, finding a behavior very close to that of GR, while the same work entertained the possibility that a very tiny discrepancy in the observed value of Mercury's perihelion precession might be explained by a value $\sigma \sim {\cal O}(10^{-9})$. However, our upper limits of order ${\cal O}(10^{-2})$ on $\sigma$ represent the first ever robust observational constraints on a non-polynomial gravity model.

\subsection{Rindler gravity}
\label{subsec:rindler}

In Ref.~\cite{Grumiller:2010bz} (see also Ref.~\cite{Grumiller:2011gg}), Gr\"{u}miller constructed \textit{ab initio} the most general theory for gravity in the IR (i.e.\ at large distances) consistent with an assumed set of symmetries and assumptions, including spherical symmetry, power counting renormalizability, analyticity, and so on. While the model was constructed based on well-founded theoretical assumptions, it also enjoys phenomenological motivation in its ability (at the time) to potentially mimic dark matter (DM) and/or dark energy, as well as explain a number of puzzles such as the flyby~\cite{Anderson:2008zza} and Pioneer anomalies~\cite{Anderson:1998jd}. While the theory is now understood to be unlikely to successfully mimic DM and dark energy (see e.g.\ Ref.~\cite{Pardo:2020epc}), and interest in these anomalies has now partially faded~\cite{Turyshev:2012mc}, here we shall simply consider it as being an interesting IR model for gravity.

In Ref.~\cite{Grumiller:2010bz}, Gr\"{u}miller constructed the Rindler gravity model from the following effective field theory for a scalar field $\Phi$ in 2 dimensions:
\begin{equation}
S = \int {\rm d}^2x\sqrt{-g}\, \left [ \Phi^2R+2(\partial\Phi)^2-6\Lambda\Phi^2+8a\Phi+2 \right ]\,,
\label{eq:actionrindler}
\end{equation}
where $\Lambda$ will eventually play the role of dark energy, and $a$ is a constant with SI units of acceleration. For a test particle located at large distance from a source (where the Newtonian acceleration is negligible), positive/negative $a$ provides a constant acceleration towards/away from the source. The new term admits a simple interpretation: a Rindler acceleration, see Ref.~\cite{Grumiller:2002nm}. While the above theory is written in 2 dimensions and thus cannot be a realistic description of our world, given the interesting phenomenology associated to the Rindler acceleration outlined in Ref.~\cite{Grumiller:2010bz}, here we shall adopt an equally phenomenological standpoint where we examine the consequences of a Rindler correction to the Schwarzschild metric (setting $\Lambda=0$), taking the above model as theoretical motivation. That is, we consider the following metric function~\cite{Grumiller:2010bz} (see also Refs.~\cite{Sakalli:2017ewb,Perivolaropoulos:2019vgl}):
\begin{eqnarray}
A(r)=1-\frac{2M}{r}+2ar \,,
\label{eq:metricrindler}
\end{eqnarray}
where $a$ characterizes an universal hair, and which is manifestly not asymptotically flat.~\footnote{We note that this same space-time can arise in a wide range of theories other than Rindler gravity, see e.g.\ Refs.~\cite{Riegert:1984zz,Mannheim:1988dj,Klemm:1998kf,Saffari:2007zt,deRham:2010kj,Soroushfar:2015wqa,Gregoris:2021plc}. It is also a particular case of the Kiselev BH~\cite{Kiselev:2002dx}, with equation of state of the surrounding fluid given by $w=-2/3$.}

\begin{figure}
\centering
\includegraphics[width=1.0\linewidth]{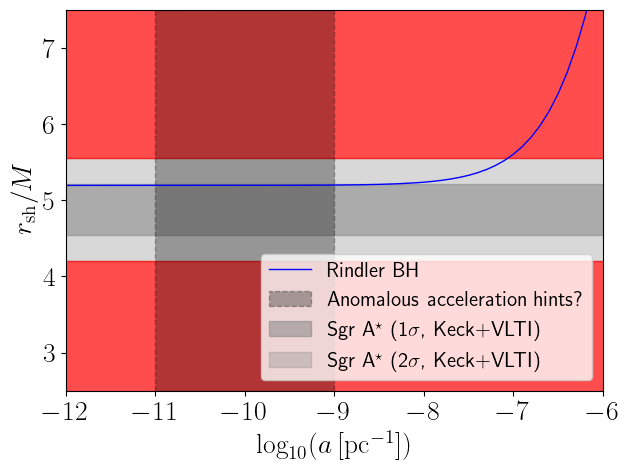}
\caption{Same as in Fig.~\ref{fig:shadow_reissner_nordstrom_bh_ns} for the Rindler BH with metric function given by Eq.~\eqref{eq:metricrindler}, as discussed in Sec.~\ref{subsec:rindler}. The dashed black box corresponds to the region possibly indicated by observational hints (explaining DM and/or the Pioneer and flyby anomalies).}
\label{fig:shadow_rindler_bh}
\end{figure}

We compute the shadow size numerically, using the full computation outlined in Sec.~\ref{subsec:bhshadowradius} for non-asymptotically flat metrics, and show its evolution against the Rindler acceleration $a$ in Fig.~\ref{fig:shadow_rindler_bh}, considering only positive values of $a$ as required to explain the observational anomalies the model was constructed to account for~\cite{Grumiller:2010bz}. For convenience, rather than reporting the shadow evolution against $aM$, we report $a$ in units of ${\rm pc}^{-1}$. We also note that a closed-form expression for the shadow radius can be found, given by:
\begin{eqnarray}
r_{\rm sh} \simeq \frac{\sqrt{3} \left ( \sqrt{1+12a}-1 \right ) \sqrt{1+2ar_O}}{2a\sqrt{2\sqrt{1+12a}-1}}
\label{eq:rshrindler}
\end{eqnarray}
which, although not obvious at first glance, does indeed reduce to $3\sqrt{3}$ as $a \to 0$, as can be seen by Taylor expanding Eq.~(\ref{eq:rshrindler}) in $a$ (we have made the valid approximation that whenever the $2ar_O$ term is dominant over $1$, it naturally also dominates over $2/r_O$). We have verified that this closed-form expression is in excellent agreement with our numerical results.

As with the Kottler space-time considered previously in Sec.~\ref{subsec:kottler}, it is clear that the effects of Rindler gravity on the BH shadow only become significant when the term $ar$ becomes sufficiently large. For smaller values of $a$ required to mimic DM and/or address anomalous acceleration hints of the order of $a \sim 10^{-10}\,{\rm m}/{\rm s}^2$,~\footnote{It is of course no coincidence that this if of the order of the MOND acceleration~\cite{Milgrom:1983ca,Milgrom:1983pn}, as Rindler gravity produces an interesting MOND-like phenomenology (see Refs.~\cite{Bertolami:2007gv,Mannheim:2010ti,Vagnozzi:2017ilo} for related work).} corresponding to the dashed black box in Fig.~\ref{fig:shadow_rindler_bh}, the effect on the shadow size is too small to be appreciated by the EHT observations. We conclude that, for phenomenologically interesting values of $a$, this model is in excellent agreement with the EHT observations but cannot be distinguished from its Schwarzschild counterpart, at least not based on the metric test we have considered here.

\subsection{Black hole surrounded by perfect fluid dark matter}
\label{subsec:pfdm}

We have spent considerable time discussing BH solutions in modified theories of gravity, broadly defined. We shall now turn our attention to BHs surrounded by various forms of matter. The influence of external matter fields on BHs has been the subject of significant study, at least since the well-known Kiselev space-time was proposed~\cite{Kiselev:2002dx}. While particular emphasis has been placed on the possibility of quintessence surrounding BHs (see e.g.\ Refs.~\cite{Chen:2005qh,Chen:2008ra,Zeng:2014xza,Azreg-Ainou:2014twa,Abdujabbarov:2015pqp,Ghosh:2015ovj,Toshmatov:2015npp,deOliveira:2018weu,Benavides-Gallego:2018odl,Saadati:2019cym,Ali:2019mxs,Belhaj:2020rdb,Sheoran:2020kmn,Zeng:2020vsj,Lambiase:2020pkc,Hendi:2020ebh,Mastrototaro:2021kmw,Cardenas:2021eri,Ndongmo:2021how,Sun:2022wya}), here we shall consider a much more strongly motivated scenario: that is, one where the BH is surrounded by dark matter (DM). In fact, it is now an observationally established fact that DM is present in (almost) all galaxies, and is therefore expected to accumulate around SMBHs inhabiting galactic centres: therefore, it is worthwhile to consider how the surrounding DM back-reacts on the BH metric.

The above question has been considered in several works~\cite{Kiselev:2003ah,Rahaman:2010xs,Li:2012zx}. Specifically, consider a perfect fluid DM component, described by an equation of state $w=0$. The metric function associated to such a configuration is then given by (see Ref.~\cite{Li:2012zx}):
\begin{eqnarray}
A(r) = 1-\frac{2M}{r}+\frac{k}{r} \ln \left ( \frac{r}{\vert k \vert} \right ) \,,
\label{eq:metricpfdm}
\end{eqnarray}
where $k$ is an integration constant, ultimately physically connected to the amount of DM surrounding the BH, and which characterizes a specific hair. We note that the above metric is \textit{not} the metric one would obtain when setting $w=0$ in the well-known Kiselev metric~\cite{Kiselev:2003ah}. The reason, as explained in Ref.~\cite{Visser:2019brz}, is that the Kiselev metric does not describe a space-time surrounded by a perfect fluid in the sense we are using here: in fact, it is easy to show that the energy-momentum tensor sourcing the Kiselev metric inevitably possesses shear stress components, and therefore cannot describe a perfect fluid~\cite{Visser:2019brz}. This is not the case, instead, for the space-time described by Eq.~(\ref{eq:metricpfdm}), which has been explicitly derived by starting from a perfect fluid source term (see Refs.~\cite{Xu:2016ylr,Xu:2017bpz,Bahamonde:2018zcq,Haroon:2018ryd,Hou:2018avu,Rizwan:2018rgs,Hendi:2020zyw,Shaymatov:2020bso,Shaymatov:2020wtj,Zhang:2020mxi,Das:2020yxw,Saurabh:2020zqg,Ma:2020dhv,Narzilloev:2020qtd,Cao:2021dcq,Atamurotov:2021hck,Rayimbaev:2021kjs,Atamurotov:2021hoq,Shaymatov:2021nff,Bhandari:2021dsh} for other relevant works).

\begin{figure}
\centering
\includegraphics[width=1.0\linewidth]{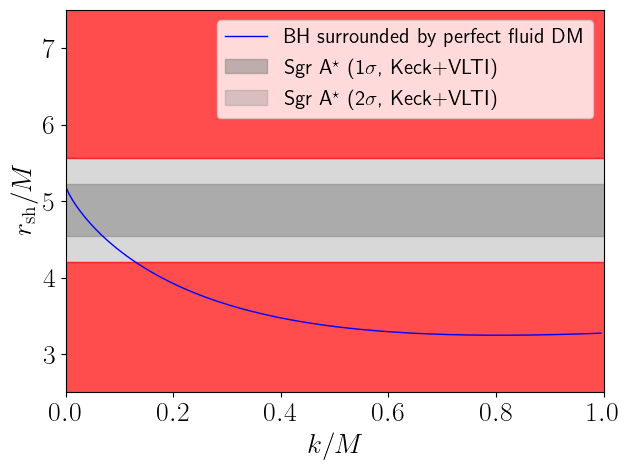}
\caption{Same as in Fig.~\ref{fig:shadow_reissner_nordstrom_bh_ns} for the BH surrounded by perfect fluid DM with metric function given by Eq.~\eqref{eq:metricpfdm}, as discussed in Sec.~\ref{subsec:pfdm}.}
\label{fig:shadow_pfdm_bh}
\end{figure}

While the size of the shadow cast by the BH described by Eq.~(\ref{eq:metricpfdm}) is easy to compute numerically, we also note that a closed-form expression exists and is given by:
\begin{eqnarray}
r_{\rm sh} = \frac{9kW \left ( \frac{2}{3}e^{\frac{1}{3}+\frac{2}{k}} \right ) ^{\frac{3}{2}}}{2\sqrt{2+3W \left ( \frac{2}{3}e^{\frac{1}{3}+\frac{2}{k}} \right ) }}
\label{eq:rshpfdm}
\end{eqnarray}
where once more $W$ denotes the principal branch of the Lambert $W$ function. As a check, we find Eq.~(\ref{eq:rshpfdm}) to be in perfect agreement with our numerical results. We show the evolution of the shadow size as a function of the perfect fluid DM parameter $k$ in Fig.~\ref{fig:shadow_pfdm_bh}, where we see that the shadow size quickly decreases as $k$ increases. We find that the EHT observations set the upper limits $k \lesssim 0.07M$ ($1\sigma$) and $k \lesssim 0.15M$ ($2\sigma$). We are not aware of an explicit discussion of the physical meaning of $k$ in the literature. On dimensional grounds, we expect the combination $c^2k/G$ to be related to the integrated amount of DM present in a shell surrounding the BH. Given the physical situation at play, we would expect the upper limits on $k/M$ to constrain the integrated amount of DM present in a shell size of size given by the photon sphere, although this does not allow us to estimate the (dimensionless) numerical factor relating these two quantities. At any rate, a more complete discussion requires a full physical interpretation of the parameter $k$, which is currently lacking in the literature: we plan to return to this issue in future work, to which we defer a more comprehensive discussion on the topic.

\subsection{Black hole with a topological defect}
\label{subsec:topologicaldefect}

Topological defects play an important role in several different context of physical interest, and likewise can be produced within a variety of different scenarios, including during phase transitions in the early Universe~\cite{Hindmarsh:1993av}, or associated to the spontaneous breaking of symmetries~\cite{Vilenkin:1981zs,Vilenkin:1981kz,Vilenkin:1982hm,Vilenkin:1984ib}. For example, global monopoles can arise in association to the breaking of a global $O(3)$ symmetry down to a $U(1)$ subgroup (this can be realized for instance with a triplet of scalar fields $\phi^i$), and their stability has been studied in various works~\cite{Goldhaber:1989na,Rhie:1990kc,Perivolaropoulos:1991du}. The gravitational field associated to a Schwarzschild BH carrying a global monopole charge was first studied in Refs.~\cite{Barriola:1989hx,Dadhich:1997mh} (see also Ref.~\cite{Abdujabbarov:2012bn}), and the corresponding BH solution is described by the following metric function:
\begin{eqnarray}
A(r)=1-k-\frac{2M}{r}\,,
\label{eq:metrictopologicaldefect}
\end{eqnarray}
where the fundamental interpretation of $k$ depends on the model being examined. For instance, we can consider a triplet of scalar fields $\phi^i$ satisfying a global $O(3)$ symmetry, described by the following Lagrangian:
\begin{eqnarray}
{\cal L} = \frac{1}{2} \left ( \partial\phi^i \right ) ^2-\frac{\lambda}{4} \left ( \phi^i\phi^i-\eta^2 \right ) ^2\,,
\label{eq:lagrangiantriplet}
\end{eqnarray}
for which the associated global monopole is given by:
\begin{eqnarray}
\phi^i = \eta f(r)\frac{x^a}{\vert \overline{\mathbf{x}} \vert}\,,
\label{eq:globalmonopole}
\end{eqnarray}
with $f(r) \to 1$ as $\vert \overline{\mathbf{x}} \vert \to 0$, and $k$ in Eq.~(\ref{eq:metrictopologicaldefect}) is given by $k=8\pi\eta^2$. In this case, $k$ is therefore related to the amplitude of the monopole field. Interestingly, the same metric can describe a BH surrounded by a cloud of strings~\cite{Letelier:1979ej}, in which case $k$ is associated to the gauge-invariant density of the cloud of strings $\sqrt{-\gamma}\rho = k/r^2$. In both cases, $k$ characterizes a specific hair. See also Refs.~\cite{Richarte:2007bx,Bahamonde:2015uwa,MoraisGraca:2016hmv,Jusufi:2017lsl,Jusufi:2018waj,Panotopoulos:2018law,Li:2020zxi,Singh:2020nwo,Gullu:2020qzu,Mustafa:2021hvq,He:2021aeo,Khodadi:2021mct,Belhaj:2022kek,Fathi:2022pqv,Fathi:2022ntj} for further works on BHs carrying global monopole charges or surrounded by clouds of strings.

\begin{figure}
\centering
\includegraphics[width=1.0\linewidth]{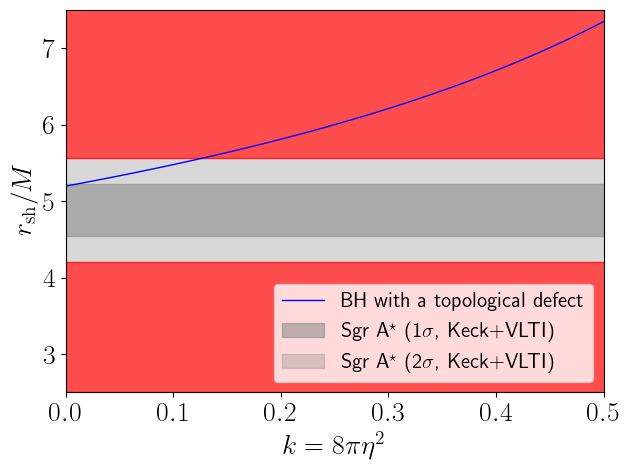}
\caption{Same as in Fig.~\ref{fig:shadow_reissner_nordstrom_bh_ns} for the Schwarzschild-like BH with a topological defect with metric function given by Eq.~\eqref{eq:metrictopologicaldefect}, as discussed in Sec.~\ref{subsec:topologicaldefect}.}
\label{fig:shadow_topological_defect_bh}
\end{figure}

The above space-time is manifestly not asymptotically flat, itself a direct consequence of the presence of the global monopole charge: in fact, it is globally conical. We can easily find that the location of the photon sphere is given by:
\begin{eqnarray}
r_{\rm ph}=\frac{3M}{1-k}\,.
\end{eqnarray}
To obtain the shadow size we can exploit the fact that $A(r_O) \to \sqrt{1-k}$ as $r_O \to \infty$ (and more precisely for $r_O \gg M$), although we need to pay attention to the mass parameter which sets the scale of the shadow radius. In fact, performing the coordinate transformations $t\to (1-k)^{-1/2} t'$, $r\to (1-k)^{1/2} r'$, $M\to (1-k)^{3/2} M'$, it is easy to see that the ADM mass is given by ${\cal M}=(1-k) M'$. We then find that the shadow size is given by:
\begin{eqnarray}
r_{\rm sh}=\frac{3\sqrt{3}{\cal M}}{\sqrt{1-k}}\,.
\label{eq:rshtopologicaldefect}
\end{eqnarray}
The difference between whether we choose to report the shadow size in units of the ADM mass ${\cal M}$ or the parameter $M$ appearing in Eq.~(\ref{eq:metrictopologicaldefect}), then, amounts to a factor of $(1-k)$. Here, we choose to report the shadow size in units of ${\cal M}$, and show its evolution against $k$ in Fig.~\ref{fig:shadow_topological_defect_bh}. We find that the EHT observations set the limits $k \lesssim 0.005$ ($1\sigma$) and $k \lesssim 0.1$ ($2\sigma$), which correspond to limits on the amplitude of the global monopole field of $\eta \lesssim 0.014$ and $\eta \lesssim 0.06$ respectively. Given the very tight limits on $k$, $1-k \approx 1$ and the ambiguity associated to reporting the shadow radius in units of ${\cal M}$ or $M$ ends up making virtually no difference in our final results.

\subsection{Hairy black hole from gravitational decoupling}
\label{subsec:hgd}

Obtaining exact solutions to Einstein's equations in theories of gravity beyond GR is usually a difficult, if not impossible, task. While one case where this is possible is that where a perfect fluid acts as gravitational source, this ceases to be true as soon as the perfect fluid is coupled to more complex forms of matter-energy. In recent years, a very interesting technique has been proposed by Ovalle to overcome these difficulties: the minimal geometric deformation approach and its latter extension, the extended gravitational decoupling approach. We shall broadly refer to these approaches as ``gravitational decoupling''. This approach was initially proposed in the context of the Randall-Sundrum brane-world model~\cite{Ovalle:2007bn}, but its power was quickly realized, to the extent that it has now been applied to an enormous variety of scenarios (see e.g.\ Refs.~\cite{Ovalle:2008se,Ovalle:2010zc,Casadio:2012pu,Casadio:2012rf,Ovalle:2013vna,Casadio:2013uma,Ovalle:2013xla,Ovalle:2014uwa,Casadio:2015jva,Casadio:2015gea,Cavalcanti:2016mbe,daRocha:2017cxu,daRocha:2017lqj,Ovalle:2017fgl,Ovalle:2017wqi,Fernandes-Silva:2017nec,Gabbanelli:2018bhs,Fernandes-Silva:2018abr,Contreras:2018vph,Contreras:2018gzd,Contreras:2018nfg,Ovalle:2018ans,Ovalle:2018gic,Contreras:2019fbk,Contreras:2019mhf,Gabbanelli:2019txr,Ovalle:2019lbs,Hensh:2019rtb,LinaresCedeno:2019aul,Leon:2019abq,Torres-Sanchez:2019wjv,Rincon:2019jal,Casadio:2019usg,Singh:2019ktp,Maurya:2019noq,Abellan:2020wjw,Contreras:2020fcj,Arias:2020hwz,Tello-Ortiz:2020ydf,Maurya:2020rny,Rincon:2020izv,Maurya:2020gjw,Meert:2021khi,Maurya:2022hav,Cavalcanti:2022cga,Yang:2022ifo,Cavalcanti:2022adb,Maurya:2022wwa,Mahapatra:2022xea,Zhang:2022niv} for an inevitably incomplete list of examples). The strength of the gravitational decoupling approach is at least three-fold: \textit{i)} it allows to construct new viable BH solutions by extending simpler seed solutions; \textit{ii)} it can be used to systematically decouple a complex source energy-momentum tensor into more simple tractable components; and \textit{iii)} it can be utilized to find exact solutions in theories beyond GR. In the context of the present discussion, our interest in this approach rests upon its potential to describe BHs surrounded by various forms of matter. For complete details on the gravitational decoupling approach, we refer the reader to Refs.~\cite{Ovalle:2007bn,Ovalle:2017fgl,Ovalle:2018gic}.

Here, we shall base ourselves upon the work of Ref.~\cite{Ovalle:2020kpd}, where the authors applied the gravitational decoupling approach to obtain a minimal deformation of the seed Schwarzschild space-time describing a hairy BH sourced by spherically symmetric ``tensor-vacuum'' $\theta_{\mu\nu}$ such that the existence of a well-defined event horizon is preserved, while the (primary) hair associated to the additional source does not violate the strong and dominant energy conditions. In short, Ref.~\cite{Ovalle:2020kpd} obtains a general solution describing a spherically symmetric space-time which violates the no-hair theorem in a non-pathological way. In particular, the work finds that the charge associated to the primary hair increases the entropy of the space-time compared to its Schwarzschild value. Specifically, this BH solution is described by the following metric function~\cite{Ovalle:2020kpd}:
\begin{eqnarray}
A(r) = 1-\frac{2M}{r}+\alpha \exp \left ( -\frac{r}{M-\alpha\ell/2} \right ) \,.
\label{eq:metrichgd}
\end{eqnarray}
where $\alpha$ and $\ell$ are the parameters which quantify deviations from the Schwarzschild metric, both characterizing specific hairs, with $0 \leq \ell \leq 2M$ representing the charge associated to the primary hair and the Schwarzschild space-time is recovered for $\alpha\ell= 2M$ (as well as in the physically less interesting case $\alpha=0$). In particular, the combination $\ell_0=\alpha\ell$ quantifies the increase in the BH entropy from its Schwarzschild value $S=4\pi M^2$. The fluid surrounding the BH which sources this space-time can be associated to DM.

\begin{figure}
\centering
\includegraphics[width=1.0\linewidth]{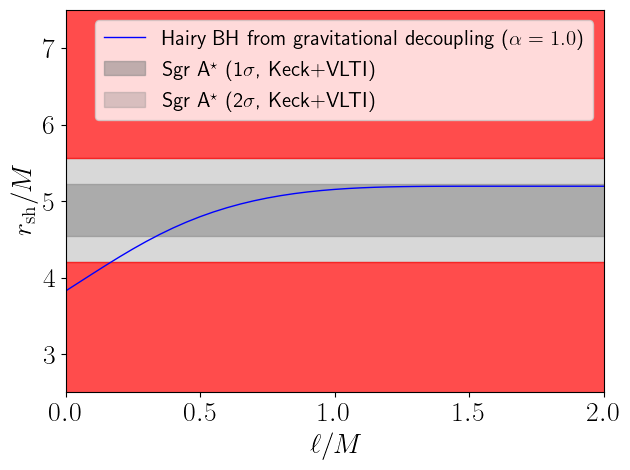}
\caption{Same as in Fig.~\ref{fig:shadow_reissner_nordstrom_bh_ns} for the hairy BH from gravitational decoupling with metric function given by Eq.~\eqref{eq:metrichgd}, as discussed in Sec.~\ref{subsec:hgd}.}
\label{fig:shadow_hgd_bh}
\end{figure}

For illustrative purposes, in order to focus our study on the charge parameter $\ell$, we fix $\alpha=1.0$, so that $\ell_0=\ell$. We compute the resulting shadow size numerically, and show its evolution against $\ell$ in Fig.~\ref{fig:shadow_hgd_bh}. We see that decreasing $\ell$ from $\ell=2M$ decreases the shadow size, consistently with earlier findings of some of us in Ref.~\cite{Afrin:2021imp}. For this particular choice of $\alpha$, we see that the EHT observations set the lower limits $\ell \gtrsim 0.35M$ ($1\sigma$) and $\ell \gtrsim 0.15M$ ($2\sigma$). These limits can be translated into upper limits on the entropy increase compared to the Schwarzschild value $S=4\pi$ (in units where $M=1$). In particular, we find that the associated hair cannot increase the entropy by more than $3\%$. Broadly speaking, this places constraints on the strength of no-hair theorem violation, in the presence of hair which does not violate the strong and dominant energy conditions. Finally, we stress that much as with a few cases studied earlier, these results only correspond to a slice of the full $\alpha$-$\ell$ parameter space, and fixing $\alpha$ to smaller (larger) values would lead to weaker (tighter) constraints on $\ell$ (for more details see Ref.~\cite{Afrin:2021imp}).

\subsection{Modified uncertainty principles}
\label{subsec:modifieduncertaintyprinciples}

The final scenarios we study can generically be referred to as ``novel fundamental physics frameworks'', which do not necessarily modify gravity (at least not directly, although as we shall see they lead to a sort of ``backreaction'' on the metrics of BHs in GR). We begin by discussing modified uncertainty principles (UPs) which go beyond the Heisenberg one.

While GR is strictly speaking not incompatible with quantum mechanics, the introduction of gravity into quantum field theories appears to spoil their renormalizability, which is one of the main reasons why the construction of a full theory of quantum gravity has eluded us so far. Various proposals for rendering gravity UV-complete have been explored, ranging from the addition of higher order curvature terms (it is known that adding terms quadratic in the curvature tensor to the Einstein-Hilbert action renders the theory renormalizable~\cite{Stelle:1976gc,Stelle:1977ry,Salvio:2018crh}, albeit at the price of a loss of unitarity), to the use of Lorentz violation as a field theory regulator~\cite{Visser:2009fg}.

However, it has long been suspected that gravity should lead to an effective UV cut-off, in turn leading to a minimum observable length and/or a maximum possible momentum, based on gedanken experiments wherein the high energies used to resolve small distances eventually ``disturb'' the space-time structure due to their gravitational effects, to the point that space-time itself cannot be resolved below some distance. Support for these ideas range from various general considerations, including within quantum gravity and quantum mechanics~\cite{Maggiore:1993rv,Maggiore:1993zu,Scardigli:1999jh,Adler:1999bu,Bruneton:2016yws}, as well as string theory~\cite{Veneziano:1986zf,Gross:1987ar,Amati:1988tn,Konishi:1989wk,Witten:1996mlx} (see also Refs.~\cite{Yoneya:1989ai,Maggiore:1993kv,Ashtekar:2002sn,Hossain:2010wy,Isi:2013cxa,Bosso:2021koi,Luciano:2021vkl,Jizba:2022icu,Bosso:2022xnm,Bosso:2022ogb}). In the presence of a minimum length scale, gravity could therefore actually help rather than hamper achieving renormalizability.

These ideas have motivated approaches wherein the Heisenberg position-momentum commutator $[x,p]=i$, or equivalently the Heisenberg UP $\Delta x\Delta p \geq 1/2$, receives (quantum gravitational) non-linear corrections~\cite{Snyder:1946qz,Yang:1947ud,Mead:1964zz}. Such approaches are typically referred to as Generalized Uncertainty Principle (GUP) or Extended Uncertainty Principle (EUP), see e.g.\ Refs.~\cite{Hossenfelder:2012jw,Tawfik:2014zca} for reviews. The notation and language when it comes to GUP and EUP is at times rather loose, but for consistency here we shall use the term ``GUP'' to refer to the case where Heisenberg's UP receives corrections non-linear in momentum, typically associated to minimum length scenarios. On the other hand, we shall use the term ``EUP'' to denote the case where the corrections are non-linear in position, as occurring in scenarios which introduce a (typically large) new fundamental scale (see e.g.\ Ref.~\cite{Bambi:2007ty}). Here we shall first discuss GUP scenarios, before turning to EUP scenarios afterwards.

\subsubsection{Generalized uncertainty principle}
\label{subsubsec:gup}

Various GUP models non-linearly deforming the Heisenberg UP have been proposed in the literature, in most cases driven by phenomenological considerations themselves based on theoretical expectations. One of the earliest proposals is that of Kempf, Mangano, and Mann, wherein Heisenberg's position-momentum commutator receives corrections non-linear in momentum, as per the following~\cite{Kempf:1994su}:
\begin{eqnarray}
[x_i, p_j] = i\delta_{ij} \left ( 1 + \beta_1\ell_{\rm Pl}^2 p^2 \right ) \,,
\label{eq:gupicommutator}
\end{eqnarray}
where $\ell_{\rm Pl}$ is the Planck length and $\beta_1$ is a dimensionless parameter. In turn, Eq.~\eqref{eq:gupicommutator} implies the following corrected UP:
\begin{eqnarray}
\Delta x\Delta p \geq \frac{1}{2} \left [ 1+ \beta_1\ell_{\rm Pl}^2(\Delta p)^2 \right ] \,,
\label{eq:gupi}
\end{eqnarray}
with minimum measurable length given by $\Delta x \approx \ell_{\rm Pl}\sqrt{\beta_1}$.

A different GUP scenario which admits both a minimum length and a maximum momentum was instead presented in Refs.~\cite{Ali:2009zq,Das:2010zf} (see also Refs.~\cite{Frassino:2011aa,Nozari:2012gd,Jusufi:2020rpw}), and modifies Heisenberg's position-momentum commutator as follows:
\begin{eqnarray}
[x_i, p_j] = i \left [ \delta_{ij}-\beta \left ( \delta_{ij}p+\frac{p_ip_j}{p} \right ) +\beta^2 \left ( \delta_{ij}p^2+3p_ip_j \right ) \right ] \,, \nonumber \\
\label{eq:gupiicommutator}
\end{eqnarray}
where $\beta = \beta_0\ell_{\rm Pl}$ and which in turn implies the following corrected UP:
\begin{eqnarray}
\Delta x\Delta p \geq \frac{1}{2} \left ( 1-2\beta \langle p \rangle+4\beta^2 \langle p^2 \rangle \right ) \,.
\label{eq:gupii}
\end{eqnarray}
The minimum measurable length implied by Eq.~(\ref{eq:gupii}) is given by $\Delta x \approx \ell_{\rm Pl}\beta_0$, whereas the maximum momentum is given by $\Delta p \approx \ell_{\rm Pl}/\beta_0$. Notice that the deformed commutators of Eqs.~(\ref{eq:gupicommutator},\ref{eq:gupiicommutator}) have both been introduced on phenomenological grounds: however, in both cases the expected value of $\beta$ is of order unity, and can be obtained using low-energy formulations of string theory~\cite{Amati:1987wq,Capozziello:1999wx}, by studying corrections to Hawking's theorem for the BH temperature obtained from two different approaches, or by computing the leading quantum corrections to Newton's gravitational potential~\cite{Scardigli:2016pjs}.

In the following, we shall consider the GUP characterized by Eqs.~(\ref{eq:gupiicommutator},\ref{eq:gupii}). GUP-deformed BH solutions have been studied in several papers, and here we shall build upon the results of Refs.~\cite{Carr:2015nqa,Vagenas:2017vsw,Carr:2020hiz}, where it has been shown that GUP effects physically lead to what one could consider quantum corrections to the BH ADM mass, which is modified to ${\cal M} = M+\beta M_{\rm Pl}^2/M^2$, where $M$ can be interpreted as the ``bare'' mass. Thus, the metric function for a GUP-corrected Schwarzschild BH reads~\cite{Carr:2015nqa,Vagenas:2017vsw,Carr:2020hiz}:
\begin{eqnarray}
A(r) = 1-\frac{2M \left ( 1+\frac{\beta}{2M^2} \right ) }{r}\,,
\label{eq:metricgup}
\end{eqnarray}
where $\beta$ characterizes an universal hair.

Despite the fact that various GUP tests have been performed in the literature using BHs and their environments (including their shadows), the form of the metric function given by Eq.~(\ref{eq:metricgup}) raises the question of whether one can actually discriminate a GUP-corrected BH from a ``genuine'' Schwarzschild BH. In our case, we believe this might indeed be possible for Sgr A$^*$, for the following reason which echoes the argument presented previously for MOG in Sec.~\ref{subsec:modifiedgravity}. We interpret $M$ as being the ``bare'' mass of the system. Quantum effects which correct $M \to {\cal M}$ effectively change the total mass/energy of the system, as measured by an observer at infinity (given the definition of ADM mass). However, a measurement of the BH mass through dynamical probes sufficiently close to the event horizon will effectively probe $M$ and not ${\cal M}$: this is indeed the case for the S-stars orbiting Sgr A$^*$ (and in particular the S2-star), whose motion has been used to obtain the exquisite mass-to-distance ratio measurements we adopt as priors. Moreover, the uncertainty on the mass (which does not disappear when we later fix $M=1$, but is simply transferred to the mass-to-distance ratio) is part of what allows us to constrain the hair parameter $\beta$: we fix the bare mass, and ask how much of a change in the total mass/energy of the system we can tolerate when varying the hair parameter. While this physical interpretation has, to the best of our knowledge, never been explicitly discussed in the literature, it is the assumption which de facto underlies the many BH metric-based GUP tests (see e.g.\ Refs.~\cite{Feng:2016tyt,Neves:2019lio,Jusufi:2020wmp,Tamburini:2021inp,Anacleto:2021qoe}).

\begin{figure}
\centering
\includegraphics[width=1.0\linewidth]{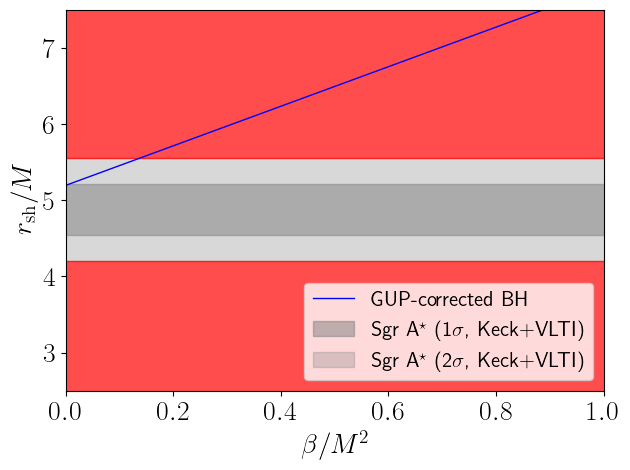}
\caption{Same as in Fig.~\ref{fig:shadow_reissner_nordstrom_bh_ns} for GUP-corrected BH with metric function given by Eq.~\eqref{eq:metricgup}, as discussed in Sec.~\ref{subsubsec:gup}.}
\label{fig:shadow_gup_bh}
\end{figure}

Keeping in mind the assumptions and caveats discussed above, it is straightforward to show that the shadow radius associated to the GUP-corrected BH with metric function given by Eq.~(\ref{eq:metricgup}) is:
\begin{eqnarray}
r_{\rm sh} = 3\sqrt{3}{\cal M} = 3\sqrt{3}M \left ( 1+\frac{\beta}{2} \right )\,.
\label{eq:shadowsizegup}
\end{eqnarray}
We show the evolution of the shadow size as a function of $\beta$ in Fig.~\ref{fig:shadow_gup_bh}. We see that the EHT observations set the upper limits $\beta \lesssim 0.01M^2$ ($1\sigma$) and $\beta \lesssim 0.14M^2$ ($2\sigma$). Restoring the Planck mass to obtain the dimensionless GUP parameter $\beta$, we require $\beta \lesssim 0.14(M/M_{\rm Pl})^2 \sim 2 \times 10^{88}$. Unsurprisingly this is approximately 6 orders of magnitude stronger than the limit obtained from the shadow of M87$^*$ in Refs.~\cite{Neves:2019lio,Jusufi:2020wmp}, as Sgr A$^*$'s mass is about 3 orders of magnitude smaller. However, this same limit is still weaker by several orders of magnitude compared to the limits obtained from independent tests such as perihelion precession, pulsar periastron shift, light deflection, gravitational waves, gravitational redshift, Shapiro time delay, geodetic precession, and quasiperiodic oscillations (see e.g.\ Refs.~\cite{Scardigli:2014qka,Jusufi:2020wmp,Okcu:2021oke}). Nonetheless, our limit remains a valuable complementary constraint on an fundamental physics scenario of great theoretical interest.

\subsubsection{Extended uncertainty principle}
\label{subsubsec:eup}

In the GUP scenarios discussed so far, momentum appears to play a preferred role. However, other possibilities discussed in the literature envisage either a ``democratic'' extension where both position and momentum play an equally important role (see e.g.\ Ref.~\cite{Bambi:2007ty} by one of us), or scenarios where the corrections to Heisenberg's UP depend only on the position operator itself. It is worth reminding the reader once more that these deformations, while inspired by theoretical scenarios, are nonetheless introduced at the phenomenological level. With these caveats in mind, in the following we shall consider an Extended Uncertainty Principle (EUP) scenario, where Heisenberg's UP is corrected as follows:
\begin{eqnarray}
\Delta x\Delta p \geq \frac{1}{2} \left [ 1+\alpha\frac{(\Delta x)^2}{L_{\star}^2} \right ] \,,
\label{eq:eup}
\end{eqnarray}
where $L_{\star}$ is a new large fundamental distance scale, and $\alpha$ is generically expected to be of order unity. It is fair to say that EUP scenarios have received much less attention than their GUP counterparts, the reason possibly being that the need for new macroscopic gravitational physics has only recently started to become appreciated. This is not unrelated to the recent realization that when dealing with BHs, quantum gravity effects may be relevant not only on microscopic scales, but perhaps more importantly on macroscopic horizon scales, and may well be tied to puzzling theoretical aspects such as the information paradox or singularity removal: see e.g.\ Refs.~\cite{Mathur:2002ie,Mathur:2005zp,Giddings:2017jts} for various proposals to address the information paradox which envisage new physics appearing at the gravitational radius of a system rather than at the Planck scale. Related recent works have put forward the possibility that the horizon itself may have a quantum origin, for example within the corpuscular framework, wherein BHs are described as Bose-Einstein condensates of $N$ soft gravitons weakly confined to a region whose size is comparable to the BH gravitational radius~\cite{Dvali:2011aa,Dvali:2012gb,Dvali:2012rt,Casadio:2018qeh,Buoninfante:2019fwr}. See Refs.~\cite{Zhu:2008cg,Moradpour:2019yiq,Moradpour:2019wpj,Aghababaei:2021gxe,Lobos:2022jsz} for further EUP-related works.

\begin{figure}
\centering
\includegraphics[width=1.0\linewidth]{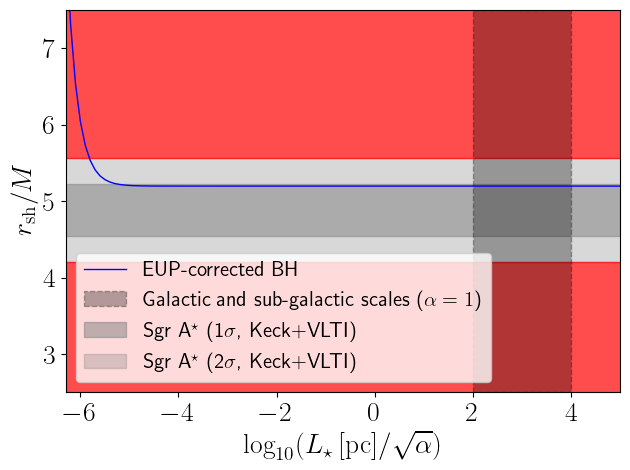}
\caption{Same as in Fig.~\ref{fig:shadow_reissner_nordstrom_bh_ns} for the EUP-inspired BH with metric function given by Eq.~\eqref{eq:metriceup}, as discussed in Sec.~\ref{subsubsec:eup}. The black shaded band denotes the range corresponding to galactic and sub-galactic scales $L_{\star} \sim {\cal O}(0.1-10)\,{\rm kpc}$ (with $\alpha=1$), where EUP-related effects could potentially mimic dark matter.}
\label{fig:shadow_eup_bh}
\end{figure}

The effect of an EUP of the form given by Eq.~(\ref{eq:eup}) on BHs was studied in Ref.~\cite{Mureika:2018gxl}. The effect of the EUP is analogous to the GUP effect discussed in Sec.~\ref{subsubsec:gup}, and amounts to (quantum) corrections to the BH ADM mass. As a result, the metric function of an EUP-corrected BH is given by~\cite{Mureika:2018gxl} (see also Ref.~\cite{Kumaran:2019qqp,Pantig:2021zqe}):
\begin{eqnarray}
A(r) = 1-\frac{2M}{r} \left ( 1+\frac{4\alpha M^2}{L_{\star}^2} \right ) \,,
\label{eq:metriceup}
\end{eqnarray}
where $\alpha$ characterizes an universal hair. As is clear from Eq.~(\ref{eq:metriceup}), the EUP-inspired corrections are large only when $L_{\star}$ is sufficiently small (this makes sense if we think of $L_{\star}$ as corresponding to an inverse energy scale). We show the evolution of the shadow size as a function of $\log_{10}(L_{\star}/\sqrt{\alpha})$, with $L_{\star}$ measured in pc, in Fig.~\ref{fig:shadow_eup_bh}. We find it convenient to express $L_{\star}$ in these units since the possibility of partially mimicking the galactic effects of dark matter has been discussed in the literature~\cite{Mureika:2018gxl}, and relies upon $L_{\star}/\sqrt{\alpha}$ lying on ${\cal O}(0.1-10)\,{\rm kpc}$ scales, i.e.\ those relevant to dwarf and spiral galaxies. From Fig.~\ref{fig:shadow_eup_bh} we see that EUP effects do not lead to visible changes in the shadow radius as long as $L/\sqrt{\alpha} \gtrsim 10^{-5}\,{\rm pc}$. For smaller values of $L/\sqrt{\alpha}$, the shadow radius increases drastically. Values of $L/\sqrt{\alpha} \simeq 10^{-6}\,{\rm pc}$ are completely excluded by the EHT observations, as the resulting shadow is much larger than what these observations allow. In addition, the black shaded band in Fig.~\ref{fig:shadow_eup_bh} denotes the range corresponding to galactic and sub-galactic scales $L_{\star} \sim {\cal O}(0.1-10)\,{\rm kpc}$ (with $\alpha=1$) discussed previously, where EUP-related effects have been argued to potentially mimic the effects of dark matter. On these scales, EUP effects leave no visible imprint on the size of Sgr A$^*$'s shadow: this is to be expected, given that such scales are several orders of magnitude above the scale of Sgr A$^*$'s gravitational radius, which is in the $\sim {\cal O}(10^{-7})\,{\rm pc}$ range.~\footnote{Another modified uncertainty relation scenario with a minimal length and maximal momentum it could be interesting to consider has been proposed in Refs.~\cite{Pedram:2011gw,Pedram:2012my}. To the best of our knowledge, BH, WH, and naked singularity solutions within this model have not been derived. We defer a full study thereof to future work.}

\subsection{Non-commutative geometry}
\label{subsec:ncg}

Another approach (arguably enjoying stronger theoretical motivation compared to the ones discussed previously) which naturally introduces a minimum length scale occurs within non-commutative geometries (NCG). The term NCG encompasses a rather large ensemble of approaches: while we shall here shall use it in the sense of a space-time with non-zero commutation relations among the position coordinates~\cite{Doplicher:1994tu}, of enormous importance is the spectral approach to NCG pioneered by Connes~\cite{Connes:1994yd}. The non-commutativity of space-time coordinates naturally encapsulates the discrete nature of space-time expected within several high-energy frameworks, including attempts towards constructing a theory of quantum gravity. For instance, this is a natural outcome of string theory, where target space-time coordinates become non-commuting operators on a \textit{D}-brane~\cite{Witten:1995im}. It is worth noting, however, that motivation for NCG covers a wide range of frameworks going well beyond string theory.

One of the earliest approaches towards constructing a non-commutative field theory was based on the Weyl-Wigner-Moyal $\ast$-product~\cite{Weyl:1927vd,Wigner:1932eb,Moyal:1949sk,Alvarez-Gaume:2000jcd,Alvarez-Gaume:2000jji}. However, this approach left open several important questions, concerning among others the issues of Lorentz invariance and UV finiteness. This prompted the development of a different approach, the so-called coherent state formalism, wherein physical coordinates are commuting c-numbers, built as expectation values on coherent states~\cite{Smailagic:2003yb,Smailagic:2003rp}: within this approach, one can construct an explicitly UV-finite non-commutative field theory (see also Ref.~\cite{Gruppuso:2005yw}). The literature on NCG is extremely vast and it is not possible to do it justice here: for representative examples of important works on NCG and its applications to various scenarios of interest, as well as related ideas, we refer the reader to Refs.~\cite{Bimonte:1994ch,Lizzi:1995kq,Bimonte:1995jp,Balachandran:1995sx,Connes:1996gi,Lizzi:1996yf,Balachandran:1996qt,Lizzi:1996vr,Figueroa:1997dm,Lizzi:1997sg,Landi:1998ii,Lizzi:1999mw,Hashimoto:1999ut,Maldacena:1999mh,Matusis:2000jf,Seiberg:2000gc,Jurco:2000ja,Lizzi:2000bc,Lizzi:2001nd,Calmet:2001na,Lizzi:2002ib,Chaichian:2002ew,Schupp:2002up,Tsujikawa:2003gh,Lizzi:2005hm,Lizzi:2006te,Galluccio:2008wk,Galluccio:2009ss,Rinaldi:2009ba,Rinaldi:2010zu,Machado:2011fq,Perrier:2012nr,Machado:2012vy,Li:2013qga,Calcagni:2013lya,Shiraishi:2014owa,Rahaman:2014dpa,Joby:2014oee,DAndrea:2016hyl,Kobakhidze:2016cqh,Araujo:2017jap,Lizzi:2017ioq,Kurkov:2017wmx,Bouhmadi-Lopez:2019hpp,Devastato:2019grb,Lizzi:2020tci,Lizzi:2021dud,Cuzinatto:2022hfo,Cuzinatto:2022xsr,Guendelman:2022gue}.

In the approach we shall follow, the non-commutativity of space-time will be encoded in the commutation relation between space-time coordinates:
\begin{eqnarray}
[x^{\mu},x^{\nu}]=i\vartheta^{\mu\nu}\,,
\label{eq:commutationrelationncg}
\end{eqnarray}
where $\vartheta^{\mu\nu}$ is an anti-symmetric real matrix. Within the coherent state formalism, the issues of covariance and unitarity are solved if one assumes the following block-diagonal form for $\vartheta^{\mu\nu}$:
\begin{eqnarray}
\vartheta^{\mu\nu} = \vartheta \text{diag} \left ( \epsilon_{ij}\,,\epsilon_{ij}\,,...\epsilon_{ij} \right ) \,,
\label{eq:thetamunu}
\end{eqnarray}
where $\epsilon_{ij}$ is the two-dimensional Levi-Civita tensor, and $\vartheta$ is a fundamental constant with dimensions of length squared, and reflects the fact that within NCG the best possible localization of a particle is within a fundamental cell of area $\vartheta$. It has been shown in Refs.~\cite{Smailagic:2003yb,Smailagic:2003rp} that the main outcome of this approach is to replace point-like objects with smeared structures, a fact which can be seen explicitly by computing the NCG-induced modifications to the Feynman path integral. This basically amounts to a substitution rule, where Dirac delta functions are replaced by a distribution of ``typical'' (in a loose sense) width $\sqrt{\vartheta}$. The same of course can and will occur for finite mass distributions, including the mass density of a particle-like gravitational source: rather than being localized at the origin, this will be smeared and diffuse throughout a region of typical width $\sqrt{\vartheta}$, as a result of the uncertainty encoded in the non-commutativity of the space-time coordinates in Eqs.~(\ref{eq:commutationrelationncg},\ref{eq:thetamunu}). Applying this ``substitution rule'', Ref.~\cite{Nicolini:2005vd} was the first to study NCG-inspired BHs: we use the term ``inspired'' since, in this approach, the choice of the function with which to replace the Dirac delta essentially amounts to a phenomenological, but theory-inspired, choice. In the following, we shall consider two different choices for such a function describing the smearing of the mass distribution of a point-like gravitational source: a Gaussian in Sec.~\ref{subsubsec:ncggaussian} and a Lorentzian in Sec.~\ref{subsubsec:ncglorentzian}.

\subsubsection{Gaussian mass distribution}
\label{subsubsec:ncggaussian}

We begin by choosing a Gaussian mass distribution, as first considered in Ref.~\cite{Nicolini:2005vd}. We note that this is the choice which enjoys strongest theoretical motivation. In this case, the point-like mass distribution $\rho(r)=M\delta(r)$ is replaced by the following:
\begin{eqnarray}
\rho_{\vartheta}(r) = \frac{M}{\left ( 4\pi\vartheta \right )^{\frac{3}{2}}}\exp \left ( -\frac{r^2}{4\vartheta} \right )\,,
\label{eq:gaussiandistribution}
\end{eqnarray}
so that the point-like gravitational source diffuses throughout a region of size $\sqrt{\vartheta}$. Given this choice of mass distribution, the corresponding BH solution was studied in Ref.~\cite{Nicolini:2005vd} by constructing the (covariantly conserved) energy-momentum tensor corresponding to the energy density of Eq.~(\ref{eq:gaussiandistribution}). The final result is that the metric function describing NCG-inspired BHs within this scenario is given by~\cite{Nicolini:2005vd} (see also Refs.~\cite{Banerjee:2009xx,Ovgun:2015box,Nunes:2016two,Gangopadhyay:2017tlp,Haldar:2017viz,Kumar:2017hgs,Ghosh:2017odw,Ovgun:2019jdo,Panotopoulos:2019gtn,Ghosh:2020cob,Maceda:2020rpv,Zeng:2021dlj,Jha:2021bue,Graca:2021ker,Rayimbaev:2022hnp} for further works on NCG-inspired BHs):
\begin{eqnarray}
A(r) = 1-\frac{4M}{r\sqrt{\pi}} \gamma \left ( \frac{3}{2},\frac{r^2}{4\vartheta} \right ) \,,
\label{eq:metricncggaussian}
\end{eqnarray}
where $\vartheta$ characterizes an universal hair, $\gamma$ denotes the lower incomplete gamma function, and the Schwarzschild metric is recovered in the limit $r/\sqrt{\vartheta} \to \infty$. For $\sqrt{\vartheta}<0.5252$ the above space-time possesses two horizons, whereas an extremal BH with degenerate horizons at $r=1.5873$ is recovered for $\sqrt{\vartheta}=0.5252$. In the following, we shall focus on the $\sqrt{\vartheta} \leq 0.5252$ regime.

\begin{figure}
\centering
\includegraphics[width=1.0\linewidth]{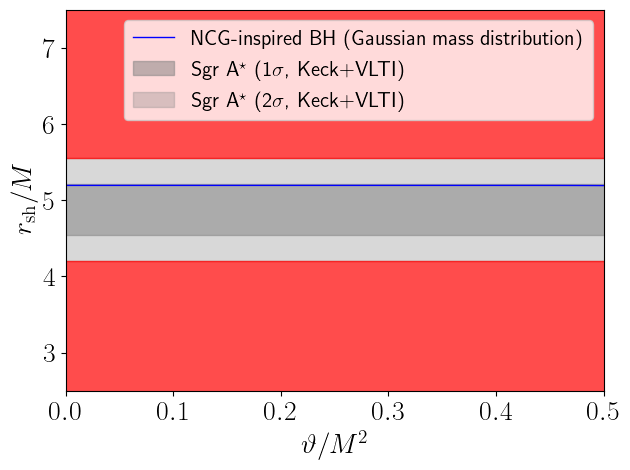}
\caption{Same as in Fig.~\ref{fig:shadow_reissner_nordstrom_bh_ns} for the NCG-inspired BH (with Gaussian mass distribution) with metric function given by Eq.~\eqref{eq:metricncggaussian}, as discussed in Sec.~\ref{subsubsec:ncggaussian}.}
\label{fig:shadow_ncg_gaussian_bh}
\end{figure}

We compute the shadow size numerically, and show its evolution as a function of the NCG parameter $\vartheta$ in Fig.~\ref{fig:shadow_ncg_gaussian_bh}. We find that the shadow size depends extremely weakly on $\vartheta$, so much so that the evolution cannot be appreciated by the naked eye. This is consistent with the earlier findings of Ref.~\cite{Wei:2015dua}. We find that for the extremal value $\sqrt{\vartheta}=0.5252$, the relative change in the shadow size compared to that of a Schwarzschild BH with the same mass is of order ${\cal O}(10^{-3})$. We therefore conclude that this particular NCG-inspired BH is in perfect agreement with the EHT observations, but distinguishing it from its Schwarzschild counterpart appears extremely unlikely, unless unrealistic angular resolutions of ${\cal O}({\rm nas})$ can be reached (which is extremely far from feasible).

\subsubsection{Lorentzian mass distribution}
\label{subsubsec:ncglorentzian}

\begin{figure}
\centering
\includegraphics[width=1.0\linewidth]{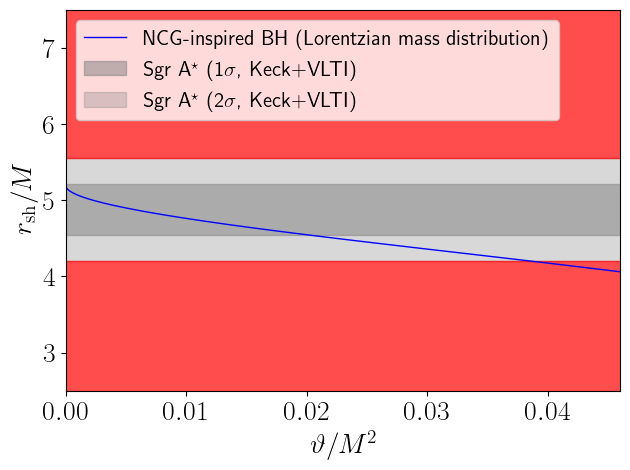}
\caption{Same as in Fig.~\ref{fig:shadow_reissner_nordstrom_bh_ns} for the NCG-inspired BH (with Lorentzian mass distribution) with metric function given by Eq.~\eqref{eq:metricncglorentzian}, as discussed in Sec.~\ref{subsubsec:ncglorentzian}.}
\label{fig:shadow_ncg_lorentzian_bh}
\end{figure}

Remaining in the realm of NCG-inspired phenomenology, another possibility which has been considered in the literature is that where a point-like gravitational source is smeared across a non-Gaussian distribution. An example which has been studied in this sense envisages a Lorentzian distribution (see e.g.\ Refs.~\cite{Nozari:2008rc,Anacleto:2019tdj,Campos:2021sff}), so that Eq.~(\ref{eq:gaussiandistribution}) is modified to the following:
\begin{eqnarray}
\rho_{\vartheta}(r) = \frac{M\sqrt{\vartheta}}{\pi^{\frac{3}{2}}(r^2+\pi\vartheta)^2}\,,
\label{eq:lorentziandistribution}
\end{eqnarray}
where again the source diffuses across a region of linear size $\sqrt{\vartheta}$. The resulting NCG-inspired BHs were studied in Ref.~\cite{Anacleto:2019tdj}, and shown to be described by the following metric function (to lowest order in $\sqrt{\vartheta}$):
\begin{eqnarray}
A(r) = 1-\frac{2M}{r}+\frac{8M\sqrt{\vartheta}}{\sqrt{\pi}r^2} \,,
\label{eq:metricncglorentzian}
\end{eqnarray}
where $\vartheta$ characterizes an universal hair, and which is clearly of the RN-like form, so we can write down the following closed-form expression for the shadow radius:
\begin{eqnarray}
r_{\rm sh} = \frac{\sqrt{2\sqrt{\pi}} \left ( 3+\sqrt{9-64\sqrt{\frac{\theta}{\pi}}} \right ) }{\sqrt{\frac{32\sqrt{\frac{\theta}{\pi}}+\sqrt{9-64\sqrt{\frac{\theta}{\pi}}}-3}{8\sqrt{\theta}}}}\,.
\label{eq:shadowsizencglorentzian}
\end{eqnarray}
We show the evolution of the shadow size as a function of the NCG parameter $\vartheta$ in Fig.~\ref{fig:shadow_ncg_lorentzian_bh}. Compared to the previous Gaussian mass distribution case, we now find a stronger effect on the BH shadow size, which decreases with increasing $\vartheta$. This is likely due to the fact that a Lorentzian distribution has heavier tails compared to a Gaussian one with the same value of $\vartheta$. We find that the EHT observations set the upper limits $\vartheta \lesssim 0.02M^2$ ($1\sigma$) and $\vartheta \lesssim 0.04M^2$ ($2\sigma$), both comparatively weak as one would typically expect NCG effects to show up at much smaller scales (see e.g.\ Ref.~\cite{Lambiase:2013dai}).

\subsection{Barrow entropy}
\label{subsec:barrow}

Since the discovery of Hawking evaporation and the fact that a temperature can be associated to BHs~\cite{Hawking:1971tu,Hawking:1975vcx}, the intriguing connections between thermodynamics and gravity have flourished and given birth to the vibrant field of BH thermodynamics~\cite{Bekenstein:1973ur,Bardeen:1973gs,Wald:1999vt}. The Bekenstein-Hawking entropy relation envisages a simple proportionality relation between the entropy and area of a BH~\cite{Hawking:1971tu,Bekenstein:1972tm}, and is unquestionably the most studied entropy relation in the BH thermodynamics field. However, over the past decades various alternative entropy relations have been explored, with a wide range of different theoretical motivations, including the fact that the non-extensive nature of gravitational systems such as BHs calls for a non-extensive entropy-area relation (unlike the Bekenstein-Hawking entropy): see e.g.\ Refs.~\cite{Tsallis:1987eu,Tsallis:2012js,Renyi:1959ghw,Kaniadakis:2005zk,Majhi:2017zao,Kaul:2000kf,Banerjee:2011jp,Nojiri:2019skr,Nojiri:2022aof,Nojiri:2022dkr} for important examples thereof.

As first explicitly noted by one of us in Ref.~\cite{Jusufi:2021fek}, in the context of a BH the presence of a modified entropy-area relation effectively \textit{back-reacts} on other thermodynamical quantities, including the BH's mass. If $M$ is the BH ``bare'' mass and $S(A)=S(M)$ is the entropy-area/mass relation in question, the ``dressed'' mass $\widetilde{M}$ is then given by $\widetilde{M} = (\partial S/\partial M)/8\pi$. The necessity of constructing a consistent thermodynamical system in the presence of a modified entropy relation is indeed an issue which has started receiving significant attention recently, see e.g.\ discussions in Refs.~\cite{Nojiri:2021czz,Nojiri:2022sfd,DiGennaro:2022grw,Cimdiker:2022ics}, as well as Ref.~\cite{Abreu:2020wbz}.

A particularly interesting modified entropy relation has been considered by Barrow~\cite{Barrow:2020tzx}, motivated by possible space-time foam effects. Barrow entropy leads to a fractal structure for the horizon with a volume which is finite (but larger than in the standard case), and area which can be either finite or infinite. In the toy model considered by Barrow, a three-dimensional fractal is built around the BH event horizon, modifying its total surface and volume.~\footnote{As Barrow himself stated, this modification was inspired by animations of the Covid-19 virus as a sphere with several attachments leaving off its surface to increase its surface area in order to increase the total number of possible latches to other cells.} In this picture, the entropy-area law of a BH is modified to:
\begin{eqnarray}
S = \left ( \frac{A}{4} \right ) ^{1+\frac{\Delta}{2}}\,,
\label{eq:barrowentropyarea}
\end{eqnarray}
where $\Delta$ characterizes an universal hair, $0\leq \Delta \leq 1$ is related to the dimension of the fractal, with $\Delta = 0$ resulting in the usual Bekenstein-Hawking entropy $S = A/4$, and $\Delta = 1$ leading the largest deviation of the BH entropy from the Bekenstein-Hawking value. Since its proposal, Barrow entropy has been at the center of several interesting applications, among which it is worth mentioning its use in driving a model of holographic dark energy~\cite{Saridakis:2020zol,Saridakis:2020lrg,Mamon:2020spa} (see also Refs.~\cite{Srivastava:2020cyk,Sharma:2020ylh,Pradhan:2021cbj,Sheykhi:2021fwh,Bhardwaj:2021chg,Dixit:2021phd,Chakraborty:2021uzp,Adhikary:2021xym,Asghari:2021bqa,Nojiri:2021jxf,Maity:2022gdy,Paul:2022doh,Sharma:2022wgi,Remya:2022frs,Srivastava:2022nex,Luciano:2022pzg,Oliveros:2022biu,Sheykhi:2022gzb,DiGennaro:2022ykp,Luciano:2022viz,Luciano:2022ffn,Luciano:2022hhy,Luciano:2023roh,Luciano:2023wtx}).

As shown in Ref.~\cite{Jusufi:2021fek} by one of us, the back-reaction of Barrow entropy on the Schwarzschild metric modifies the ``bare'' mass to ${\cal M}=(\Delta/2+1)(4\pi)^{\Delta/2}M^{\Delta+1}$, so that the metric function of the Barrow entropy-corrected BH is given by~\cite{Jusufi:2021fek}:
\begin{eqnarray}
A(r) = 1-\frac{(\Delta+2)M^{\Delta+1}(4\pi)^{\Delta/2}}{r} \,.
\label{eq:metricbarrow}
\end{eqnarray}
It is therefore trivial to show that the shadow size is given by:~\footnote{Note that Eq.~(27) in Ref.~\cite{Jusufi:2021fek} by one of us features a spurious factor of $2^{\Delta}$, which we have corrected here.}
\begin{eqnarray}
r_{\rm sh} = 3\sqrt{3}{\cal M}=3\sqrt{3} \left ( \frac{\Delta}{2}+1 \right ) M^{1+\Delta}\pi^{\Delta/2}\,.
\label{eq:rshbarrow}
\end{eqnarray}
The shadow size increases with increasing $\Delta$, reflecting the fact that the BH surface area increases due to the quantum gravity effects envisaged in Ref.~\cite{Barrow:2020tzx}.

\begin{figure}
\centering
\includegraphics[width=1.0\linewidth]{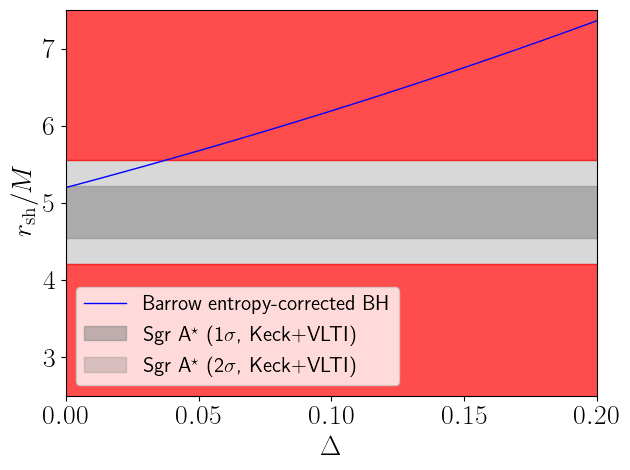}
\caption{Same as in Fig.~\ref{fig:shadow_reissner_nordstrom_bh_ns} for the Barrow entropy-corrected BH with metric function given by Eq.~\eqref{eq:metricbarrow}, as discussed in Sec.~\ref{subsec:barrow}.}
\label{fig:shadow_barrow_bh}
\end{figure}

We show the evolution of the shadow size as a function of $\Delta$ in Fig.~\ref{fig:shadow_barrow_bh}. With the same caveats already discussed previously for MOG in Sec.~\ref{subsec:modifiedgravity}, as well as GUP and EUP scenarios in Sec.~\ref{subsec:modifieduncertaintyprinciples} (see also Ref.~\cite{DiGennaro:2022grw} for a different point of view), we clearly see that the EHT observations set extremely strong constraints on the parameter $\Delta$. In particular, we find the upper limit $\Delta \lesssim 0.001$ at $1\sigma$, tighter than cosmological constraints requiring $\Delta \lesssim 0.1888$~\cite{Anagnostopoulos:2020ctz,Leon:2021wyx}, but weaker than Big Bang Nucleosynthesis constraints requiring $\Delta \lesssim 1.4 \times 10^{-4}$~\cite{Barrow:2020kug}. At $2\sigma$, we have the weaker constraint $\Delta \lesssim 0.035$, which is nonetheless still stronger than the cosmological one. One particularly interesting consequence of these limits is that the behavior of a holographic DE component based on Barrow entropy is prevented from deviating significantly from the cosmological constant $\Lambda$, particularly in the phantom regime, and for this reason we expect that Barrow entropy cannot play a significant role in the context of the Hubble tension~\cite{Knox:2019rjx,DiValentino:2021izs,Perivolaropoulos:2021jda,Schoneberg:2021qvd,Abdalla:2022yfr}: see e.g.\ Refs.~\cite{Vagnozzi:2018jhn,Yang:2018euj,El-Zant:2018bsc,Li:2019yem,Vagnozzi:2019ezj,Visinelli:2019qqu,DiValentino:2019jae,Zumalacarregui:2020cjh,Alestas:2020mvb,DiValentino:2020naf,Yang:2021hxg,Kumar:2021eev,Bag:2021cqm,Roy:2022fif,Sharma:2022ifr} for relevant discussions in relation to phantom dark energy. This highlights an interesting complementarity between BH shadows and cosmology, already hinted to in some of our earlier results, and which we plan to explore more in future works.

\subsection{Other models and space-times}
\label{subsec:others}

Although in this work we have covered a large number of scenarios, our limited resources meant that we were inevitably forced to make a choice as to which models to study, leaving out an even larger number of potentially interesting ones. Here, we very briefly comment on a few interesting models which have been left out.

\subsubsection{Vector-tensor Horndeski gravity}
\label{subsubsec:vectortensorhorndeski}

While Horndeski's scalar-tensor theory~\cite{Horndeski:1974wa} is probably among the best known modified gravity frameworks, much less known is the fact that Horndeski also wrote down the most general 4D vector-tensor theory with second order equations of motion~\cite{Horndeski:1976gi}. While not sharing the same success as its scalar-tensor counterpart, the theory has started re-gaining interest recently (see e.g.\ Refs.~\cite{Barrow:2012ay,BeltranJimenez:2013btb,Heisenberg:2014rta,Davydov:2015epx,DeFelice:2016cri,DeFelice:2016uil,Heisenberg:2016eld,BeltranJimenez:2017cbn,Heisenberg:2017hwb,Nakamura:2017lsf,Davydov:2017kxz,Brihaye:2020oxh,GallegoCadavid:2020dho,Verbin:2020fzk,Brihaye:2021ich,Gurses:2021ogc,Garnica:2021fuu,Brihaye:2021qvc,GallegoCadavid:2021ljh,GallegoCadavid:2022uzn}). The theory allows for the coupling of vector fields to gravity through their field-strength tensors, and this can be done for \textit{any} vector field, including the gauge fields of the Standard Model. With $F^{\mu\nu}$ (possibly carrying additional gauge and/or family indices) denoting the field-strength tensor of a gauge field, the gauge-gravity coupling is of the form:
\begin{eqnarray}
{\cal L} \supset \frac{\lambda}{4M_{\rm Pl}^2} \left ( RF_{\mu\nu}F^{\mu\nu}-4R_{\mu\nu}F^{\mu\alpha}F_{\alpha}^{\nu}+R_{\mu\nu\alpha\beta}F^{\mu\nu}F^{\alpha\beta} \right ) \,, \nonumber \\
\label{eq:vectortensorhorndeski}
\end{eqnarray}
with $\lambda$ the dimensionless gauge-gravity coupling, which of course can take different values for each gauge field.

In Ref.~\cite{Allahyari:2020jkn}, one of us derived upper limits on $\lambda$ by estimating the order-of-magnitude of the gauge-gravity coupling as $\sim \lambda{\cal R}{\cal F}^2/M_{\rm Pl}^2$, with ${\cal R}$ and ${\cal F}$ the typical curvature and field-strength tensor strengths respectively, and requiring that the Horndeski interaction term be weaker than both the Einstein-Hilbert and Maxwell Lagrangian terms, in order for the associated effects to be subdominant. This leads to upper limits of the order of:
\begin{eqnarray}
\vert \lambda \vert \lesssim \left ( \frac{\ell_{\cal R}}{\ell_{\rm Pl}} \right ) ^2\,,
\label{eq:upperlimitsvectortensorhorndeski}
\end{eqnarray}
where $\ell_{\cal R}$ is the length scale of the system under consideration. For instance, in the case of BH shadows one expects the correction to the shadow size due to the gauge-gravity interaction to be of the order of:
\begin{eqnarray}
\frac{\delta r_{\rm sh}}{r_{\rm sh}} \sim \lambda \left ( \frac{\ell_{\rm Pl}}{\ell_{\cal R}} \right ) ^2\,,
\label{eq:correctionshadow}
\end{eqnarray}
where $\ell_{\cal R} \sim M_{\rm bh}/M_{\rm Pl}^2$, with $M_{\rm bh}$ the BH mass. Requiring that the fractional correction to the shadow size be much smaller than unity results in the upper limit:
\begin{eqnarray}
\vert \lambda \vert \lesssim \left ( \frac{M_{\rm bh}}{M_{\rm Pl}} \right ) ^2\,.
\label{eq:upperlimitvectortensorhorndeskibhshadow}
\end{eqnarray}
For M87$^*$, Eq.~(\ref{eq:upperlimitvectortensorhorndeskibhshadow}) gives the limit $\vert \lambda \vert \lesssim 10^{88}$. Of course, with Sgr A$^*$ we can improve this limit by about 6 orders of magnitude as a direct consequence of the fact that we are probing a regime of much higher curvature (see Fig.~\ref{fig:potential_curvature_plot}), obtaining $\vert \lambda \vert \lesssim 10^{82}$. This is still 7 order of magnitude short of the strongest constraint obtained in Ref.~\cite{Allahyari:2020jkn} from neutron stars, and is overall weak in a broad sense. It is plausible that cosmological or GW constraints on this theory may yield much stronger bounds but such constraints, to the best of our knowledge, are not yet available.

\subsubsection{Bumblebee gravity}
\label{subsubsec:bumblebee}

Bumblebee gravity, first proposed by Kosteleck\'{y} and Samuel in Ref.~\cite{Kostelecky:1988zi} (see also Refs.~\cite{Bluhm:2004ep,Bluhm:2008yt}), is a class of effective field theories of gravity exhibiting spontaneous Lorentz symmetry breaking driven by the dynamics of a vector field acquiring a VEV. As such, the model constitutes an interesting playground for studying effects associated to spontaneous breaking of Lorentz symmetry, and has been the subject of a large number of studies (see e.g.\ Refs.~\cite{Bertolami:2005bh,Seifert:2009gi,Maluf:2014dpa,Santos:2014nxm,Hernaski:2014jsa,Capelo:2015ipa,Maluf:2015hda,Ovgun:2018xys,Ovgun:2018ran,Gomes:2018oyd,Oliveira:2018oha,Assuncao:2019azw,Kanzi:2019gtu,Li:2020dln,Maluf:2020kgf,Jha:2020pvk,Gullu:2020qzu,KumarJha:2020ivj,Ding:2021iwv,Khodadi:2021owg,Jha:2021eww,Maluf:2021lwh,Jiang:2021whw,Wang:2021gtd,Khodadi:2022dff,Gogoi:2022wyv,Gu:2022grg,Uniyal:2022xnq}). An exact non-rotating BH solution in bumblebee gravity was constructed in Ref.~\cite{Casana:2017jkc}, and is described by a single additional parameter $\ell$, which is related to the VEV of the bumblebee field, with the full metric given by:
\begin{eqnarray}
{\rm d}s^2 = - \left ( 1 - \frac{2M}{r} \right ) {\rm d}t^2+ \left ( 1+\ell \right ) \frac{{\rm d}r^2}{1-\frac{2M}{r}}+r^2{\rm d}\Omega^2 \,,\nonumber \\
\label{eq:metricbumblebee}
\end{eqnarray}
However, in this metric the only modifications to the Schwarzschild space-time appear in the $g_{rr}$ term, and do not therefore change the size of the shadow, which remains $3\sqrt{3}M$. Therefore, for the purposes of the metric test presented in this work, BHs in bumblebee gravity are not distinguishable from a Schwarzschild BH with the same mass. A rotating version of this solution was derived in Ref.~\cite{Ding:2019mal} (see also Ref.~\cite{Ding:2020kfr}): for what concerns the shadow size computation, this BH has the same form of the Kerr metric, with an effective spin given by $a_{\star}\sqrt{1+\ell}$, and therefore can only be sensibly constrained if one has an independent spin measurement. We note, however, that the correctness of this BH solution was criticized in Ref.~\cite{Maluf:2022knd} (see also Refs.~\cite{Kanzi:2022vhp,Kuang:2022xjp}).

\subsubsection{Dynamical Chern-Simons gravity}
\label{subsubsec:chernsimons}

Dynamical Chern-Simons gravity modifies the Einstein-Hilbert action by adding a scalar field $\vartheta$ with a standard kinetic term and a potential, and coupling this field to the Pontryagin density obtained by contracting the Riemann tensor with its dual, as follows~\cite{Jackiw:2003pm,Alexander:2009tp}:
\begin{eqnarray}
{\cal L} \supset \frac{\alpha}{4}\vartheta R_{\mu\nu\rho\sigma} {^{*}R^{\mu\nu\rho\sigma}}\,,
\label{eq:lagrangianchernsimons}
\end{eqnarray}
where the dual of the Riemann tensor is given by:
\begin{eqnarray}
^{*}R^{\mu\nu\rho\sigma} = \frac{1}{2}\epsilon^{\rho\sigma\alpha\beta}R^{\mu\nu}_{\alpha\beta}\,.
\label{eq:dual}
\end{eqnarray}
Such a coupling enjoys strong motivation from various theoretical scenarios, including string theory, LQG, effective field theories of inflation, and so on, and has been extensively studied in the literature~\cite{Achucarro:1986uwr,Alexander:2004xd,Weinberg:2008hq,Taveras:2008yf,Calcagni:2009xz,Bartolo:2017szm,Visinelli:2018utg,Bartolo:2018elp,Nojiri:2019nar,Nojiri:2020pqr,Mirzagholi:2020irt,Odintsov:2022hxu,Hou:2022wfj,Philcox:2022hkh,Lambiase:2022ucu}.

As shown in Ref.~\cite{Jackiw:2003pm}, non-rotating BHs in dynamical Chern-Simons gravity are still described by the Schwarzschild space-time. Physically speaking, this is a consequence of the fact that the stationary Cotton tensor vanishes on a spherically symmetric background. Therefore, at the level of non-rotating solutions as considered in the present work, the EHT observations cannot be used to test dynamical Chern-Simons gravity. Rotating BH solutions in this theory have been considered in various works. While exact solutions valid for all values of the spin are not known, a number of slowly-rotating solutions have been derived in the literature~\cite{Grumiller:2007rv,Yunes:2009hc,Ali-Haimoud:2011zme,Yagi:2012ya,Maselli:2017kic}.

\subsubsection{Casimir wormhole}
\label{subsubsec:casimirwh}

The Casimir effect, i.e.\ the force exerted between two plane parallel, close, uncharged, metallic plates in vacuum, was predicted theoretically almost a century ago~\cite{Casimir:1948dh}, and later verified experimentally. The associated Casimir energy has long been known to be a potential candidate source to generate and support a traversable WH (see e.g.\ Refs.~\cite{Morris:1988tu,Visser:1995cc}, as well as Refs.~\cite{Khabibullin:2005ad,Butcher:2014lea,Tripathy:2020ehi,Santos:2020taq,Garattini:2020kqb,Javed:2020mjb,Garattini:2021kca,Samart:2021tvl,Sokoliuk:2022jcq,Hassan:2022hcb} for later works). In Ref.~\cite{Garattini:2019ivd}, Garattini explicitly constructed various interesting traversable WH solutions supported by Casimir energy. One example is described by the following metric:
\begin{eqnarray}
{\rm d}s^2 = - \left ( \frac{3r}{3r+a} \right ) ^2 {\rm d}t^2+\frac{{\rm d}r^2}{1-\frac{2a}{3r}+\frac{a^2}{3r^2}}+r^2{\rm d}\Omega^2 \,, \nonumber \\
\label{eq:casimirwh}
\end{eqnarray}
where $a$ is the throat radius. From the above we can immediately read off that the WH's ADM mass is $M=3a$, so that the metric function can be rewritten as:
\begin{eqnarray}
A(r) = \left ( \frac{r}{r+M} \right ) ^2\,.
\label{eq:metriccasimirwh}
\end{eqnarray}
It is easy to see that the above metric function is identical to the metric function of the null naked singularity considered in Sec.~\ref{subsec:nullns}. The key difference between the two, rendering one a WH and the other a NS, lies in the $g_{rr}$ metric element. However, this does not impact the computation of the photon sphere and associated shadow size. Just as for the null NS, we find that the Casimir WH does not have a photon sphere (formally the solution to the equation determining the photon sphere is $r_{\rm ph}=0$). We would nonetheless expect it to cast a shadow of size $r_{\rm sh}=M$, which is immediately ruled out at very high significance by the EHT observations. However, a definitive answer as to whether or not the Casimir WH casts a shadow would require a light trajectory analysis similar to the one performed by some of us in Ref.~\cite{Joshi:2020tlq} for the null NS (see Fig.~1f thereof), given the presence of the WH throat at $r=M/3$, which may prevent the results of Ref.~\cite{Joshi:2020tlq} from fully carrying over here. We defer this analysis to a future work.

\subsubsection{T-duality}
\label{subsubsec:tduality}

In string theory, T-duality is an important symmetry relating the physical properties of two otherwise unrelated backgrounds (loosely speaking, one at large radii and the other one at small radii), which however behave identically as far as physical observables are concerned~\cite{Green:1982sw}. However, T-duality is not an exclusive property of string theory, and has been observed to emerge quite generally in a number of other systems~\cite{Padmanabhan:1996ap}, wherein the path integral is invariant under exchange of the infinitesimal path length ${\cal D}x \to 1/{\cal D}x$. Recently, an important application of T-duality consisted in the formulation of a finite electrodynamics theory driven by a T-dual photon propagator~\cite{Gaete:2022une,Mondal:2021izm}, which naturally incorporates a minimum length $\ell_0$. String-inspired T-duality corrections to BH space-times have been studied in Ref.~\cite{Nicolini:2019irw}, and subsequently in Ref.~\cite{Gaete:2022ukm} by one of us. There, it was found that the resulting BH solution is identical to the Bardeen regular BH considered in Sec.~\ref{subsec:bardeen}, with the minimum length $\ell_0$ playing the role of Bardeen charge $q_m$ in Eq.~(\ref{eq:metricbardeen}). Therefore, we can automatically translate the results obtained in Sec.~\ref{subsec:bardeen} to the T-duality scenario, finding that the EHT observations are unable to place meaningful constraints on the minimum length $\ell_0$, at least based on the metric test considered here.

\section{Brief discussion}
\label{sec:briefdiscussion}

\begin{table*}[!t]
\centering
\scalebox{0.75}{
\begin{tabular}{|c||c|c|c|c|}       
\hline\hline
Model & Section & References & $A(r)$, $C(r)$ & Constraints ($1\sigma$, $2\sigma$) \\ \hline
Reissner-Nordstr\"{o}m BH & Sec.~\ref{subsec:rn} & \cite{1916AnP...355..106R,1917AnP...359..117W,1918KNAB...20.1238N,1921RSPSA..99..123J} & Eq.~(\ref{eq:metricrn}), $Q/M \leq 1$ & $Q \lesssim 0.8M,\,0.95M$ \\
Reissner-Nordstr\"{o}m NS & Sec.~\ref{subsec:rn} & \cite{1916AnP...355..106R,1917AnP...359..117W,1918KNAB...20.1238N,1921RSPSA..99..123J} & Eq.~(\ref{eq:metricrn}), $1 < Q/M \leq \sqrt{9/8}$ & Excluded \\
Bardeen BH & Sec.~\ref{subsec:bardeen} & \cite{Bardeen:1968ghw} & Eq.~(\ref{eq:metricbardeen}) & $q_m$ unconstrained (consistent) \\
Hayward BH & Sec.~\ref{subsec:hayward} & \cite{Hayward:2005gi} & Eq.~(\ref{eq:metrichayward}) & $\ell$ unconstrained (consistent) \\
Frolov BH & Sec.~\ref{subsec:frolov} & \cite{Frolov:2016pav} & Eq.~(\ref{eq:metricfrolov}), $\ell=0.3$ & $q \lesssim 0.8M,\,0.9M$ \\
Einstein-Bronnikov BH & Sec.~\ref{subsec:eb} & \cite{Bronnikov:2000vy} & Eq.~(\ref{eq:metriceb}) & $q_m \lesssim 0.8M,\,M$ \\
GCSV BH & Sec.~\ref{subsec:gcsv} & \cite{Ghosh:2014pba,Culetu:2014lca,Simpson:2019mud} & Eq.~(\ref{eq:metricgcsv}) & $g \lesssim 0.8M,\,M$ \\
Kazakov-Solodukhin BH & Sec.~\ref{subsec:ks} & \cite{Kazakov:1993ha} & Eq.~(\ref{eq:metricks}) & $\ell \lesssim 0.2M,\,M$ \\
Ghosh-Kumar BH & Sec.~\ref{subsec:ghoshkumar} & \cite{Ghosh:2021clx} & Eq.~(\ref{eq:metricghoshkumar}) & $k \lesssim 1.4M,\,1.6M$ \\
Simpson-Visser BH & Sec.~\ref{subsec:simpsonvisser} & \cite{Simpson:2018tsi} & Eq.~(\ref{eq:metricsimpsonvisser}), $a/M<2$ & $a$ unconstrained (consistent) \\
Simpson-Visser WH & Sec.~\ref{subsec:simpsonvisser} & \cite{Simpson:2018tsi} & Eq.~(\ref{eq:metricsimpsonvisser}), $2 \leq a/M<3$ & $a$ unconstrained (consistent) \\
Morris-Thorne WH & Sec.~\ref{subsec:morristhorne} & \cite{Morris:1988cz} & Eqs.~(\ref{eq:metricmorristhorne},\ref{eq:metrictraversablewormhole}) & Excluded \\
Damour-Solodukhin WH & Sec.~\ref{subsec:damoursolodukhin} & \cite{Damour:2007ap} & Eq.~(\ref{eq:metricdamoursolodukhin}) & $\lambda \lesssim 0.4,\,0.5$ \\
Janis-Newman-Winicour NS & Sec.~\ref{subsec:jnw} & \cite{Janis:1968zz} & Eq.~(\ref{eq:metricjnw}) & $\overline{\nu} \lesssim 0.4M,\,0.45M$ \\
Joshi-Malafarina-Narayan-1 NS & Sec.~\ref{subsec:jmn1} & \cite{Joshi:2011zm} & Eq.~(\ref{eq:metricjmn1interior}), $2/3 \leq \chi < 1$ & $\chi$ unconstrained (consistent) \\
Thin shell matter NS & Sec.~\ref{subsec:tsmns} & \cite{Dey:2020haf} & Eqs.~(\ref{eq:metricjmn1i},\ref{eq:metricjmn1e}), $\chi_e=0.6$, $R_{b2} \approx 0.33$ & $R_{b1} \gtrsim 1.5M,\,2M$ \\
Null NS & Sec.~\ref{subsec:nullns} & \cite{Joshi:2020tlq} & Eq.~(\ref{eq:metricnullns}) & Excluded \\
Minimally coupled scalar with potential & Sec.~\ref{subsec:minimallycoupledscalar} & \cite{Gonzalez:2013aca,Gonzalez:2014tga,Khodadi:2020jij} & Eq.~(\ref{eq:metricmch}) & $\nu \lesssim 0.01M,\,0.4M$ \\
BH with conformal scalar hair & Sec.~\ref{subsec:conformallycoupledscalar} & \cite{Astorino:2013sfa,Khodadi:2020jij} & Eq.~(\ref{eq:metriccc}), $0 \leq S<M^2$ & $S \lesssim 0.65M^2,\,0.9M^2$ \\
WH with conformal scalar hair & Sec.~\ref{subsec:conformallycoupledscalar} & \cite{Astorino:2013sfa,Khodadi:2020jij} & Eq.~(\ref{eq:metriccc}), $S<0$ & $S \gtrsim -0.04M^2,\,-0.4M^2$ \\
Clifton-Barrow $f(R)$ gravity & Sec.~\ref{subsec:cliftonbarrow} & \cite{Clifton:2005aj} & Eq.~(\ref{eq:metriccliftonbarrow}) & $\delta \gtrsim -0.15,\,-0.20$ \\
BH in Horndeski gravity (case 1) & Sec.~\ref{subsubsec:horndeskicase1} & \cite{Babichev:2017guv} & Eq.~(\ref{eq:metrichorndeskicase1peff}), $-M^2/8\pi<p_{\rm eff} \leq 0$ & $p_{\rm eff} \gtrsim -0.005M^2,\,-0.01M^2$ \\
WH in Horndeski gravity (case 1) & Sec.~\ref{subsubsec:horndeskicase1} & \cite{Babichev:2017guv} & Eq.~(\ref{eq:metrichorndeskicase1peff}), $p_{\rm eff}>0$ & $p_{\rm eff} \lesssim 0.002M^2,\,0.01M^2$ \\
BH in Horndeski gravity (case 2) & Sec.~\ref{subsubsec:horndeskicase2} & \cite{Bergliaffa:2021diw} & Eq.~(\ref{eq:metrichorndeskicase2}) & $-0.01,\,-0.15 \lesssim h \lesssim 0.40,\,0.60$ \\
MOdified Gravity & Sec.~\ref{subsec:modifiedgravity} & \cite{Moffat:2005si} & Eq.~(\ref{eq:metricmodifiedgravity}) & $\alpha \lesssim 0.01,\,0.1$ \\
Brane-world BH & Sec.~\ref{subsec:rsii} & \cite{Dadhich:2000am} & Eq.~(\ref{eq:metricrsii}), $q \geq 0$ & $q \lesssim 0.15,\,0.2$ \\
Brane-world WH & Sec.~\ref{subsec:rsii} & \cite{Dadhich:2000am} & Eq.~(\ref{eq:metricrsii}), $q<0$ & $q \gtrsim -0.01,\,-0.1$ \\
Einstein-Euler-Heisenberg BH & Sec.~\ref{subsec:eeh} & \cite{Allahyari:2019jqz} & Eq.~(\ref{eq:metriceeh}), $\mu=0.3$, $q_m<1$ & $q_m \lesssim 0.7M,\,0.8M$ \\
Sen BH & Sec.~\ref{subsec:sen} & \cite{Sen:1992ua} & Eq.~(\ref{eq:metricsen}), $q_m \leq \sqrt{2}$ & $q_m \lesssim 0.6M,\,0.75M$ \\
Kalb-Ramond BH & Sec.~\ref{subsec:kalbramond} & \cite{Lessa:2019bgi} & Eq.~(\ref{eq:metrickalbramond}), $\Gamma=0.5$ & $k \lesssim 1.2,\,1.5$ \\
Einstein-Maxwell-dilaton-1 BH & Sec.~\ref{subsec:emd1} & \cite{Gibbons:1987ps,Garfinkle:1990qj,Garcia:1995qz} & Eq.~(\ref{eq:metricemd1}), $q \leq \sqrt{2}M$ & $q \lesssim 0.8M,\,M$ \\
Einstein-\ae ther type 1 BH & Sec.~\ref{subsec:einsteinaether} & \cite{Eling:2006ec,Barausse:2011pu} & Eq.~(\ref{eq:metriceinsteinaether1}) & $c_{13} \lesssim 0.1,\,0.75$ \\
Einstein-\ae ther type 2 BH & Sec.~\ref{subsec:einsteinaether} & \cite{Eling:2006ec,Barausse:2011pu} & Eq.~(\ref{eq:metriceinsteinaether2}), $c_{13}=0.99$ & $1.3 \lesssim c_{14} \lesssim 1.4$ \\
4D Einstein-Gauss-Bonnet gravity & Sec.~\ref{subsec:4degb} & \cite{Glavan:2019inb} & Eq.~(\ref{eq:metric4degb}), $\alpha \leq M^2/16\pi$ & $\alpha$ unconstrained (consistent) \\
Asymptotically safe gravity (IR limit) & Sec.~\ref{subsec:asg} & \cite{Bonanno:2000ep} & Eq.~(\ref{eq:metricasg}), $\widetilde{\omega}<M^2$ & $\widetilde{\omega} \lesssim 0.9M^2$, unconstrained (consistent) \\
Rastall gravity & Sec.~\ref{subsec:rastall} & \cite{Heydarzade:2017wxu} & Eq.~(\ref{eq:metricrastall}), $N_s=0.005$, $\omega=0$, $\kappa \leq 1/6$ & $\kappa$ unconstrained (consistent) \\
Loop quantum gravity & Sec.~\ref{subsec:lqg} & \cite{Modesto:2008im} & Eq.~(\ref{eq:metriclqg}), $a_0=0$ & $P \lesssim 0.05,\,0.08$ \\
Kottler BH & Sec.~\ref{subsec:kottler} & \cite{Kottler:1918ghw} & Eq.~(\ref{eq:metrickottler}), $0<\Lambda<\sqrt{3M}$ & $\log_{10}(\Lambda [{\rm m}^{-2}] \lesssim -41.0,\,-40.5$ \\
Electromagnetic-Weyl coupling (PPL) & Sec.~\ref{subsec:weylcorrected} & \cite{Chen:2015cpa} & Eq.~(\ref{eq:metricweylcorrected}) & $\alpha \lesssim 0.01M^2,\,0.2M^2$ \\
Electromagnetic-Weyl coupling (PPM) & Sec.~\ref{subsec:weylcorrected} & \cite{Chen:2015cpa} & Eq.~(\ref{eq:metricweylcorrected}) & $\alpha \lesssim 0.25M^2,\,0.3M^2$ \\
DST BH & Sec.~\ref{subsec:dst} & \cite{Deser:2007za} & Eq.~(\ref{eq:metricdst}) & $\sigma \lesssim 0.04,\,0.05$ \\
Rindler gravity & Sec.~\ref{subsec:rindler} & \cite{Grumiller:2010bz} & Eq.~(\ref{eq:metricrindler}) & $\log_{10}(a[{\rm pc}^{-1}]) \lesssim -8,\,-7$ \\
BH surrounded by PFDM & Sec.~\ref{subsec:pfdm} & \cite{Li:2012zx} & Eq.~(\ref{eq:metricpfdm}) & $k \lesssim 0.07M,\,0.15M$ \\
BH with a topological defect & Sec.~\ref{subsec:topologicaldefect} & \cite{Barriola:1989hx,Dadhich:1997mh} & Eq.~(\ref{eq:metrictopologicaldefect}) & $k \lesssim 0.005,\,0.1$ \\
Hairy BH from gravitational decoupling & Sec.~\ref{subsec:hgd} & \cite{Ovalle:2020kpd} & Eq.~(\ref{eq:metrichgd}), $\alpha=1.0$ & $\ell \gtrsim 0.35M,\,0.15M$ \\
GUP BH & Sec.~\ref{subsubsec:gup} & \cite{Carr:2015nqa,Vagenas:2017vsw,Carr:2020hiz} & Eq.~(\ref{eq:metricgup}) & $\beta \gtrsim 0.01M^2,\,0.14M^2$ \\
EUP BH & Sec.~\ref{subsubsec:eup} & \cite{Mureika:2018gxl} & Eq.~(\ref{eq:metriceup}) & $\log_{10}(L_{\star}[{\rm pc}]/\sqrt{\alpha}) \gtrsim -5,\,-4$ \\
Non-commutative geometry (Gaussian) & Sec.~\ref{subsubsec:ncggaussian} & \cite{Nicolini:2005vd} & Eq.~(\ref{eq:metricncggaussian}), $\vartheta \leq 0.5252$ & $\vartheta$ unconstrained (consistent) \\
Non-commutative geometry (Lorentzian) & Sec.~\ref{subsubsec:ncglorentzian} & \cite{Anacleto:2019tdj} & Eq.~(\ref{eq:metricncglorentzian}) & $\vartheta \lesssim 0.02M^2,\,0.04M^2$ \\
Barrow entropy & Sec.~\ref{subsec:barrow} & \cite{Jusufi:2021fek} & Eq.~(\ref{eq:metricbarrow}) & $\Delta \lesssim 0.001,\,0.035$ \\ \thickhline
\textit{Vector-tensor Horndeski gravity} & Sec.~\ref{subsubsec:vectortensorhorndeski} & \cite{Horndeski:1976gi,Allahyari:2020jkn} & -- & $\vert \lambda \vert \lesssim 10^{82}$? \\
\textit{Bumblebee gravity} & Sec.~\ref{subsubsec:bumblebee} & \cite{Ding:2019mal,Ding:2020kfr,Maluf:2022knd} & -- & $\ell$ unconstrained (consistent)? \\
\textit{Dynamical Chern-Simons gravity} & Sec.~\ref{subsubsec:chernsimons} & \cite{Jackiw:2003pm} & -- & $\alpha$ unconstrained (consistent)? \\
\textit{Casimir WH} & Sec.~\ref{subsubsec:casimirwh} & \cite{Garattini:2019ivd} & Eq.~(\ref{eq:metriccasimirwh}) & Excluded? \\
\textit{String-inspired T-duality-corrected BH} & Sec.~\ref{subsubsec:tduality} & \cite{Gaete:2022ukm} & -- & $\ell_0$ unconstrained (consistent)? \\
\hline \hline                                                  
\end{tabular}}
\caption{Summary of our results. The first four columns indicate respectively the model's name (``BH'', ``WH'', and ``NS'' stand for ``black hole'', ``wormhole'', and ``naked singularity''), the corresponding Section,  the main references for the corresponding solution (these are not necessarily the references where the underlying model was first proposed or discussed), and the equation(s) where the metric function(s) are presented, alongside further restrictions on the values of the hair parameter(s). The final column summarizes the $1\sigma$ and $2\sigma$ constraints on the hair parameters: ``unconstrained (consistent)'' indicates that the parameter is unconstrained because for all parameter values the shadow size is consistent with the EHT observations (for asymptotically safe gravity this occurs only for the $2\sigma$ constraints), whereas ``excluded'' indicates that the model is excluded for all parameter values. The final five rows are separated from the previous ones and formatted differently to indicate that these models were only briefly discussed in Sec.~\ref{subsec:others}.}
\label{tab:summarytable}                                              
\end{table*}

Having tested a wide variety of models and fundamental physics scenarios, our goal here is to provide a very brief bird's-eye overall discussion of our results, while also providing an outlook on complementary constraints and future prospects. We note that a much more complete discussion, at least as far as complementary constraints from additional probes is concerned, is provided in Sec.~6 of Ref.~\cite{EventHorizonTelescope:2022xqj}, to which we refer the reader who wishes to examine the topic more in depth. Several of the scenarios we have considered featured dimensionful parameters, either tied to fundamental Lagrangian parameters, appearing as integration constants (possibly tied to conserved charges/hair parameters), or introduced on phenomenological grounds. For these parameters, the constraints we have obtained (upper or lower limits) are typically of order ${\cal O}(M)$, with $M$ Sgr A$^*$'s mass. This is not unexpected, given the fact that BH shadow imaging probes horizon scales of order $M$, that the EHT resolution is comparable to Sgr A$*$'s gravitational radius, and also given the results of related earlier works based on M87$^*$'s shadow (whose constraints we have nonetheless improved typically by at least 3 orders of magnitude). Where available, we have reported constraints on the same scenarios based on complementary probes such as cosmology, GWs, and X-ray reflection spectroscopy to mention a few: typically, these constraints are much stronger than the ones we have obtained from the shadow of Sgr A$^*$.

The above points clearly raise the question of whether there is value in performing shadow-based tests of fundamental physics, if the resulting limits are so weak. We believe the answer is a resounding yes! Although in Fig.~\ref{fig:potential_curvature_plot} we have only shown a small region of the Baker-Psaltis-Skordis potential-curvature plot, it is clear that BH shadows probe a regime which cannot otherwise be tested by other observations: therefore, regardless of the weakness of the resulting limits, verifying that GR holds (or at the very least is not in strong tension with observations) in this region of parameter space \textit{as well as within other widely distant regimes} provides a very non-trivial consistency check/null test of the theory. This point was strongly emphasized in Ref.~\cite{EventHorizonTelescope:2022xqj}, to which we refer the reader for further discussions (see in particular Fig.~23 thereof). Nonetheless, we have also noted that for the case of dimensionless parameters we have been able to obtain comparatively strong constraints, in some cases competitive with or even stronger than cosmological ones, particularly within scenarios where the extra parameter makes the BH shadow larger.

The above discussion, however, raises another question: can one truly straightforwardly compare constraints based on BH shadows to those obtained from these other probes? This is not a simple question. If one has a specific model in mind, then this is in principle possible, provided one is able to exactly compute the signature expected in these complementary probes, within a given model, from first principles. However, this is not always simple: in fact, in most cases, this is not even possible. To model cosmological observations, leaving aside the spiny issue of non-linearities, one needs to solve the full Einstein-Boltzmann system, whose derivation is in general not simple for an arbitrary fundamental model.~\footnote{In fact, one often works with parametrizations which capture the effect of new physics in a model-agnostic way, but are not necessarily tied to any fundamental model.} For GWs, if the field equations cannot be solved exactly, in general one can only obtain conservative limits on new physics provided the BH metric is known. Finally, for X-ray observations, one needs to known the corresponding rotating BH solution, which again is not always known or even easy to obtain.

Even if we leave aside the above problems, there remains the issue that several tests of gravity are not based on specific models, but rather on model-agnostic parametrized tests, each based on different frameworks and optimal only for the particular systems studied: among the many examples, we mention the PPN expansion~\cite{Will:2001mx}, post-Keplerian parametrizations~\cite{Wex:2020ald}, parametrized post-Einstein frameworks~\cite{Yunes:2009ke}, effective-one-body frameworks~\cite{Buonanno:1998gg}, post-Newtonian frameworks~\cite{Khan:2015jqa}, and so on. The differences between these types of tests do not allow for a simple cross-comparison (if such a comparison is possible at all) with the results obtained from BH shadows, particularly without reference to any specific fundamental model. However, as emphasized in Sec.~6 of Ref.~\cite{EventHorizonTelescope:2021dqv} (to which we once more encourage the reader to refer to), the wide range of assumptions and conditions involved allow at the very least for important null tests/consistency checks of GR. In fact, if no deviation from GR is observed in \textit{any} of these probes, barring extremely unlikely fine-tuning or fortuitous cancellations, large deviations from the Kerr-Newman family of metrics are very hard to accommodate. Of course, going from such a consistency check to constraints on specific theories remains difficult, but there is no doubt that such consistency checks are highly valuable in their own right.

A related concern was raised in Ref.~\cite{Glampedakis:2021oie}, where it was pointed out that the ${\cal O}(M)$ upper limits on fundamental physics obtained from BH shadows, when extrapolated to the potential-curvature regime probed by GW observations, would allow for extremely large deviations in the latter, which are not observed. This is simply another way of stating the fact that GW constraints are in principle much tighter than BH shadows-based ones. However, most viable models beyond GR (including many of the ones we have considered) are equipped with screening mechanisms which allow them to pass Solar System tests of gravity~\cite{Khoury:2003aq,Sakstein:2018fwz,Brax:2021wcv}. The computation of the expected GW signal in the presence of such screening mechanisms can be significantly more complicated, and is an issue which has not yet been addressed in full generality (see e.g.\ Refs.~\cite{Barausse:2014tra,Honardoost:2019rha,terHaar:2020xxb,Bezares:2021yek,Bezares:2021dma} for important works).

Still along the above line, it is also worth noting that GWs are sensitive to the dynamics of the theory, and in particular to the propagating gravitational modes, whereas BH shadows provide a static test of \textit{photons} propagating along the BH background (see e.g.\ Refs.~\cite{Psaltis:2007cw,Barausse:2008xv,Gair:2011ym} for important discussions in this sense). This distinction can become particularly important in the presence of screening mechanisms, and further hinders a direct, straightforward comparison between constraints obtained from BH shadows and GWs or, for that matter, other dynamical probes of gravity in the strong-field regime, particularly if based on model-agnostic parametrizations which do not necessarily (if at all) take the effect of screening into account. These difficulties notwithstanding, a more detailed exploration of these points, and more generally of how to cross-compare and combine BH shadow tests with complementary tests, is clearly warranted, and we certainly plan to return to these issues in future works.

In closing, we briefly comment on how the constraints we have obtained can be improved by future probes. We expect improvements at the very least on two fronts: \textit{i)} improvements in the precision of the (already exquisitely precise) mass-to-distance ratio priors for Sgr A$^*$, and \textit{ii)} improvements in the angular resolution achievable by VLBI arrays. Regarding the first point, significant improvements can be expected from GRAVITY+, the upgrades to GRAVITY and the VLTI instrument, featuring wide-separation fringe tracking, new adaptive optics, and laser guide stars. These will enable all-sky, high contrast, $\mu{\rm as}$ interferometry, which can significantly improve the precision to which sources orbiting around Sgr A$^*$ can be tracked, thereby improving the determination of Sgr A$^*$'s mass. The first phase of GRAVITY+, named GRAVITY Wide, has just begun, and we can therefore expect more precise measurements in the near future~\cite{Abuter:2022ghw}. Further improvements can also be expected from MATISSE, a second generation mid-infrared imaging spectro-interferometer for VLTI, which has recently seen first light~\cite{Lopez:2021ghw}.

On the VLBI side, significant improvements can be expected from the next-generation EHT (ngEHT), the planned upgrade to the EHT array, which will double the number of array antennas and incorporate dual-frequency receivers, while more than doubling the sensitivity and dynamical range compared to the current EHT capabilities~\cite{Blackburn:2019bly}. The ngEHT is expected to begin the commissioning phase no earlier than 2027, so improvements in this sense may be expected in the next decade. Even more futuristic upgrades to the EHT involving space-based telescopes can further improve the achievable angular resolution by more than an order of magnitude, potentially reaching ${\cal O}(\mu{\rm as})$ resolutions~\cite{Fish:2019epg}. Looking even further into the future (most likely no earlier than 2060), X-ray interferometry (XRI) through constellations of satellites is an interesting possibility which may achieve sub-$\mu{\rm as}$ angular resolution~\cite{Uttley:2019ngm}. However, XRI is best suited for imaging SMBHs with optically thick disks, and the corresponding images would therefore not probe the photon sphere, but the inner edge of the accretion disk, whose location however is expected to depend more strongly on the BH spin, making it absolutely necessary to move beyond the spherically symmetric space-times considered here.

We expect that the improvements on the mass measurements and VLBI sides discussed above will allow for substantial improvements in the limits we have obtained, which should be strengthened by at least a couple of orders of magnitude. Of course, by the time these limits are obtained, we expect complementary limits (e.g.\ from GWs and cosmology) to have become tighter as well. Additional improvements on the BH shadow side can be achieved if a robust measurement of the spin of Sgr A$^*$ (or, for that matter, of other BHs being imaged) can be obtained. In this case, the use of techniques based on orbital angular momentum and twisted light in the vicinity of the BH appears promising, although for Sgr A$^*$ the use of these techniques may be challenging due to its very short dynamical timescale~\cite{Tamburini:2019vrf,Tamburini:2021jok,Tamburini:2021lyi}. In closing, we further note that VLBI images allow for a wide variety of tests of fundamental physics beyond those we have considered here, including but not limited to tests of the equivalence principle based on multi-frequency images~\cite{Li:2019lsm}, to probes of ultra-light particles and other aspects of fundamental physics based on shadow evolution~\cite{Roy:2019esk,Frion:2021jse,Roy:2021uye,Chen:2022nbb}, spin measurements~\cite{Davoudiasl:2019nlo}, and polarimetric measurements~\cite{Yuan:2020xui,Chen:2021lvo}, and various tests of fundamental physics based on the BH photon ring, its auto-correlation, and other probes~\cite{Himwich:2020msm,Hadar:2020fda,Chen:2022nbb,Tsupko:2022kwi,Chen:2022nlw,Zhu:2023omf,Wang:2023nwd}. Studies on these and related possibilities are underway, and we expect to report on these in future works.

\section{Conclusions}
\label{sec:conclusions}

Horizon-scale images of supermassive black holes and their shadows have opened a new unparalleled window onto tests of gravity and fundamental physics in the very strong-field regime, including the possibility that astrophysical BHs may be described by alternatives to the Kerr metric. In this work, we have used the horizon-scale images of Sgr A$^*$ provided by the Event Horizon Telescope~\cite{EventHorizonTelescope:2022xnr,EventHorizonTelescope:2022vjs,EventHorizonTelescope:2022wok,EventHorizonTelescope:2022exc,EventHorizonTelescope:2022urf,EventHorizonTelescope:2022xqj,EventHorizonTelescope:2022gsd,EventHorizonTelescope:2022ago,EventHorizonTelescope:2022okn,EventHorizonTelescope:2022tzy} to test some of the most popular and well-motivated scenarios deviating from the Kerr metric.~\footnote{See Ref.~\cite{Shahzadi:2022rzq} for a similar set of tests making use of flare data from Sgr A$^*$.} Compared to horizon-scale images of M87$^*$, there are significant advantages in the use of images of Sgr A$^*$, as we discussed towards the end of Sec.~\ref{sec:introduction}, with the most notable advantage being that Sgr A$^*$'s proximity to us results in an exquisite calibration of its mass-to-distance ratio, essential for connecting the angular size of its shadow to theoretical predictions within a given model. Our tests have been performed by connecting the angular size of the bright ring of emission to that of the underlying BH shadow, and utilizing prior information on Sgr A$^*$'s exquisitely calibrated mass-to-distance ratio: this is a robust and well-tested methodology, adopted for instance by the EHT collaboration themselves~\cite{EventHorizonTelescope:2021dqv,EventHorizonTelescope:2022xqj}. We also note that our tests are parametric, and are complementary to theory-agnostic tests which can simultaneously constrain several theories, and which have been considered elsewhere (see e.g.\ Refs.~\cite{Nampalliwar:2021oqr,Nampalliwar:2021ytm,Kocherlakota:2022jnz}).

We have studied over 50 well-motivated scenarios, including both fundamental theoretical scenarios and more phenomenological ones, considering in particular (see Tab.~\ref{tab:summarytable} for a summary of the models considered as well as the corresponding results): various regular BH space-times; string-inspired space-times; metrics arising within non-linear electrodynamics theories; space-times violating the no-hair theorem due to the presence of additional fields (including but not limited to scalar fields); alternative gravity theories; fundamental frameworks leading to new ingredients such as modified commutation or entropy relations; and finally BH mimickers including various wormhole and naked singularity solutions. For almost all of the space-times considered, we have found that the EHT observations set limits on fundamental dimensional parameters of  order ${\cal O}(0.1)$ times some power of mass $M$, which is not surprising considering that the EHT angular resolution is comparable to Sgr A$^*$'s gravitational radius, so constraints of this order should be expected. For dimensionless parameters, we have found constraints of the order or ${\cal O}(0.1)$ or stronger (in some cases surpassing complementary constraints from cosmology). Overall, we have shown that the EHT horizon-scale images of Sgr A$^*$ place particularly tight constraints on fundamental physics scenarios which predict shadows \textit{larger} than that of a Schwarzschild BH of the same mass, as the EHT observations on average prefer a shadow size which is slightly smaller than the latter, given the \textrm{Keck} and \textrm{VLTI} priors on the mass-to-distance ratio (this had also been noted in Ref.~\cite{EventHorizonTelescope:2022xqj}). Finally, we have shown that the metric tests considered in this work do not exclude the extremely intriguing possibility of Sgr A$^*$ being a BH mimicker, in the context of various wormhole and naked singularity space-times we have studied.

There is certainly ample opportunity for further work in this very promising direction. First of all, for obvious reasons we were forced to make a selection when considering which fundamental scenarios to study, and many interesting models were inevitably left out: we plan to return to these in future works. These scenarios could include viable models for the dark matter and dark energy permeating the Universe, as well as well-motivated alternatives to GR~\footnote{Concerning alternatives to GR, it could be particularly interesting to examine possible constraints on torsion and non-metricity, especially in light of the so-called ``geometrical trinity of gravity''~\cite{BeltranJimenez:2019esp}, which offers three complementary descriptions of gravity in terms of the effects of curvature, torsion, and non-metricity. Static spherically symmetric within teleparallel $f(T)$ gravity and non-metricity (symmetric teleparallel) $f(Q)$ gravity have been obtained in recent years, see e.g.\ Refs.~\cite{Nashed:2002tz,Capozziello:2012zj,Nashed:2018cth,Pfeifer:2021njm,Bahamonde:2021gfp,DAmbrosio:2021zpm,Bahamonde:2022lvh,Bahamonde:2022esv,Calza:2022mwt,Bahamonde:2022kwg}: we defer a dedicated study of torsion and non-metricity constraints from BH shadows to a follow-up work. Other solutions it would be interesting to study include those of Refs.~\cite{Dymnikova:1992ux,Anabalon:2012ih,Nashed:2019tuk,Dimakis:2020dqs}.}: in this sense, a detailed investigation of the synergy between BH shadows, GWs, cosmological and astrophysical probes, and laboratory tests of gravity and fundamental physics is certainly a direction worth pursuing (see e.g.\ Ref.~\cite{Baker:2014zba}), as one can expect a very strong complementarity across these probes, although with all the caveats discussed in Sec.~\ref{sec:briefdiscussion}. A future study could also include the effects of rotation on all the space-times considered.  In closing, BH shadows have opened an unprecedented window onto gravity and fundamental physics in the era of multi-messenger astrophysics, and we have only just begun exploring their potential (see e.g.\ Refs.~\cite{Islam:2022ybr,Chen:2022lct,Jusufi:2022loj,Wang:2022ivi,Uniyal:2022vdu,Carballo-Rubio:2022imz,Pantig:2022ely,Ghosh:2022kit,Kuang:2022ojj,Wang:2022fgj,Khodadi:2022pqh,Banerjee:2022iok,KumarWalia:2022aop,KumarWalia:2022ddq,Banerjee:2022bxg,Mustafa:2022xod,Banerjee:2022jog,Carballo-Rubio:2022aed,Shaikh:2022ivr,Pantig:2022qak,Saha:2022hcd,Pantig:2022gih,Zakharov:2022gwk,Atamurotov:2022nim,Oikonomou:2022tjm,Kumar:2022fqo,Sengo:2022jif,Ghosh:2022gka,Afrin:2022ztr,Kumar:2022vfg,Kumaran:2022soh,Pantig:2022sjb,Badia:2022phg,Hu:2022lek,Frizo:2022cfq,Wu:2022ydk,Antoniou:2022dre,Islam:2022wck,Atamurotov:2022knb,Sau:2022afl,Joshi:2022azj,Singh:2023zmy,Tan:2023ngk,Gomez:2023wei,Anjum:2023axh,Roder:2023oqa,Singh:2023ops,Xu:2023xqh,Nguyen:2023clb,Pantig:2023yer} for other works along similar lines, first appearing or updated around the same time or after our preprint had been posted to arXiv).

\section*{Note added}
We had independently started this work before the Event Horizon Telescope press conference announcing the new observations of Sgr A$^*$ on May 12, 2022. Some of the models we tested (specifically the Reissner-Nordstr\"{o}m, Bardeen, Hayward, Frolov, Janis-Newman-Winicour, Joshi-Malafarina-Narayan, Kazakov-Solodukhin, and Einstein-Maxwell-dilaton space-times) have also inevitably been tested by the EHT collaboration in Ref.~\cite{EventHorizonTelescope:2022xqj}. However, in this work we have considered other models and scenarios beyond those considered by the EHT collaboration. We have verified that, for the overlapping models, both results agree, therefore providing a valuable and independent cross-check.

\section*{Data availability}
No new data were created or analysed in this study.

\begin{acknowledgments}
\noindent We thank Richard Brito, Yifan Chen, and Naoki Tsukamoto for useful discussions. S.V.\ was partially supported by the Isaac Newton Trust and the Kavli Foundation through a Newton-Kavli Fellowship, and by a grant from the Foundation Blanceflor Boncompagni Ludovisi, n\'{e}e Bildt. R.R.\ is supported by the Shanghai Government Scholarship (SGS). Y.-D.T. is supported by U.S. National Science Foundation (NSF) Theoretical Physics Program, Grant PHY-1915005, and also in part by the NSF under Grant No. NSF PHY-1748958. M.A.\ is supported by a DST-INSPIRE Fellowship, Department of Science and Technology, Government of India. S.G.G.\ is supported by SERB-DST through project No.~CRG/2021/005771. R.K.W.\ acknowledges support from the United States-India Educational Foundation (USIEF) through a Fulbright-Nehru Postdoctoral Research Fellowship (award 2847/FNPDR/2022), and from the University of KwaZulu-Natal and Prof.\ Sunil Maharaj through a postdoctoral fellowship of the National Research Foundation of South Africa (NRF). A.\"{O}.\ thanks Prof.\ Durmu\c{s} Ali Demir for hospitality at Sabanc{\i} University while this work was being conducted. A. {\"O}.\ acknowledges the contribution of the COST Action CA18108 -- Quantum gravity phenomenology in the multi-messenger approach (QG-MM). C.B.\ is supported by the National Natural Science Foundation of China (NSFC) through grant No.~11973019, the Natural Science Foundation of Shanghai through grant No.~22ZR1403400, the Shanghai Municipal Education Commission through grant No.~2019-01-07-00-07-E00035, and Fudan University through grant No.~JIH1512604.
\end{acknowledgments}

\newpage

\bibliography{SgrA_tests_extended}

\end{document}